\documentclass[10pt,conference]{IEEEtran}

\usepackage[english]{babel}

\usepackage[cmex10]{amsmath}
\usepackage{amssymb}
\usepackage{graphicx}
\usepackage{stmaryrd}
\usepackage{ifthen}
\usepackage{url}
\usepackage{cite}
\usepackage{dblfloatfix}

\newcommand{\macrospath}{macros}

\usepackage[usenames,dvipsnames,svgnames,table]{xcolor}
\usepackage{bussproofs}
\usepackage{xspace}

\newcommand{\ignore}[1]{}
\newcommand{\sep}{\hspace*{0.5cm}}

\newcommand{\myinput}[1]{\ifthenelse{\boolean{withimages}}{\input{#1}}{}}

\newcommand{\reflemma}[1]{Lemma~\ref{l:#1}}
\newcommand{\reflemmap}[2]{Lemma~\ref{l:#1}.\ref{p:#1-#2}}
\newcommand{\reflemmaeq}[1]{{L.\ref{l:#1}}}
\newcommand{\reflemmaeqp}[2]{{L.\ref{l:#1}.\ref{p:#1-#2}}}
\newcommand{\refcorollary}[1]{Corollary~\ref{c:#1}}

\newcommand{\refth}[1]{Theorem~\ref{th:#1}}

\newcommand{\refthm}[1]{Theorem~\ref{thm:#1}}
\newcommand{\refthmp}[2]{Theorem~\ref{thm:#1}.\ref{p:#1-#2}}
\newcommand{\reftm}[1]{Theorem~\ref{tm:#1}}

\newcommand{\refprop}[1]{Proposition~\ref{prop:#1}}
\newcommand{\refsect}[1]{Sect.~\ref{sect:#1}}

\newcommand{\refssect}[1]{Subsect.~\ref{ssect:#1}}
\newcommand{\refapp}[1]{Appendix~\ref{sect:#1}}
\newcommand{\reftab}[1]{Table~\ref{tab:#1}}
\newcommand{\refeq}[1]{(\ref{eq:#1})}

\newcommand{\refcoro}[1]{Corollary~\ref{coro:#1}}
\newcommand{\refdef}[1]{Definition~\ref{def:#1}}
\newcommand{\refrem}[1]{Remark~\ref{rem:#1}}

\newcommand{\casealt}[1]{{\bf #1.}}
\newcommand{\caselight}[1]{\textit{#1}}

\newcommand{\ie}{\textit{i.e.}}
\newcommand{\eg}{\textit{e.g.}\xspace}
\newcommand{\ih}{\textit{i.h.}}

\newcommand{\Evaluable}{Evaluable}

\newcommand{\deff}[1]{\textbf{#1}}

\newcommand{\ben}[1]{{\red{#1}}}
\renewcommand{\ben}[1]{{#1}}

\newcommand{\claudio}[1]{#1}
\newcommand{\cclaudio}[2]{#2}

\newcommand{\defeq}{\mathrel{:=}}
\newcommand{\grameq}{\mathrel{::=}}
\newcommand{\set}[1]{\{#1\}}

\newcommand{\size}[1]{|#1|}

\newcommand{\LeftRightarrow}{\Lleftarrow\!\!\!\!\Rrightarrow}

\newcommand{\dbv}{{\tt dBv}}

\newcommand{\lsvsym}{{\tt lsv}}

\newcommand{\admsym}{{\mathtt c}}

\newcommand{\mulsym}{m}
\newcommand{\expsym}{e}

\newcommand{\bsym}{{\mathtt b}}

\renewcommand{\l}{\lambda}
\newcommand{\isub}[2]{\{#1/#2\}}
\renewcommand{\isub}[2]{\{#1{\shortleftarrow}#2\}}
\newcommand{\esub}[2]{[#1/#2]}
\renewcommand{\esub}[2]{[#1{\shortleftarrow}#2]}
\newcommand{\fv}[1]{{\tt fv}(#1)}

\newcommand{\rootRew}[1]{\mapsto_{#1}}
\newcommand{\Rew}[1]{\rightarrow_{#1}}

\newcommand{\rtodbv}{\rootRew{\db\vsym}}
\newcommand{\rtolsv}{\rootRew{\lssym\vsym}}

\newcommand{\towhl}{\stackrel{\mathtt{wh}}{\multimap}}

\renewcommand{\towhl}{\togen}

\newcommand{\towhlcek}{\stackrel{\mathtt{ns}}{\multimap}_\vsym}

\renewcommand{\towhlcek}{\togen}

\newcommand{\tostructsym}{\LeftRightarrow}

\newcommand{\esym}{{\mathtt e}}
\newcommand{\msym}{{\mathtt m}}
\newcommand{\fsym}{{\mathtt f}}
\newcommand{\psym}{{\mathtt p}}
\newcommand{\ssym}{{\mathtt s}}

\newcommand{\togen}{\multimap}
\newcommand{\togenm}{\multimap_\msym}
\newcommand{\togene}{\multimap_\esym}

\newcommand{\togenx}{\multimap_{\mathtt x}}

\newcommand{\tom}{\Rew{\msym}}

\newcommand{\eqstruct}{\equiv}
\newcommand{\tostruct}{\eqstruct}

\newcommand{\tostructap}{\tostruct_{@}}
\newcommand{\tostructapl}{\tostruct_{@l}}
\newcommand{\tostructapr}{\tostruct_{@r}}
\newcommand{\tostructes}{\tostruct_{[\cdot]}}
\newcommand{\tostructcom}{\tostruct_{com}}

\newcommand{\tm}{t}
\newcommand{\tmtwo}{u}
\newcommand{\tmthree}{w}
\newcommand{\tmfour}{r}
\newcommand{\tmfive}{q}
\newcommand{\tmsix}{p}

\newcommand{\tmp}{\tm'}
\newcommand{\tmtwop}{\tmtwo'}
\newcommand{\tmthreep}{\tmthree'}

\newcommand{\tmfivep}{\tmfive'}

\newcommand{\tmpp}{\tm''}

\newcommand{\var}{x}
\newcommand{\vartwo}{y}
\newcommand{\varthree}{z}
\newcommand{\varfour}{w}

\newcommand{\val}{v}
\newcommand{\valtwo}{\val'}

\newcommand{\valp}{v}
\newcommand{\valptwo}{\valp'}
\newcommand{\valpthree}{\valp''}

\newcommand{\valt}{v}

\newcommand{\ctxholep}[1]{\langle #1\rangle}
\newcommand{\ctxhole}{\ctxholep{\cdot}}

\newcommand{\sctx}{L}
\newcommand{\sctxtwo}{\sctx'}
\newcommand{\sctxthree}{\sctx''}
\newcommand{\sctxp}[1]{\sctx\ctxholep{#1}}
\newcommand{\sctxtwop}[1]{\sctxtwo\ctxholep{#1}}
\newcommand{\sctxthreep}[1]{\sctxthree\ctxholep{#1}}
\newcommand{\sctxOne}{\sctx_1}
\newcommand{\sctxTwo}{\sctx_2}
\newcommand{\sctxOnep}[1]{\sctxOne\ctxholep{#1}}
\newcommand{\sctxTwop}[1]{\sctxTwo\ctxholep{#1}}

\newcommand{\wctx}{W}
\newcommand{\wctxtwo}{\wctx'}

\newcommand{\wctxp}[1]{\wctx\ctxholep{#1}}

\newcommand{\arbctxp}[1]{\arbctxp{#1}}
\newcommand{\arbctxtwop}[1]{\arbctxtwop{#1}}

\newcommand{\genevctx}{F}
\newcommand{\genevctxtwo}{\genevctx'}
\newcommand{\genevctxp}[1]{\genevctx\ctxholep{#1}}
\newcommand{\genevctxtwop}[1]{\genevctxtwo\ctxholep{#1}}

\newcommand{\tctx}{S}
\newcommand{\tctxtwo}{\tctx'}
\newcommand{\tctxthree}{\tctx''}
\newcommand{\tctxfour}{\tctx'''}
\newcommand{\tctxfive}{\tctx_4}
\newcommand{\tctxp}[1]{\tctx\ctxholep{#1}}
\newcommand{\tctxtwop}[1]{\tctxtwo\ctxholep{#1}}
\newcommand{\tctxthreep}[1]{\tctxthree\ctxholep{#1}}

\newcommand{\tctxfivep}[1]{\tctxfive\ctxholep{#1}}

\newcommand{\tpctx}[2]{\startctx{#1}{#2}}
\newcommand{\tpctxp}[3]{\tpctx{#1}{#2}\ctxholep{#3}}

\newcommand{\tctxONE}{\tctx_1}
\newcommand{\tctxTWO}{\tctx_2}
\newcommand{\tctxONEtwo}{\tctx_1'}
\newcommand{\tctxTWOtwo}{\tctx_2'}

\newcommand{\tctxONEtwop}[1]{\tctxONEtwo\ctxholep{#1}}

\newcommand{\evctx}{\genevctx}
\newcommand{\evctxtwo}{\genevctxtwo}
\newcommand{\evctxthree}{V''}
\newcommand{\evctxp}[1]{\genevctxp{#1}}
\newcommand{\evctxtwop}[1]{\genevctxtwop{#1}}
\newcommand{\evctxthreep}[1]{\evctxthree\ctxholep{#1}}

\newcommand{\evctxONE}{\evctx_1}
\newcommand{\evctxTWO}{\evctx_2}
\newcommand{\evctxONEtwo}{\evctx_1'}
\newcommand{\evctxTWOtwo}{\evctx_2'}
\newcommand{\evctxONEp}[1]{\evctxONE\ctxholep{#1}}

\newcommand{\evctxONEtwop}[1]{\evctxONEtwo\ctxholep{#1}}

\newcommand{\sctxal}{\widehat{\sctx}}

\newcommand{\sctxpal}[1]{\sctxal\ctxholep{#1}}

\newcommand{\evctxal}{\widehat{\evctx}}

\newcommand{\evctxpal}[1]{\evctxal\ctxholep{#1}}

\newcommand{\tomachhole}[1]{\leadsto_{#1}}
\newcommand{\tomach}{\tomachhole{}}
\newcommand{\tomachm}{\tomachhole{\mulsym}}

\newcommand{\tomache}{\tomachhole{\expsym}}
\newcommand{\tomachei}{\tomachhole{\osym\esym\csym}}
\newcommand{\tomachee}{\tomachhole{\osym\esym\ssym}}
\newcommand{\tomacha}{\tomachhole{\admsym}}

\newcommand{\tomachx}{\tomachhole{\mathtt{x}}}

\newcommand{\admnf}[1]{\mathtt{nf}_{\admsym}(#1)}
\newcommand{\tomachc}{\tomachhole{\admsym}}
\newcommand{\tomachcp}[1]{\tomachhole{\admsym#1}}
\newcommand{\tomachcone}{\tomachhole{\admsym_1}}
\newcommand{\tomachctwo}{\tomachhole{\admsym_2}}
\newcommand{\tomachcthree}{\tomachhole{\admsym_3}}
\newcommand{\tomachcfour}{\tomachhole{\admsym_4}}
\newcommand{\tomachcfive}{\tomachhole{\admsym_5}}
\newcommand{\tomachcsix}{\tomachhole{\admsym_6}}

\newcommand{\tomachp}{\tomachhole{\psym}}
\newcommand{\tomachsm}{\tomachhole{\msym}}
\newcommand{\tomachse}{\tomachhole{\esym}}
\newcommand{\tomachum}{\tomachhole{\usym\msym}}
\newcommand{\tomachue}{\tomachhole{\usym\esym}}
\newcommand{\tomachom}{\tomachhole{\osym\msym}}

\newcommand{\code}{\overline{\tm}}
\newcommand{\codetwo}{\overline{\tmtwo}}

\newcommand{\codeval}{\overline{\val}}
\newcommand{\codevalt}{\overline{\valt}}
\newcommand{\codevalp}{\overline{\valp}}

\newcommand{\genv}{E}
\newcommand{\genvtwo}{E'}

\newcommand{\decgenv}{\decode{\genv}}

\newcommand{\decgenvpx}[2]{\ctxholep{#2}\decode{#1}}
\newcommand{\decgenvp}[1]{\decgenvpx\genv{#1}}

\newcommand{\econsx}[2]{#1\cons#2} 
\newcommand{\econs}[2]{#1#2} 

\newcommand{\stempty}{\epsilon}
\newcommand{\cons}{:}

\newcommand{\nfnst}[2]{(#1,#2)}

\newcommand{\stack}{\pi}
\newcommand{\stacktwo}{\pi'}

\newcommand{\decstack}{\decode{\pi}}

\newcommand{\decodestack}[2]{\ctxholep{#1}\decode{#2}}
\newcommand{\decstackp}[1]{\decodestack{#1}{\pi}}
\newcommand{\decstacktwop}[1]{\decodestack{#1}{\stacktwo}}

\newcommand{\heapempty}{\epsilon}

\newcommand{\decheapp}[1]{{#1}^{\heap}}
\newcommand{\decheappp}[2]{{#1}^{#2}}

\newcommand{\state}{s}
\newcommand{\statetwo}{s'}

\newcommand{\heap}{H}
\newcommand{\heaptwo}{H'}

\newcommand{\rename}[1]{#1^\alpha}

\newcommand{\exec}{\rho}

\newcommand{\decode}[1]{\llbracket #1\rrbracket}
\renewcommand{\decode}[1]{\underline{#1}}

\newcommand{\sizeb}[1]{\size{#1}_{\bsym}}

\newcommand{\sizeue}[1]{\size{#1}_{\usym\esym}}
\newcommand{\sizeuei}[1]{\size{#1}_{\osym\esym\csym}}
\newcommand{\sizeuee}[1]{\size{#1}_{\osym\esym\ssym}}
\newcommand{\sizeum}[1]{\size{#1}_{\osym\msym}}
\newcommand{\sizeoe}[1]{\size{#1}_{\osym\esym}}

\newcommand{\deriv}{d}
\newcommand{\derivtwo}{e}

\newcommand{\sizee}[1]{|#1|_{\esym}}
\newcommand{\sizem}[1]{|#1|_{\msym}}

\newcommand{\sizep}[1]{|#1|_p}

\newcommand{\distil}{{\tt D}}
\newcommand{\calculus}{{\tt C}}
\renewcommand{\calculus}{\togen}
\newcommand{\mach}{{\tt M}}

\newcommand{\dumpempty}{\epsilon}
\renewcommand{\dump}{D}
\newcommand{\dumptwo}{D'}

\newcommand{\decdump}{\decode \dump}
\newcommand{\decdumpp}[1]{\decdump \ctxholep{#1}}

\renewcommand{\rtodbv}{\rootRew{\msym}}
\renewcommand{\rtolsv}{\rootRew{\esym}}

\newcommand{\const}{a}
\newcommand{\consttwo}{b}
\newcommand{\constthree}{c}

\newcommand{\gconst}{A}
\newcommand{\gconsttwo}{B}
\newcommand{\gconstthree}{C}

\newcommand{\fire}{f}
\newcommand{\firetwo}{g}
\newcommand{\firethree}{h}

\newcommand{\rtof}{\rootRew{\fsym}}
\newcommand{\tof}{\Rew{\fsym}}

\newcommand{\pevctx}{C}
\newcommand{\pevctxtwo}{C'}

\newcommand{\pevctxp}[1]{\pevctx\ctxholep{#1}}
\newcommand{\pevctxtwop}[1]{\pevctxtwo\ctxholep{#1}}

\newcommand{\psctx}[1]{\pevctx_{#1}}

\newcommand{\psctxp}[2]{\psctx{#1}\ctxholep{#2}}

\newcommand{\ssctx}[1]{\sctx_{#1}}

\newcommand{\ssctxp}[2]{\ssctx{#1}\ctxholep{#2}}

\newcommand{\unfsym}{\rotatebox[origin=c]{-90}{$\rightarrow$}}
\newcommand{\unf}[1]{#1\unfsym}
\newcommand{\relunf}[2]{\unf{#1}_{#2}}

\newcommand{\alive}{\val}
\newcommand{\dead}{\gconst}

\newcommand{\lab}{l}

\newcommand{\fireball}{fireball}
\newcommand{\Fireball}{Fireball}

\newcommand{\dentry}[2]{(#1,#2)}

\newcommand{\usym}{{\mathtt u}}
\newcommand{\csym}{{\mathtt c}}

\newcommand{\osym}{{\mathtt o}}
\newcommand{\togens}{\multimap_{\fsym}}
\newcommand{\togensm}{\multimap_{\msym}}
\newcommand{\togense}{\multimap_{\esym}}

\newcommand{\togenfm}{\multimap_{\osym\msym}}
\newcommand{\togenfe}{\multimap_{\osym\esym}}
\newcommand{\togenfes}{\multimap_{\osym\esym\ssym}}
\newcommand{\togenfec}{\multimap_{\osym\esym\csym}}

\newcommand{\togenu}{\multimap_{\usym\fsym}}
\newcommand{\togenum}{\multimap_{\usym\msym}}
\newcommand{\togenue}{\multimap_{\usym\esym}}
\newcommand{\togenlu}{\multimap_{\osym\fsym}}
\newcommand{\togenlum}{\multimap_{\osym\msym}}
\newcommand{\togenlue}{\multimap_{\osym\esym}}
\newcommand{\togenluee}{\multimap_{\osym\esym\ssym}}
\newcommand{\togenluei}{\multimap_{\osym\esym\csym}}

\newcommand{\proper}{{proper}}
\newcommand{\Proper}{{Proper}}
\newcommand{\properness}{{properness}}

\newcommand{\refpoint}[1]{Point~\ref{p:#1}}
\newcommand{\refpointmute}[1]{\ref{p:#1}}

\newcommand{\la}[1]{\lambda #1.}

\newcommand{\ictx}{I}
\newcommand{\ictxtwo}{I'}

\newcommand{\ictxp}[1]{\ictx\ctxholep{#1}}
\newcommand{\ictxtwop}[1]{\ictxtwo\ctxholep{#1}}

\newcommand{\maca}[5]{#2 & #3 & #4 & #5 & #1}
\newcommand{\usefmaca}[4]{#2 & #3 & #4 & #1}
\newcommand{\mac}[5]{(#2,#3,#4,#5,#1)}
\newcommand{\usefmac}[4]{(#2,#3,#4,#1)}
\newcommand{\glam}[4]{(#2,#3,#4,#1)}
\newcommand{\glamst}[4]{(#1,#2,#3,#4)}

\newcommand{\sizecom}[1]{|#1|_c}
\newcommand{\sizecomphone}[1]{|#1|_{c_{1-5}}}
\newcommand{\sizecomsix}[1]{|#1|_{c_6}}

\newcommand{\basic}{basic}

\newcommand{\stackitem}{\phi}
\newcommand{\pair}[2]{(#1,#2)}
\newcommand{\herval}[1]{#1^\alive}
\newcommand{\decodeheap}[2]{\decheappp{#2}{#1}}

\newcommand{\decodep}[2]{\decode{#1}\ctxholep{#2}}

\newcommand{\myproof}[1]{
\ifthenelse{\boolean{omitproofs}}{\begin{IEEEproof} Proof available but omitted for readability. \end{IEEEproof}}{#1}}

\newcommand{\Micro}{Explicit}

\newcommand{\FC}{FBC}
\newcommand{\fb}{FBC}

\newcommand{\fast}{Unchaining}
\newcommand{\ShFC}{\Micro\ \FC}

\newcommand{\UsFC}{Useful \FC}
\newcommand{\FFC}{\fast\ \FC}

\newcommand{\GLAM}{GLAM}
\newcommand{\glamour}{\GLAM oUr}
\newcommand{\Fglamour}{\fast\ \glamour}

\newcommand{\decodefun}{\decode{{ }\cdot{ }}}

\renewcommand{\dbv}{\msym}
\renewcommand{\lsvsym}{\esym}

\newcommand{\symb}{symbol\xspace}
\newcommand{\Symb}{Symbol\xspace}
\newcommand{\symbs}{symbols\xspace}

\newcommand{\levy}{{L{\'e}vy}\xspace}

\newcommand{\startctx}[2]{\overleftarrow{#1}^{#2}}

\newcommand{\appl}{@}
\newcommand{\abst}{\Lambda}
\newcommand{\es}{\mathtt{ES}}
\newcommand{\node}{\mathtt{n}}

\usepackage{tikz}
\usetikzlibrary{matrix,arrows}
\usetikzlibrary{decorations.shapes}
\usetikzlibrary{decorations.text}
\usetikzlibrary{decorations.pathmorphing}
\usetikzlibrary{decorations.markings}

\tikzset{
node distance=1.3cm, auto,
every node/.style={font=\tiny },
ocenter/.style={baseline={([yshift=-.5ex, xshift=-.5ex]current bounding box)}},  
labelBeginAbove/.style={postaction={decorate,decoration={markings,mark=at position 0 with {\node[inner sep= 0.6pt, above=1pt]{\tiny #1};}} } },
labelBeginBelow/.style={postaction={decorate,decoration={markings,mark=at position 0 with {\node[inner sep= 0.6pt, below=1pt]{\tiny #1};}}}},
labelEndAbove/.style={postaction={decorate,decoration={markings,mark=at position 1 with {\node[inner sep= 0.6pt, above=1pt]{\tiny #1};}}}},
labelEndBelow/.style={postaction={decorate,decoration={markings,mark=at position 1 with {\node[inner sep= 0.6pt, below=1pt]{\tiny #1};}}}},
labelEndRight/.style={postaction={decorate,decoration={markings,mark=at position 1 with {\node[inner sep= 0.6pt, right=1pt]{\tiny #1};}}}},
labelEndLeft/.style={postaction={decorate,decoration={markings,mark=at position 1 with {\node[inner sep= 0.6pt, left=1pt]{\tiny #1};}}}}
}

\newcommand{\nodeHorDist}{2cm}
\newcommand{\nodeVerDist}{0.8cm}

\newcommand{\commutesredgenEEABCD}[8]{
      \begin{tikzpicture}[ocenter]
       \node (s) {\normalsize\ensuremath{#3}};
       \node at (s.center)  [right=1.7*\nodeHorDist](s1){\normalsize\ensuremath{#4}};
       \node at (s.center)  [below=\nodeVerDist](s2) {\normalsize\ensuremath{#5}};
       \node at (s2.center) [right=1.7*\nodeHorDist](t) {\normalsize\ensuremath{#6}};
       \node at (s.center)  [below=\nodeVerDist/2](eq1){\normalsize\ensuremath{#1}};
       \node at (s1.center) [below=\nodeVerDist/2](eq2){\normalsize\ensuremath{#2}};
       \draw[-o] (s) to node {$#7$} (s1);
       \draw[-o, dashed] (s2) to node {$#8$} (t);
      \end{tikzpicture} 
}

\newcommand{\commutesredEEABCD}[6]{
\commutesredgenEEABCD{#1}{#2}{#3}{#4}{#5}{#6}{}{}
}

\newcommand{\commutesdbvEEABCD}[6]{
\commutesredgenEEABCD{#1}{#2}{#3}{#4}{#5}{#6}{\dbv}{\dbv}
}

\newcommand{\commuteslsvEEABCD}[6]{
\commutesredgenEEABCD{#1}{#2}{#3}{#4}{#5}{#6}{\lsvsym}{\lsvsym}
}
\input{\macrospath/moveproofs.tex}

\renewcommand{\qedhere}{\hfill\IEEEQEDhere}

\newboolean{omitproofs}
\setboolean{omitproofs}{false}

\newboolean{withproofs}
\setboolean{withproofs}{true}

\begin{document}

\title{On the Relative Usefulness of Fireballs}

\author{\IEEEauthorblockN{Beniamino Accattoli}
\IEEEauthorblockA{INRIA \& LIX/\'Ecole Polytechnique\\
}
\and
\IEEEauthorblockN{Claudio Sacerdoti Coen}
\IEEEauthorblockA{
University of Bologna\\
}}

\maketitle

\begin{abstract}  
In CSL-LICS 2014, Accattoli and Dal Lago \cite{DBLP:conf/csl/AccattoliL14} showed that there is an implementation of the ordinary (i.e. strong, pure, call-by-name) $\lambda$-calculus into models like RAM machines which is polynomial in the number of $\beta$-steps, answering a long-standing question. The key ingredient was the use of a calculus with useful sharing, a new notion whose complexity was shown to be polynomial, but whose implementation was not explored. This paper, meant to be complementary, studies useful sharing in a call-by-value scenario and from a practical point of view. We introduce the \emph{Fireball Calculus}, a natural extension of call-by-value to open terms\ben{, that is an intermediary step towards the strong case, }and we present three results. First, we adapt and refine useful sharing, refining the meta-theory. Then, we introduce the \glamour\, a simple abstract machine implementing the Fireball Calculus extended with useful sharing. Its key feature is that usefulness of a step is tested---surprisingly---in constant time. Third, we provide a further optimisation that leads to an implementation having only a linear overhead with respect to the number of $\beta$-steps.
\end{abstract}

\section{Introduction}
The $\lambda$-calculus is an interesting computational model because it is machine-independent, simple to define, and it compactly models functional and higher-order logic programming languages. Its definition has only one rule, the $\beta$ rule, and no data structures. The catch is the fact that the $\beta$-rule---which by itself is Turing-complete---is not an atomic rule. Its action, namely $(\la\var\tm)\tmtwo \to_{\beta} \tm\isub\var\tmtwo$, can make many copies of an arbitrarily big sub-program $\tmtwo$. In other computational models like Turing or RAM machines, an atomic operation can only move the head on the ribbon or access a register. Is $\beta$ atomic in that sense? Can one count the number of $\beta$-steps to the result and then claim that it is a reasonable bound on the complexity of the term? Intuition says no, because $\beta$ can be nasty, and make the program grow at an exponential rate. This is the \emph{size explosion problem}.

\paragraph*{Useful Sharing} nonetheless, it is possible to take the number of $\beta$-steps as an invariant cost model, \ie\ as a complexity measure polynomially related to RAM or Turing machines. While this was known for some notable sub-calculi \cite{DBLP:conf/fpca/BlellochG95,DBLP:conf/birthday/SandsGM02,DBLP:journals/tcs/LagoM08,DBLP:journals/corr/abs-1208-0515,DBLP:conf/rta/AccattoliL12}, the first proof for the general case is a recent result by Accattoli and Dal Lago \cite{DBLP:conf/csl/AccattoliL14}. Similarly to the literature, they circumvent size explosion by factoring the problem via an intermediary model in between $\lambda$-calculus and machines. Their model is the \emph{linear substitution calculus} (LSC) \cite{DBLP:conf/popl/AccattoliBKL14,DBLP:conf/csl/AccattoliL14}, that is a simple $\lambda$-calculus with sharing annotations (also known as explicit substitutions) where the substitution process is decomposed in micro steps, replacing one occurrence at a time. In contrast with the literature, the general case is affected by a stronger form of size explosion, requiring an additional and sophisticated layer of sharing, called \emph{useful sharing}. Roughly, a micro substitution step is \emph{useful} if it contributes somehow to the creation of a $\beta$-redex, and \emph{useless} otherwise. Useful reduction then selects only useful substitution steps, avoiding the useless ones. In \cite{DBLP:conf/csl/AccattoliL14}, the Useful LSC is shown to be polynomially related to both $\l$-calculus (in a quadratic way) and RAM machines (with polynomial overhead, conjectured linear). It therefore follows that there is a polynomial relationship $\lambda \to \mbox{RAM}$. Pictorially:
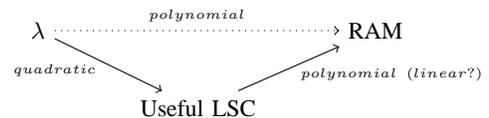
\begin{center}
      \begin{tikzpicture}[ocenter]
       \node (l) {\small $\lambda$};
       \node at (l.center)  [right=2*\nodeHorDist](ram){\small RAM};
       \node at (l.center)  [right=\nodeHorDist](ghost) {};
       \node at (ghost.center)  [below =\nodeVerDist](ulsc) {\small Useful LSC};
       \draw[->, dotted] (l) to node {$polynomial$} (ram);
       \draw[->] (l) to node[below = 2pt, left =1pt] {$quadratic$} (ulsc);
       \draw[->] (ulsc) to node[below = 2pt, right =1pt, overlay] {{$polynomial\ \ (linear?)$}} (ram);
      \end{tikzpicture} 
\end{center}

Coming back to our questions, the answer of \cite{DBLP:conf/csl/AccattoliL14} is yes, $\beta$ is atomic, up to a polynomial overhead. It is natural to wonder how big this overhead is. Is $\beta$ reasonably atomic? Or is the degree of the polynomial big and  does the invariance result only have a theoretical value? In particular, in \cite{DBLP:conf/csl/AccattoliL14} the definition of useful steps relies on a \emph{separate} and \emph{global} test for usefulness, that despite being tractable might not be feasible in practice. Is there an efficient way to implement the Useful LSC? Does useful sharing---\ie\ the avoidance of useless duplications---bring a costly overhead? This paper answers these questions. But, in order to stress the practical value of the study, it shifts to a slightly different setting. 

\paragraph*{The Fireball Calculus} we recast the problem in terms of the new \emph{fireball calculus} (\FC), essentially the weak call-by-value $\l$-calculus generalised to handle open terms. It is an intermediary step towards a strong call-by-value $\l$-calculus, that can be seen as iterated open weak evaluation. A similar approach to strong evaluation is followed also by Gr\'egoire and Leroy in \cite{DBLP:conf/icfp/GregoireL02}. \ben{It avoids some of the complications of the strong case, and at the same time exposes all the subtleties of dealing with open terms}.


Free variables are actually formalised using a distinguished syntactic class, that of \emph{\symbs}, noted $\const,\consttwo,\constthree$. This approach is technically convenient because it allows restricting to closed terms, so that any variable occurrence $\var$ is bound somewhere, while still having a representation of free variables (as \symbs).

The basic idea is that---in the presence of \symbs---restricting $\beta$-redex to \emph{fire} only in presence of values is problematic. Consider indeed the following term:
$$\tm\defeq ((\la\var\la\vartwo\tmtwo)(\const\const))\tmthree$$
where $\tmthree$ is normal. For the usual call-by-value operational semantics $\tm$ is normal (because $\const\const$ is not a value) while for theoretical reasons (see \cite{DBLP:journals/ita/PaoliniR99,DBLP:conf/flops/AccattoliP12,DBLP:conf/fossacs/CarraroG14}) one would like to be able to fire the blocked redex, reducing to $(\la\vartwo\tmtwo\isub\var{\const\const})\tmthree$, so that a new redex is created and the computation can continue. According to the standard classification of redex creations due to \levy \cite{thesislevy}, this is a creation of type 1\footnote{\ben{The reader unfamiliar with redex creations should not worry. Creations are a key concept in the study of usefulness---which is why we mention them---but for the present discussion it is enough to know that there exists two kinds of creations (type 1 and the forthcoming type 3, other types will not play a role), no expertise on creations is required}.}.

The solution we propose here is to relax the constraint about values, allowing $\beta$-redexes to fire whenever the argument is a more general structure, a so-called \emph{fireball}, defined recursively by extending values with \emph{inerts}, \ie\ applications of \symbs to fireballs. In particular, $\const\const$ is inert, so that \eg $\tm \to (\la\vartwo\tmtwo\isub\var{\const\const})\tmthree$, as desired.

Functional programming languages are usually modelled by weak and \emph{closed} calculi, so it is natural to wonder about the practical relevance of the \FC. Applications are along two axes. On the one hand, the evaluation mechanism at work in proof assistants has to deal with open terms for comparison and unification. For instance, Gregoire and Leroy's \cite{DBLP:conf/icfp/GregoireL02}, meant to improve the implementation of Coq, relies on inerts (therein called \emph{accumulators}). On the other hand, \symbs may also be interpreted as \emph{constructors}, meant to represent data as lists or trees. The dynamics of fireballs is in fact consistent with the way constructors are handled by Standard ML~\cite{citeulike:113339} and in several formalisation of core ML, as in~\cite{Clement:1986:SAL:319838.319847}. In this paper we omit destructors, whose dynamics is orthogonal to the one of $\beta$-reduction, and we expect all results presented here to carry-over with minor changes to a calculus with destructors. Therefore firing redexes involving inerts is also justified from a practical perspective. 

\paragraph*{The Relative Usefulness of Fireballs} as we explained, the generalisation of values to fireballs is motivated by creations of type 1 induced by the firing of inerts. There is a subtlety, however. While \emph{substituting} a value can create a new redex (\eg\ as in $(\la\var(\var I)) I \to (\var I) \isub\var I = II$, where $I$ is the identity---these are called creations of type 3), substituting a inert can not. Said differently, duplicating inerts is useless, and leads to size explosion. Note the tension between different needs: redexes involving inerts have to be fired (for creations of type 1), and yet the duplication and the substitution of inerts should be avoided (since they do not give rise to creations of type 3). \ben{We solve the tension by turning to sharing, and use the simplicity of the framework to explore the implementation of useful sharing}. Both values and inerts (\ie\ fireballs) in argument position will trigger reduction, and both will be shared after the redex is reduced, but only the substitution of values might be useful, because inerts are always useless. This is what we call \emph{the relative usefulness of fireballs}. It is also why---in contrast to Gregoire and Leroy---we do not identify fireballs and values.

\paragraph*{The Result} our main result is an implementation of \FC\ relying on useful sharing and such that it has only a linear overhead with respect to the number of $\beta$-steps. To be precise, the overhead is \emph{bilinear}, \ie\ linear in the number of $\beta$-steps \emph{and} in the size of the initial term (roughly the size of the input). The dependency from the size of the initial term is induced by the action of $\beta$ on whole subterms, rather than on atomic pieces of data as in RAM or Turing machines. Therefore, $\beta$ is not exactly as atomic as accessing a register or moving the head of a Turing machine, and this is the price one must pay for embracing higher-order computations. Bilinearity, however, guarantees that such a price is mild and that the number of $\beta$ step---\ie\ of function calls in a functional program---is a faithful measure of the complexity of a program\cclaudio{SI POTREBBE CITARE UNO DEI TANTI ARTICOLI, PER ESEMPIO CE NE ERA UNO A POPL, CHE USA QUESTO COST MODEL E CHE, ORA, \`E GIUSTIFICATO}{}. To sum up, our answer is yes, $\beta$ is also reasonably atomic.

\paragraph*{A Recipe for Bilinearity, with Three Ingredients} our proof technique is a \emph{tour de force} progressively combining together and adapting to the \FC\ three recent works involving the LSC, namely the already cited invariance of useful sharing of \cite{DBLP:conf/csl/AccattoliL14}, the tight relationship with abstract machines developed by Accattoli, Barenbaum, and Mazza in \cite{DBLP:conf/icfp/AccattoliBM14}, and the optimisation of the substitution process studied by the present authors in \cite{DBLP:conf/wollic/AccattoliC14}. The next section will give an overview of these works and of how they are here combined, stressing how the proof is more than a simple stratification of techniques. In particular, it was far from evident that the orthogonal solutions introduced by these works could be successfully combined together.

\paragraph*{This Paper} the paper is meant to be self-contained, and mostly follows a didactic style. For the first half we warm up by discussing  design choices, the difficulty of the problem, and the abstract architecture. The second half focuses on the results. We also suggest reading the introductions of \cite{DBLP:conf/csl/AccattoliL14,DBLP:conf/icfp/AccattoliBM14,DBLP:conf/wollic/AccattoliC14}, as they provide intuitions about concepts that here are only hinted at. Although not essential, they will certainly soften the reading of this work. Omitted proofs are in the appendix and related work is discussed in \refsect{related-work}.

\section{A Recipe with Three Ingredients}
\label{sect:recipe}
This section gives a sketch of how the bilinear implementation is built by mixing together tools from three different studies on the LSC.

\paragraph*{1) Useful Fireballs} we start by introducing the Useful Fireball Calculus (Useful \FC), akin to the Useful LSC, and provide the proof that the relationship \FC\ $\to$ \UsFC, analogously to the arrow $\l$ $\to$ Useful LSC, has a quadratic overhead. Essentially, this step provides us with the following diagram:
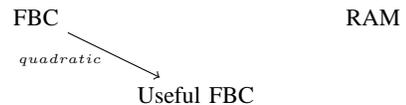
\begin{center}
      \begin{tikzpicture}[ocenter]
       \node (l) {\small \FC};
       \node at (l.center)  [right=2*\nodeHorDist](ram){\small RAM};
       \node at (l.center)  [right=\nodeHorDist](ghost) {};
       \node at (ghost.center)  [below =\nodeVerDist](ulsc) {\small Useful \FC};
       \draw[->] (l) to node[below = 2pt, left =1pt] {$quadratic$} (ulsc);
      \end{tikzpicture} 
\end{center}
We go beyond simply adapting the study of \cite{DBLP:conf/csl/AccattoliL14}, as the use of evaluation contexts (typical of call-by-value scenarios) leads to the new notion of \emph{useful evaluation context}, that simplifies the technical study of useful sharing. Another key point is the \emph{relative usefulness of fireballs}, according to their nature: only values are properly subject to the useful discipline, \ie\ are duplicated only when they contribute somehow to $\beta$-redexes, while inerts are never duplicated.

\paragraph*{2) Distilling Useful Fireballs} actually, we do not follow \cite{DBLP:conf/csl/AccattoliL14} for the study of the arrow \UsFC\ $\to$ RAM. We rather refine the whole picture, by introducing a further intermediary model, an \emph{abstract machine}, mediating between the Useful \FC\ and RAM. We adopt the \emph{distillation technique} of \cite{DBLP:conf/icfp/AccattoliBM14}, that establishes a fine-grained and modular view of abstract machines as strategies in the LSC up to a notion of structural equivalence on terms. The general pattern arising from \cite{DBLP:conf/icfp/AccattoliBM14} is that for call-by-name/value/need weak and closed calculi the abstract machine adds only a bilinear overhead with respect to the shared evaluation within the LSC: 
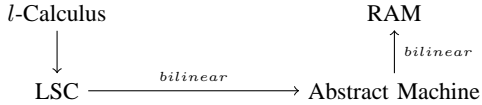
\begin{center}
      \begin{tikzpicture}[ocenter]
       \node (fc) {\small $\l$-Calculus};
       \node at (fc.center)  [right=2*\nodeHorDist](ram){\small RAM};
       \node at (fc.center)  [below =\nodeVerDist](ufc) {\small LSC};
       \node at (ram.center)  [below =\nodeVerDist](glam) {\small Abstract Machine};
       \draw[->] (ufc) to node {$bilinear$} (glam);
       \draw[->] (fc) to (ufc);
       \draw[->] (glam) to node[right] {$bilinear$} (ram);
      \end{tikzpicture} 
\end{center}
Such \emph{distilleries} owe their name to the fact that the LSC retains only part of the dynamics of a machine. Roughly, it isolates the relevant part of the computation, distilling it away from the search for the next redex implemented by abstract machines. The search for the redex is mapped to a notion of  structural equivalence, a particular trait of the LSC, whose key property is that it can be postponed. Additionally, the transitions implementing the search for the next redex are proved to be bilinear in those simulated by the LSC: the LSC then turns out to be a complexity-preserving abstraction of abstract machines.

The second ingredient for the recipe is then a new abstract machine, called \glamour, that we prove implements the Useful \FC\ within a bilinear overhead. Moreover, the \glamour\ itself can be implemented within a bilinear overhead. Therefore, we obtain the following diagram:
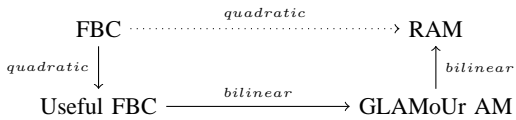
\begin{center}
      \begin{tikzpicture}[ocenter]
       \node (fc) {\small \FC};
       \node at (fc.center)  [right=2*\nodeHorDist](ram){\small RAM};
       \node at (fc.center)  [below =\nodeVerDist](ufc) {\small Useful \FC};
       \node at (ram.center)  [below =\nodeVerDist](glam) {\small \glamour\ AM};
       \draw[->, dotted] (fc) to node {$quadratic$} (ram);
       \draw[->] (ufc) to node {$bilinear$} (glam);
       \draw[->] (fc) to node[left] {$quadratic$} (ufc);
       \draw[->] (glam) to node[right] {$bilinear$} (ram);
      \end{tikzpicture} 
\end{center}
This is the most interesting and original step of our study. First, it shows that distilleries are compatible with open terms and useful sharing. Second, while in \cite{DBLP:conf/icfp/AccattoliBM14} distilleries were mainly used to revisit machines in the literature, here the distillation principles are used to guide the design of a new abstract machine. Third, useful sharing is handled via a refinement of an ordinary abstract machine relying on a basic form of labelling. The most surprising fact is that such a labelling (together with invariants induced by the call-by-value scenario) allows a straightforward and very efficient implementation of useful sharing. While the calculus is based on \emph{separate} and \emph{global} tests for the usefulness of a substitution step, the labelling allows the machine to do \emph{on-the-fly} and \emph{local} tests, requiring only constant time (!). It then turns out that implementing usefulness is much easier than analysing it. Summing up, useful sharing is easy to implement and thus a remarkable case of a theoretically born concept with relevant practical consequences.

\paragraph*{3) Unchaining Substitutions}
at this point, it is natural to wonder if the bottleneck given by the side of the diagram \FC\ $\to$ \UsFC, due to the overhead of the decomposition of substitutions, can be removed. The bound on the overhead is in fact tight, and yet the answer is yes, if one refines the actors of the play. Our previous work \cite{DBLP:conf/wollic/AccattoliC14}, showed that (in ordinary weak and closed settings) the quadratic overhead is due to malicious chains of \emph{renamings}, \ie\ of substitutions of variables for variables, and that the substitution overhead reduces to linear if the evaluation is modified so that variables are never substituted, \ie\ if values do not include variables.

For the fireball calculus the question is tricky. First of all a disclaimer: with \emph{variables} we refer to occurrences of bound variables and not to \symbs/free variables. Now, our initial definition of the calculus will exclude variables from fireballs, but useful sharing will force us to somehow reintroduce them. Our way out is an optimised form of substitution that \emph{unchains} renaming chains, and whose overhead is proved linear by a simple amortised analysis. Such a third ingredient is first mixed with both the \UsFC\ and the \glamour, obtaining the \FFC\ and the \Fglamour, and then used to prove our main result, an implementation \FC\ $\to$ RAM having an overhead linear in the number of $\beta$ steps and in the size of the initial term: 

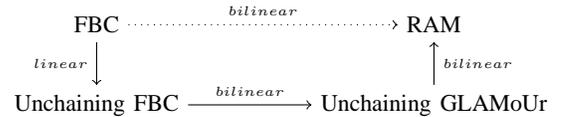
\begin{center}
      \begin{tikzpicture}[ocenter]
       \node (fc) {\small \FC};
       \node at (fc.center)  [right=2*\nodeHorDist](ram){\small RAM};
       \node at (fc.center)  [below =\nodeVerDist](ufc) {\small \FFC};
       \node at (ram.center)  [below =\nodeVerDist](glam) {\small \Fglamour};
       \draw[->, dotted] (fc) to node {$bilinear$} (ram);
       \draw[->] (ufc) to node {$bilinear$} (glam);
       \draw[->] (fc) to node[left] {$linear$} (ufc);
       \draw[->] (glam) to node[right] {$bilinear$} (ram);
      \end{tikzpicture} 
\end{center}
In this step, the original content is that the unchaining optimisation---while inspired by \cite{DBLP:conf/wollic/AccattoliC14}---is subtler to define than in \cite{DBLP:conf/wollic/AccattoliC14}, as bound variables cannot be simply removed from the definition of fireballs, because of usefulness. Moreover, we also show how such an optimisation can be implemented at the machine level.\medskip

The next section discusses related work. Then there will be a long preliminary part providing basic definitions, an abstract decomposition of the implementation, and a quick study of both a calculus, the \ShFC, and a machine, the \GLAM, without useful sharing. Both the calculus and the machine will not have any good asymptotical property, but they will be simple enough to familiarise the reader with the framework and with the many involved notions.

\section{Related Work}
\label{sect:related-work}
In the literature, invariance results for the weak
call-by-value $\l$-calculus have been proved three times,
independently. First, by Blelloch and Greiner
\cite{DBLP:conf/fpca/BlellochG95}, while studying cost models for
parallel evaluation. Then by Sands, Gustavsson and Moran
\cite{DBLP:conf/birthday/SandsGM02}, while studying speedups for
functional languages, and finally by Dal Lago and Martini
\cite{DBLP:journals/tcs/LagoM08}, who addressed the invariance thesis
for $\l$-calculus. Among them, \cite{DBLP:conf/birthday/SandsGM02} is the closest one, as it also provides an abstract machine and bounds its overhead. These works however concern closed terms, and so they deal with a much simpler case. Other simple call-by-name cases are studied in \cite{DBLP:journals/corr/abs-1208-0515} and \cite{DBLP:conf/rta/AccattoliL12}. The difficult case of the strong $\l$-calculus has been studied in \cite{DBLP:conf/csl/AccattoliL14}, which is also the only reference for useful sharing. 

The LSC is a variation over a $\l$-calculus with ES by Robin Milner \cite{DBLP:journals/entcs/Milner07,KesnerOConchuir}, obtained by plugging in some of the ideas of the structural $\l$-calculus by Accattoli and Kesner \cite{DBLP:conf/csl/AccattoliK10}, introduced  as a syntactic reformulation of linear
logic proof nets. The LSC is similar to calculi studied by De Bruijn~\cite{deBruijn87} and Nederpelt~\cite{Ned92}. Its first appearances in the literature are in \cite{DBLP:conf/rta/AccattoliL12,DBLP:conf/rta/Accattoli12}, but its inception is actually due to Accattoli and Kesner.

Many abstract machines can be rephrased as strategies in \emph{$\l$-calculi with explicit substitutions} (ES), see at least \cite{DBLP:journals/tcs/Curien91,DBLP:journals/jfp/HardinM98,DBLP:journals/tocl/BiernackaD07,DBLP:journals/lisp/Lang07,DBLP:journals/lisp/Cregut07,DBLP:journals/toplas/AriolaBS09}.

The related work that is by far closer to ours is the already cited study by Gr\'egoire and Leroy of an abstract machine for call-by-value weak and open reduction in \cite{DBLP:conf/icfp/GregoireL02}. We developed our setting independently, and yet the \FC\ is remarkably close to their calculus, in particular our \emph{inerts} are essentially their \emph{accumulators}. The difference is that our work is complexity-oriented while theirs is implementation-oriented. On the one hand they do not recognise the relative usefulness of fireballs, and so their machine is not invariant, \ie\ our machine is more efficient and on some terms even exponentially faster. On the other hand, they extend the language up to the calculus of constructions, present a compilation to bytecode, and certify in Coq the correctness of the implementation.

The abstract machines in this paper use \emph{global} environments, an approach followed only by a minority of authors (\eg\  
\cite{DBLP:conf/birthday/SandsGM02,DBLP:journals/entcs/FernandezS09,DBLP:conf/ppdp/DanvyZ13,DBLP:conf/icfp/AccattoliBM14}) and essentially identifying the environment with a store.

The distillation technique was developed to better understand the relationship between the KAM and weak linear head reduction pointed out by Danos \& Regnier \cite{Danos04headlinear}. The idea of distinguishing between \emph{operational content} and \emph{search for the redex} in an abstract machine is not new, as it underlies in particular the \emph{refocusing semantics} of Danvy and Nielsen \cite{Danvy04refocusingin}. Distilleries however bring an original refinement where logic, rewriting, and complexity enlighten the picture, leading to formal bounds on machine overheads.

Our unchaining optimisation is a lazy variant of an optimisation that repeatedly appeared in the literature on abstract machines, often with reference to space consumption and \emph{space leaks}, for instance in \cite{DBLP:conf/birthday/SandsGM02} as well as in Wand's \cite{DBLP:journals/lisp/Wand07} (section 2), Friedman et al.'s \cite{DBLP:journals/lisp/FriedmanGSW07} (section 4), and Sestoft's \cite{DBLP:journals/jfp/Sestoft97} (section 4).

\section{The \Fireball\ Calculus}\label{sect:fireballcalc}
\begin{atend}\section{Proofs Omitted From \refsect{fireballcalc}\\ (The \Fireball\ Calculus)}\end{atend}
The setting is the one of the call-by-value $\lambda$-calculus extended with \symbs $\const,\consttwo,\constthree$, meant to denote free variables (or constructors). The syntax is:
\begin{center}
$\begin{array}{lllllllllllll}
	\mbox{Terms} & \tm,\tmtwo,\tmthree,\tmfour &\grameq& \var\mid \const \mid \l\var.\tm \mid \tm\tmtwo\\
	\mbox{Values} & \val,\valtwo & \grameq & \l \var.\tm\\
\end{array}
$\end{center}
with the usual notions of free and bound variables, capture-avoiding substitution $\tm\isub\var\tmtwo$, and closed (\ie\ without free variables) term. We will often restrict to consider closed terms, the idea being that an open term as $\var (\la\vartwo\varthree\vartwo)$ is rather represented as the closed term $\const (\la\vartwo\consttwo\vartwo)$.


The ordinary (\ie\ without \symbs) call-by-value $\l$-calculus has a nice operational characterisation of values: 
\begin{center}
\emph{closed normal forms are values}
\end{center}

Now, the introduction of \symbs breaks this property, because there are closed normal forms as $\const (\la\var\var)$ that are not values. In order to restore the situation, we generalise values to \emph{\fireball s\footnote{About \emph{\fireball}: the first choice was \emph{fire-able}, but then the spell checker suggested \emph{\fireball}.}}, that are either values $\val$ or \emph{inerts} $\gconst$, \ie\ \symbs possibly applied to fireballs. Associating to the left, fireballs and inerts are compactly defined by 
$$\begin{array}{lllllllllllll}
	\mbox{Fireballs} & \fire, \firetwo, \firethree & \grameq & \val \mid \gconst\\
	\mbox{Inerts} & \gconst,\gconsttwo, \gconstthree & \grameq &  \const \fire_1\ldots \fire_n & n\geq 0
\end{array}$$
For instance, $\la\var\vartwo$ and $\const$ are fireballs, as well as $\const(\la\var\var)$, $\const\consttwo$, and $(\const(\la\var\var))(\consttwo\constthree) (\la\vartwo(\varthree\vartwo))$. Fireballs can also be defined more atomically by mixing values and inerts as follows:
$$\begin{array}{lllllllllllll}
	\fire & \grameq & \val \mid \gconst&&&&
	\gconst & \grameq & \const \mid \gconst \fire
\end{array}$$
Note that $\gconst\gconsttwo$ and $\gconst\gconst$ are always inerts.

Next, we generalise the call-by-value rule $(\la\var\tm)\val \to_{\beta_\val} \tm\isub\var\val$ to substitute fireballs $\fire$ rather than values $\val$. First of all, we define a notion of evaluation context (noted $\evctx$ rather than $E$, reserved to forthcoming global environments), mimicking right-to-left CBV evaluation:
\begin{center}$\begin{array}{lllllllllllll}
	\mbox{Evaluation Contexts} & \evctx &\grameq & \ctxhole\mid \tm\evctx\mid \evctx\fire
\end{array}$\end{center}
note the case $\evctx\fire$, that in CBV would be $\evctx\val$. Last, we define the $\fsym$(fireball) rule $\tof$ as follows
$$\begin{array}{c@{\hspace{1.5cm}}c}
  \textsc{Rule at Top Level} & \textsc{Contextual closure} \\
	(\l\var.\tm)\fire\rtof \tm\isub\var\fire &
        \evctxp \tm \tof \evctxp \tmtwo \textrm{~~~if } \tm \rtof \tmtwo \\
\end{array}$$

Our definitions lead to:

\begin{atend}

The following lemmas are required to prove \refthm{firstmain}.

\begin{lemma}\label{l:closedtofireball}
Let $\tm$ be a closed $\tof$-normal term. Then $\tm$ is a \fireball.
\end{lemma}

\begin{IEEEproof}
	 by induction on $\tm$. Cases:
	\begin{enumerate}
		\item \emph{Variable}. Impossible, because $\tm$ is closed.
		\item \emph{\Symb} and \emph{Abstraction}. Then $\tm$ is a \fireball.
		\item \emph{Application}. Then $\tm = \tmtwo\tmthree$, with $\tmtwo$ and $\tmthree$ both closed and $\tof$-normal. By \ih\ they are both \fireball s. Moreover, $\tmtwo$ cannot be a value, otherwise $\tm$ would not be $\tof$-normal. Then it is a inert and $\tm$ is a fireball.\qedhere
	\end{enumerate}
\end{IEEEproof}

\begin{lemma}
	\label{l:gconst-or-fire-fnorm}
	Let $\tm$ be a inert or a \fireball. Then $\tm$ is $\tof$-normal.
\end{lemma}

\begin{IEEEproof}
by induction on $\tm$. If $\tm$ is a value $\val$ or a \symb $\const$ then it is $\tof$-normal. Otherwise $\tm = \gconst \fire$ and by \ih\ both $\gconst$ and $\fire$ are $\tof$-normal. Since $\gconst$ cannot be an abstraction, $\tm$ is $\tof$-normal.
\end{IEEEproof}

\begin{lemma}[Determinism of $\tof$]
	\label{l:determ-tof}
	Let $\tm$ be a term. Then $\tm$ has at most one $\tof$ redex.
\end{lemma}

\begin{IEEEproof}
	 by induction on $\tm$. Cases:
	\begin{enumerate}
		\item \emph{Variable}, \emph{\Symb}, or \emph{Abstraction}. No redexes.
		\item \emph{Application $\tm = \tmtwo\tmthree$}. By \ih, there are two cases for $\tmthree$:
		\begin{enumerate}
			\item \emph{$\tmthree$ has exactly one $\tof$ redex}. Then $\tm$ has a $\tof$ redex, because $\tmtwo\ctxhole$ is an evaluation context. Moreover, no $\tof$ redex for $\tm$ can lie in $\tmtwo$, because by \reflemma{gconst-or-fire-fnorm} $\tmthree$ is not a \fireball, and so $\ctxhole\tmthree$ is not an evaluation context.
			\item \emph{$\tmthree$ has no $\tof$ redexes}. If $\tmthree$ is not a \fireball\ then $\tm$ has no redexes, because $\ctxhole\tmthree$ is not an evaluation context. If $\tmthree$ is a \fireball\ we look at $\tmtwo$. By \ih, there are two cases:
			\begin{enumerate}
				\item \emph{$\tmtwo$ has exactly one $\tof$ redex}. Then $\tm$ has a $\tof$ redex, because $\ctxhole\tmthree$ is an evaluation context and $\tmthree$ is a \fireball. Uniqueness comes from the fact that $\tmthree$ has no $\tof$ redexes.
				\item \emph{$\tmtwo$ has no $\tof$ redexes}. If $\tmtwo$ is not a \fireball\ (and thus not a value) then $\tm$ has no redexes. If $\tmtwo$ is a \fireball\ there are two cases:
				\begin{itemize}
					\item \emph{$\tmtwo$ is a inert $\gconst$}. Then $\tm$ is a \fireball. 
					\item \emph{$\tmtwo$ is a value $\l\var.\tmfour$}. Then $\tm = (\l\var.\tmfour) \tmthree$ is a $\tof$ redex, because $\tmthree$ is a \fireball. Moreover, there are no other $\tof$ redexes, because evaluation does not go under abstractions and $\tmthree$ is a \fireball.\qedhere
				\end{itemize}
			\end{enumerate}
		\end{enumerate}
	\end{enumerate}
\end{IEEEproof}

\end{atend}

\begin{theorem}\label{thm:firstmain}\hfill
\begin{enumerate}
	\item \label{p:firstmain-two}Closed normal forms are fireballs.
	\item $\tof$ is deterministic.	
\end{enumerate}
\end{theorem}
\begin{proofatend}
by \reflemma{determ-tof} and \reflemma{closedtofireball}.
\end{proofatend}

In the introduction we motivated the notion of fireball both from theoretical and practical points of view. \refthmp{firstmain}{two} provides a further, strong justification: it expresses a sort of internal harmony of the \FC, allowing to see it as the canonical completion of call-by-value to the open setting.

\begin{atend}

The following easy properties of substitution will be needed later.

\begin{lemma}\hfill
	\label{l:tof-and-subs}
	\begin{enumerate}
		\item \label{p:tof-and-subs-one}\emph{Substitutions Commute}:\\ $\tm\isub\var\tmtwo\isub\vartwo\tmthree=\tm\isub\vartwo\tmthree\isub\var{\tmtwo\isub\vartwo\tmthree}$;
		\item \label{p:tof-and-subs-two} \emph{\Fireball s are Stable by Substitution}: 
		\begin{enumerate}
			\item \label{p:tof-and-subs-two-one} If $\tmtwo$ is a inert then $\tmtwo\isub\var\tm$ is a inert, and
			\item \label{p:tof-and-subs-two-two} if  $\tmtwo$ is a \fireball\ then $\tmtwo\isub\var\tm$ is a \fireball.
		\end{enumerate}
		\item \emph{$\tof$ and Substitution Commute}: if $\evctxp\tm \tof \evctxp\tmtwo$ with $\tm \rtof \tmtwo$ then $\evctxp\tm\isub\var\tmthree \tof \evctxp\tmtwo\isub\var\tmthree$ with $\tm\isub\var\tmthree \rtof \tmtwo\isub\var\tmthree$.

	\end{enumerate}
\end{lemma}

\begin{IEEEproof}\hfill
	\begin{enumerate}
		\item By induction on $\tm$.

		\item By induction on $\tmtwo$.		
		\begin{enumerate}
			\item $\tmtwo$ is a inert. Cases:
			\begin{enumerate}
				\item If $\tmtwo = \const$ then $\tmtwo\isub\var\tm = \const\isub\var\tm = \const$ is a inert.
				\item If $\tmtwo = \sctxtwop\gconst \sctxthreep\fire$ then by \ih\ $\sctxtwop\gconst\isub\var\tm$ is a inert and $\sctxthreep\fire\isub\var\tm$ is a \fireball, and so $\tmtwo\isub\var\tm = (\sctxtwop\gconst \sctxthreep\fire)\isub\var\tm = \sctxtwop\gconst\isub\var\tm \sctxthreep\fire\isub\var\tm$ is a inert.
			\end{enumerate}
			\item \emph{$\tmtwo$ is a \fireball}. Cases:
			\begin{enumerate}
				\item \emph{$\tmtwo$ is a value} $\la\var\tmthree$. Then $\tmtwo\isub\var\tm = \la\var\tmthree\isub\var\tm$, which is a value, \ie\ a \fireball.
				\item \emph{$\tmtwo$ is a inert} $\gconst$. Then by \refpoint{tof-and-subs-two-one}\ $\tmtwo\isub\var\tm$ is a inert, \ie\ a \fireball.
			\end{enumerate}
		\end{enumerate}
		
		\item By induction on $\evctx$. Cases:
			\begin{enumerate}
				\item \emph{Empty context $\evctx = \ctxhole$}. If $\tm = (\la\vartwo\tmfour)\fire \rtof \tmfour\isub\vartwo\fire = \tmtwo$ then
				\[\begin{array}{llll}
					\tm\isub\var\tmthree & = \\
                                        ((\la\vartwo\tmfour)\fire)\isub\var\tmthree & =  &\mbox{(def. of $\cdot\isub\cdot\cdot$)}\\
					(\la\vartwo\tmfour\isub\var\tmthree)\fire\isub\var\tmthree & \tof \\
					\tmfour\isub\var\tmthree \isub\vartwo{\fire\isub\var\tmthree} & = &\mbox{(\refpoint{tof-and-subs-one})}\\
					\tmfour \isub\vartwo{\fire}\isub\var\tmthree & =\\
                                \tmtwo\isub\var\tmthree
				\end{array}\]
				\item \emph{Application Right $\evctx = \tmfour \evctxtwo$}. Then 
				\[\begin{array}{llll}
					\evctxp\tm\isub\var\tmthree & =\\
                                        (\tmfour\evctxtwop\tm)\isub\var\tmthree & = \\
					\tmfour\isub\var\tmthree\evctxtwop\tm\isub\var\tmthree & \tof  &\mbox{(\ih)}\\
					\tmfour\isub\var\tmthree\evctxtwop\tmtwo\isub\var\tmthree & =\\
					(\tmfour\evctxtwop\tmtwo)\isub\var\tmthree  & =\\
                                        \evctxp\tmtwo\isub\var\tmthree 
				\end{array}\]
				\item \emph{Application Left $\evctx = \evctxtwo\fire$}. Then
				\[\begin{array}{llll}
					\evctxp\tm\isub\var\tmthree  & =\\
                                        (\evctxtwop\tm\fire)\isub\var\tmthree & = \\
					\evctxtwop\tm\isub\var\tmthree\fire\isub\var\tmthree & \tof  &\mbox{(\ih\ \&\ \refpoint{tof-and-subs-two-two})}\\
					\evctxtwop\tmtwo\isub\var\tmthree\fire\isub\var\tmthree & =\\
					(\evctxtwop\tmtwo\fire)\isub\var\tmthree  & =\\
                                        \evctxp\tmtwo\isub\var\tmthree
				\end{array}\]

			\end{enumerate}
		\end{enumerate}
\end{IEEEproof}
\end{atend}

\begin{table*}
\centering
\caption{Syntax, Rewriting Rules, and Structural Equivalence of the \ShFC}
\label{tab:explicit-rewritingrules}
\begin{tabular}{c|cccc}
\centering
$\begin{array}{lllllllllllll}
	\tm,\tmtwo,\tmthree,\tmfour &\grameq& \var\mid \const \mid \l\var.\tm \mid \tm\tmtwo\mid  \tm\esub\var\tmtwo\\
	\val,\valtwo & \grameq & \l \var.\tm\\
	\sctx,\sctxtwo & \grameq  & \ctxhole \mid \sctx\esub{\var}{\tm}\\
	\gconst,\gconsttwo, \gconstthree & \grameq & \const \mid \sctxp\gconst \sctxp\fire\\
	\fire, \firetwo, \firethree & \grameq & \val \mid \gconst \\
	
	\evctx &\grameq & \ctxhole\mid \tm\evctx\mid \evctx\sctxp\fire\mid \evctx\esub{\var}{\tm}
\end{array}$
&
\begin{tabular}{cc}
\centering
$\begin{array}{c@{\hspace{1cm}}l}
  \textsc{Rule at Top Level} & \textsc{Contextual Closure} \\
	\sctxp{\l\var.\tm}\sctxtwop\fire \rtodbv \sctxp{\tm\esub\var{\sctxtwop\fire}} &
\evctxp \tm \togensm \evctxp \tmtwo \textrm{~~~if } \tm \rtodbv \tmtwo \\\\
		\evctxp\var\esub\var{\sctxp{\fire}} \rtolsv  \sctxp{\evctxp\fire\esub\var\fire} &
\evctxp \tm \togense \evctxp \tmtwo \textrm{~~~if } \tm \rtolsv \tmtwo

\end{array}$
&\\\\\cline{1-2}\\
\centering
	$\begin{array}{rll@{\hspace*{0.75cm}}l}
		\tm\esub{\var}{\tmtwo}\esub{\vartwo}{\tmthree} &\tostructcom& \tm\esub{\vartwo}{\tmthree}\esub{\var}{\tmtwo}&\mbox{if $\vartwo\notin\fv{\tmtwo}$ and $\var\notin\fv{\tmthree}$}\\
		(\tm\tmthree)\esub{\var}{\tmtwo} &\tostructapr&  \tm\tmthree\esub{\var}{\tmtwo} & \textrm{if }\var\not\in\fv\tm \\
		(\tm\tmthree)\esub\var\tmtwo &\tostructapl& \tm\esub\var\tmtwo\tmthree & \textrm{if }\var\not\in\fv\tmthree \\
		\tm\esub{\var}{\tmtwo}\esub{\vartwo}{\tmthree} &\tostructes& \tm\esub{\var}{\tmtwo\esub{\vartwo}{\tmthree}} & \mbox{if $\vartwo\not\in\fv{\tm}$}  
	\end{array}$
\end{tabular}
\end{tabular}
\end{table*}
\section{Size Explosion}
\label{sect:size-exp}
Size Explosion is the side effect of a discrepancy between the dynamics and the representation of terms. The usual substitution $\tm\isub\var\tmtwo$ makes copies of $\tmtwo$ for all the occurrences of $\var$, even if $\tmtwo$ is \emph{useless}, \ie\ it is normal and it does not create redexes after substitution. These copies are the burden leading to the exponential growth of the size. To illustrate the problem, let's build a size exploding family of terms. 

Note that a inert $\gconst$, when applied to itself still is a inert $\gconst\gconst$. In particular, it still is a fireball, and so it can be used as an argument for redexes. We can then easily build a term of size linear in $n$ that in $n$ steps evaluates a complete binary tree $\gconst^{2^n}$. Namely, define the family of terms $\tm_n$ for $n\geq1$:
\begin{center}$\begin{array}{lllllllll}
	\tm_1 & \defeq & \la{\var_{1}}(\var_{1}\var_{1})\\
	\tm_{n+1} & \defeq & \la{\var_{n+1}}(\tm_{n}(\var_{n+1}\var_{n+1}))
\end{array}$\end{center}

Now consider $\tm_n \gconst$, that for a fixed $\gconst$ has size linear in $n$. The next proposition shows that $\tm_n \gconst$ reduces in $n$ steps to $ \gconst^{2^n}$, causing size-explosion.

\begin{proposition}[Size Explosion in the \fb]
	$\tm_n \gconst \tof^n \gconst^{2^n}$.
\end{proposition}
\begin{IEEEproof}
	by induction on $n$. Let $\gconsttwo \defeq \gconst^2 = \gconst \gconst$. Cases:
$$\begin{array}{llllll}
	\tm_1 & = & (\la{\var_{1}}(\var_{1}\var_{1})) \gconst &	\tof & \gconst^2	\\
	\tm_{n+1} & = & (\la{\var_{n+1}}(\tm_{n}(\var_{n+1}\var_{n+1}))) \gconst & \tof \\
	& & \tm_n \gconst^2 = \tm_n \gconsttwo & \tof^n&&\mbox{(\ih)}\\
	& & \gconsttwo^{2^n} & = & \gconst^{2^{n+1}} & \hfill\qedhere
\end{array}$$
\end{IEEEproof}

\section{Fireballs and Explicit Substitutions}
\label{sect:fireb-end-es}
\begin{atend}\section{Proofs Omitted From \refsect{fireb-end-es}\\ (Fireballs and Explicit Substitutions)} \label{sect:app-fireb-end-es}\end{atend}

In a ordinary weak scenario, sharing of subterms prevents size explosion. In the \fb\ however this is no longer true, as we show in this section. Sharing of subterms is here represented in a variation over the Linear Substitution Calculus, a formalism with explicit substitutions coming from a linear logic interpretation of the $\lambda$-calculus. At the dynamic level, the \emph{small-step} operational semantics of the \FC\ is refined into a \emph{micro-step} one, where explicit substitutions replace one variable occurrence at a time, similarly to abstract machines. 

The language of the \emph{\Micro\ Fireball Calculus} (\ShFC) is:
\begin{center}$\begin{array}{lllllllllllll}
	\tm,\tmtwo,\tmthree,\tmfour &\grameq& \var\mid \const \mid \l\var.\tm \mid \tm\tmtwo\mid  \tm\esub\var\tmtwo
\end{array}$\end{center}
where  $\tm\esub\var\tmtwo$ is the explicit substitution (ES) of $\tmtwo$ for $\var$ in $\tm$, that is an alternative notation for ${\tt let}\ \var = \tmtwo\ {\tt in}\ \tm$, and where $\var$ becomes bound (in $\tm$). We
silently work modulo $\alpha$-equivalence of these bound variables, \eg\ $(\var\vartwo)\esub\vartwo\tm\isub\var\vartwo =
(\vartwo\varthree)\esub\varthree\tm$. We use $\fv\tm$ for the set of
free variables of $\tm$. 

\paragraph*{Contexts} the dynamics of explicit substitutions is defined using (one-hole) contexts.
\emph{Weak contexts} subsume all the kinds of context in the paper, and are defined by
\begin{center}$\begin{array}{lllllllllllll}
	\wctx,\wctxtwo 		& \grameq & \ctxhole\mid \tm\wctx\mid \wctx\tm \mid \wctx\esub\var\tm \mid \tm\esub\var\wctx
\end{array}$\end{center}
The plugging $\wctxp\tm$ of a term $\tm$ into a context $\wctx$ is defined as
$\ctxhole\ctxholep\tm\defeq\tm$, $(\l\var.\wctx)\ctxholep\tm \defeq
\l\var.(\wctx\ctxholep\tm)$, and so on. As usual, plugging in a
context can capture variables,
\eg\ $((\ctxhole\vartwo)\esub\vartwo\tm)\ctxholep\vartwo =
(\vartwo\vartwo)\esub\vartwo\tm$. The plugging $\wctxp\wctxtwo$ of a
context $\wctxtwo$ into a context $\wctx$ is defined analogously. Since all kinds of context we will deal with will be weak, the definition of plugging applies uniformly to all of them.

A special and frequently used class of contexts is that of
\emph{substitution contexts} $\sctx\grameq\ctxhole\mid
\sctx\esub{\var}{\tm}$.

Switching from the \FC\ to the \ShFC\ the syntactic categories of \emph{inerts} $\gconst$, \emph{\fireball s} $\fire$, and \emph{evaluation contexts} $\evctx$ are generalised in \reftab{explicit-rewritingrules} as to include substitution contexts $\sctx$.
Note that fireballs may now contain substitutions, but \emph{not at top level}, because it is technically convenient to give a separate status to a fireball $\fire$ in a substitution context $\sctx$: terms of the form $\sctxp\fire$ are called \emph{answers}. An \emph{initial term} is a closed term with no explicit substitutions.

\paragraph*{Rewriting Rules} the fireball rule $\tof$ is replaced by $\togens$, defined as the union of the two rules $\togensm$ and $\togense$ in \reftab{explicit-rewritingrules}:
\begin{enumerate}
	\item \emph{Multiplicative} $\togensm$: is a version of $\tof$ where $\l\var.\tm$ and $\fire$ can have substitution contexts $\sctx$ and $\sctxtwo$ around, and the substitution is delayed. 
	\item \emph{Exponential} $\togense$: the substitution or exponential rule $\togense$ replaces exactly one occurrence of a variable $\var$ currently under evaluation (in $\evctx$) with its definiendum $\fire$ given by the substitution. Note the apparently strange position of $\sctx$ in the reduct. It is correct: $\sctx$ has to commute outside to bind both copies of $\fire$, otherwise the rule would create free variables.
\end{enumerate}
The name of the rules are due to the linear logic interpretation of the LSC. 

\paragraph*{Unfolding} the shared representation is related to the usual one via the crucial notion of \emph{unfolding}, producing the $\l$-term $\unf\tm$ denoted by $\tm$ and defined by:
\begin{center}$\begin{array}{lllllllllllll}
	\unf\var 				& \defeq & \var&&&& 	
	\unf{(\tm \tmtwo)} 		& \defeq & \unf{\tm} \unf{\tmtwo}\\
	\unf{(\l\var.\tm)}		& \defeq & \l\var.\unf\tm&&&&
	\unf{\tm \esub\var\tmtwo} 		& \defeq & \unf\tm \isub\var{\unf\tmtwo}\\
\end{array}$\end{center}
 Note that $\unf{\tmfour_n} = \gconst^{2^n}$.
 
As for the \FC, evaluation is well-defined:

\begin{atend}

\subsection{Closed Normal Forms and Determinism}
\label{ssect:explicit-nfs-det}
The first step is to identify the reduction invariants, the most
important one being the shape of terms---called \emph{proper}---that
are produced by the strategy $\togens$ starting with a term without ES.

\begin{definition}[\Proper\ Term]
	A term $\tm$ is \emph{\proper} if 
	\begin{enumerate}
		\item \emph{ES}: any explicit substitution in $\tm$ contains an answer, and
		\item \emph{Value}: any value in $\tm$ does not contain ES.
	\end{enumerate}
	We also say that an ES is \proper\ when it contains a \proper\ answer.
\end{definition}

Note that initial terms (having no ES) are \proper\ and so the next lemma applies in particular when the starting term is initial.

\begin{lemma}[\Proper\ Invariant]
	Let $\tm$ be a \proper\ term. If $\tm\togens^*\tmtwo$ then $\tmtwo$ is \proper.
\end{lemma}

\begin{IEEEproof}
	by induction on the length $k$ of the derivation $\tm\togens^*\tmtwo$. If $k=0$ the statement is just the hypothesis. Otherwise $\tm\togens^{k-1}\tmthree\togens\tmtwo$ and by \ih\ $\tmthree$ is \proper. Note that 1) multiplicative steps create \proper\ ES, and 2) exponential steps copy \proper\ \fireball s only out of values and ES, preserving \properness.
\end{IEEEproof}

We now characterize normal forms: the next three lemmas conclude that
normal terms are answers, and that answers are fireballs up to unfolding.

Point 2.a of the next statement is given with respect to unfolding relative with shallow context (defined in \refsect{rel-unfolding}, page \pageref{sect:rel-unfolding}) because it will be used in this more general form in later sections.

\begin{lemma}[Properties of Answers]
	\label{l:es-answers-are-normal}
	Let $\tm=\sctxp\tmtwo$. Answers are $\togens$-normal, do not decompose as $\evctxp\var$, and (relatively) unfold to \fireball s. More precisely,
	\begin{enumerate}
	\item If $\tmtwo$ is a inert or a \fireball\ then $\tm$ is $\togens$-normal and it does not decompose as $\evctxp\var$, 
	\item Moreover,
	\begin{enumerate}
		\item If  $\tmtwo$ is a inert then $\relunf\tm\tctx$ is a inert. 
		\item If  $\tmtwo$ is a \fireball\ then $\relunf\tm\tctx$ is a \fireball.
	\end{enumerate}
	\end{enumerate}
\end{lemma}

\myproof{
\begin{IEEEproof}
	by induction on $\sctx$. Cases:
	\begin{itemize}
		\item \emph{Empty List} $\ctxhole$. By induction on $\tmtwo$.		
		\begin{enumerate}
			\item $\tmtwo$ is a inert. Cases:
			\begin{enumerate}
				\item \emph{$\tmtwo$ is a \symb $\const$}. Then it is normal and clearly does not decompose as $\evctxp\var$. Moreover, $\relunf\tm\tctx=\relunf\const\tctx = \const$ is a inert.
				\item \emph{$\tmtwo$ is a inert $\sctxtwop\gconst \sctxthreep\fire$}. Then by \ih\ both $\sctxtwop\gconst$ and $\sctxthreep\fire$ are normal. Since $\gconst$ cannot be an abstraction, the topmost application cannot be a $\togensm$-redex, and so $\tmtwo$ is normal. For each of $\sctxtwop\gconst$ and $\sctxthreep\fire$ \ih\ gives that it does not decompose as $\evctxp\var$. Then $\tmtwo$ does not decompose either. Moreover, by \ih\ $\relunf{\sctxtwop\gconst}\tctx$ is a inert and $\relunf{\sctxthreep\fire}\tctx$ is a \fireball, and so $\relunf\tm\tctx =_{\reflemmaeqp{relunf-properties}{zero}} \relunf{\sctxtwop\gconst}\tctx \relunf{\sctxthreep\fire}\tctx$ is a inert.

			\end{enumerate}
			\item \emph{$\tmtwo$ is a \fireball}. Cases:
			\begin{enumerate}
				\item \emph{$\tmtwo$ is a value} $\la\var\tmthree$. Then it is normal and does not decompose as $\evctxp\var$. Moreover, $\unf\tm =_{\reflemmaeqp{relunf-properties}{zero}} \la\var\unf\tmthree$, which is a value, \ie\ a \fireball.
				\item \emph{$\tmtwo$ is a inert} $\gconst$. Given by the \ih.
			\end{enumerate}
		\end{enumerate}
		\item \emph{Non-Empty List} $\sctx = \sctxtwo\esub\var\tmthree$. By \ih, $\sctxtwop\tmtwo$ is normal and cannot be decomposed as $\evctxp\var$, and so there cannot be $\togense$ redexes involving $\esub\var\tmthree$. Thus $\tm$ is normal. 
		
		For the absence of a decomposition, note that---apart from the trivial decomposition $\ctxholep{\sctxp\tmtwo}$, that is not of the form $\evctxp\var$---every decomposition of $\sctxp\tmtwo$ is obtained from a decomposition of $\sctxtwop\tmtwo$ by appending $\esub\var\tmthree$, and so $\sctxp\tmtwo$ does not decompose as $\evctxp\var$. 
		
		For the \emph{moreover} part, by \ih\ $\sctxtwop\tmtwo$ verifies the statement for no matter which shallow context. Then $\relunf\tm\tctx = \relunf{\sctxtwop\tmtwo\esub\var\tmthree}\tctx =_{\reflemmaeqp{relunf-properties}{four}} \relunf{\sctxtwop\tmtwo}{\tctxp{\ctxhole\esub\var\tmthree}}$ also verifies the statement.\qedhere
	\end{itemize}
\end{IEEEproof}
}

\begin{lemma}[Normal Form Characterization]
	\label{l:es-normal-form-char}
	Let $\tm$ be a $\togens$-normal term.
	\begin{enumerate}
		\item \label{p:es-normal-form-char-answer} Either $\tm$ is an answer,
		\item \label{p:es-normal-form-char-freev} or $\tm = \evctxp\var$.
	\end{enumerate}
\end{lemma}

\myproof{
\begin{IEEEproof}
	by induction on $\tm$. Cases:
	\begin{enumerate}
		\item \emph{Variable} $\tm = \var$. Here \refpoint{es-normal-form-char-freev} holds, while evidently \refpoint{es-normal-form-char-answer} is false.
		\item \emph{\Symb} $\tm = \const$. Here \refpoint{es-normal-form-char-answer} holds, and \refpoint{es-normal-form-char-freev} is false.
		\item \emph{Abstraction} $\tm = \la\var\tmtwo$. Here \refpoint{es-normal-form-char-answer} holds, and \refpoint{es-normal-form-char-freev} is false.
		\item \emph{Application} $\tm = \tmtwo\tmthree$. By \ih\ we are in one of the following two cases for the right sub-term $\tmthree$:
		\begin{enumerate}
			\item \emph{\refpoint{es-normal-form-char-answer} holds but not \refpoint{es-normal-form-char-freev}}. By \reflemma{es-answers-are-normal}, $\tmthree$ is normal. Note that $\ctxhole\tmthree$ is an evaluation context. The \ih\ gives one of the following two cases for the left sub-term $\tmtwo$:
			\begin{enumerate}
				\item \emph{\refpoint{es-normal-form-char-answer} holds but not \refpoint{es-normal-form-char-freev}}. Given that both $\tmtwo$ and $\tmthree$ do not satisfy \refpoint{es-normal-form-char-freev}, neither does $\tm$. Being an answer, $\tmthree$ has the form $\sctxp\fire$. Two cases:
				\begin{enumerate}
					\item \emph{$\fire$ is a inert} $\gconst$. Then $\tm = \sctxp\gconst \tmthree$ is the application of a inert to an answer, which is a inert---\ie\ an answer---and \refpoint{es-normal-form-char-answer} holds.
					\item \emph{$\fire$ is a value}. Then $\tm$ is a $\togensm$ redex, absurd.
				\end{enumerate}
				
				\item \emph{\refpoint{es-normal-form-char-freev} holds but not \refpoint{es-normal-form-char-answer}}. Then \refpoint{es-normal-form-char-freev} holds for $\tm$, because $\ctxhole\tmthree$ is an evaluation context. Since $\tmtwo$ is not a inert, $\tm$ is not answer, and \refpoint{es-normal-form-char-answer} does not hold.
			\end{enumerate}
			\item \emph{\refpoint{es-normal-form-char-freev} holds but not \refpoint{es-normal-form-char-answer}}. Then \refpoint{es-normal-form-char-freev} holds for $\tm$, because $\tmtwo\ctxhole$ is an evaluation context, and \refpoint{es-normal-form-char-answer} does not, because $\tmthree$ is not an answer.
		\end{enumerate}
		
		\item \emph{Substitution} $\tm = \tmtwo\esub\var\tmthree$. Since $\ctxhole\esub\var\tmthree$ is an evaluation context we can apply the \ih, and fall into one of the two following cases:
		\begin{enumerate}
			
			\item \emph{\refpoint{es-normal-form-char-answer} holds but not \refpoint{es-normal-form-char-freev}}. Then $\tm$ is an answer, \ie\ \refpoint{es-normal-form-char-answer} holds. Note that since any non-empty evaluation context for $\tm$ comes from an evaluation context for $\tmtwo$, \refpoint{es-normal-form-char-freev} holds for $\tm$ iff it holds for $\tmtwo$, \ie\ it does not.
			\item \emph{\refpoint{es-normal-form-char-freev} holds but not \refpoint{es-normal-form-char-answer}}. 
			Then $\tmtwo =\evctxtwop\vartwo$ and we conclude taking $\evctx \defeq \evctxtwo\esub\var\tmthree$. Note that it may be that $\var = \vartwo$, but in that case $\tmthree$ is not an answer (otherwise there would be a redex). There is no contradiction, because we are not assuming $\tm$ to be proper (case in which one necessarily has $\var\neq\vartwo$).\qedhere
		\end{enumerate}
	\end{enumerate}
\end{IEEEproof}
}

\begin{corollary}
	\label{coro:es-closed-proper-normal-eq-answer}
	Let $\tm$ be a closed proper $\togens$-normal term. Then $\tm$ is an answer and $\unf\tm$ is $\tof$-normal.
\end{corollary}

\begin{IEEEproof}
	if $\tm$ is $\togens$-normal then by \reflemma{es-normal-form-char} either $\tm$ is an answer or it has the form $\evctxp\var$. Suppose that it has the form $\evctxp\var$. Since $\tm$ is closed, $\evctx$ has a substitution on $\var$, and since $\tm$ is proper, that substitution contains an answer. Then $\tm$ has a $\togense$-redex, absurd. Then $\tm$ is an answer. By \reflemma{es-answers-are-normal}, $\unf\tm$ is a \fireball. By \reflemma{gconst-or-fire-fnorm}, $\unf\tm$ is $\tof$-normal.
\end{IEEEproof}

In order to prove determinism of the calculus, we need the notion of \emph{position} and a final property of answers.

The position of a multiplicative redex is the context $\evctx$ in which the rule takes place, and this is standard. The position of an exponential redex $\evctxtwop{\evctxp\var\esub\var{\sctxp{\fire}}} \togense \tctxtwop{\sctxp{\evctxp\fire\esub\var\fire}}$ is the context around the substituted variable, \ie\ $\evctxtwop{\evctx\esub\var{\sctxp{\fire}}}$.

Given a term $\tm$, a redex is \emph{contained} in a sub-term $\tmtwo$ if the whole rewriting pattern is contained in $\tmtwo$. An exponential redex is \emph{partially contained} in $\tmtwo$ if $\tmtwo$ contains the substituted variable (and then the position of the redex) but not the acting substitution.

\begin{lemma}[Answers do not (Partially) Contain Redexes]
	\label{l:es-answers-no-position}
	Let $\tm = \evctxp{\tmtwo}$ be a term with $\tmtwo$ an answer. Then no redex of $\tm$ can have its position in $\tmtwo$.
\end{lemma}

\begin{IEEEproof}
	by \reflemma{es-answers-are-normal}, $\tmtwo$ is $\togens$-normal and so no $\togensm$-redex of $\tm$ can have its position in $\tmtwo$. Moreover, $\tmtwo$ is not of the form $\evctxp\vartwo$ and so no $\togense$-redex of $\tm$ can be entirely nor partially contained in $\tmtwo$. 
\end{IEEEproof}

\begin{lemma}[Determinism]
	\label{l:es-determinism}
	Let $\tm$ be a term and $\evctxONE$ and $\evctxTWO$ the positions of two redexes in $\tm$. Then $\evctxONE = \evctxTWO$.
\end{lemma}

\myproof{
\begin{IEEEproof}
	let $\tm = \evctxONEp\tmtwo$. By induction on $\evctxONE$. Cases:
	\begin{enumerate}
		\item \emph{Empty} $\evctxONE = \ctxhole$. Cases:
		\begin{enumerate}
			\item \emph{Multiplicative Redex}, \ie\ $\tm = \sctxp{\la\var\tmfour}\tmfive$ with $\tmfive$ an answer. By \reflemma{es-answers-no-position}, $\evctxTWO$ cannot lie in $\sctxp{\la\var\tmfour}$ nor in $\tmfive$. Then necessarily $\evctxTWO = \evctxONE = \ctxhole$.
			\item \emph{Exponential Redex}. This case is impossible because the position of an exponential redex is the context around the substituted variable and if $\evctxONE = \ctxhole$ then $\tm = \var$ and there is no substitution acting on $\var$.
		\end{enumerate}
			
		\item \emph{Right Application} $\evctxONE = \tmfour\evctxONEtwo$ and $\tm =\tmfour\evctxONEtwop\tmtwo$. By \reflemma{es-answers-no-position}, $\evctxONEtwop\tmtwo$ is not an answer and so $\evctxTWO$ does not lie in $\tmfour$, nor $\evctxTWO$ can be empty (\ie\ $\tm = \tmfour\evctxONEtwop\tmtwo$ cannot be a $\togensm$-redex). Then, $\evctxTWO = \tmtwo\evctxTWOtwo$, and the statement follows from the \ih\ applied to $\evctxONEtwo$ and $\evctxTWOtwo$.

		\item \emph{Left Application} $\evctxONE = \evctxONEtwo \sctxp\fire$ and and $\tm = \evctxONEtwop\tmtwo\sctxp\fire$. By \reflemma{es-answers-no-position}, $\evctxTWO$ does not lie in $\sctxp\fire$. And $\evctxTWO$ cannot be empty (\ie\ the position of a $\togensm$-redex), because then $\evctxONEtwop\tmtwo$ would have the form $\sctxp{\la\var\tmsix}$, \ie\ it would be an answer, and so by \reflemma{es-answers-no-position} no redex positions can lie in $\evctxONEtwop\tmtwo$, against the hypothesis of the case. Then, $\evctxTWO = \evctxTWOtwo\tmthree$, and the statement follows from the \ih\ applied to $\evctxONEtwo$ and $\evctxTWOtwo$. 
		
		\item \emph{Substitution} $\evctxONE = \evctxONEtwo\esub\var\tmthree$. Then necessarily $\evctxTWO = \evctxTWOtwo\esub\var\tmthree$ (remember the position of a $\togense$-redex is given by the context around the substituted variable, and not by the one around the acting substitution) and the statement follows from the \ih.\qedhere
	\end{enumerate}
\end{IEEEproof}
}

\begin{corollary}
	\label{coro:es-closed-determ}
	Let $\tm$ be a proper closed term. Then either $\tm$ contains exactly one $\togens$-redex, or $\tm$ is an answer.
\end{corollary}

\begin{IEEEproof}
	by \reflemma{es-determinism}, $\tm$ contains at most one redex. If it contains no redexes, then by \refcoro{es-closed-proper-normal-eq-answer} it is an answer.
\end{IEEEproof}

\end{atend}

\begin{theorem}\label{thm:ES-determ-and-nfs}\hfill
\begin{enumerate}	
	\item Closed normal forms are answers, \ie\ fireballs in substitution contexts.
	\item $\togens$ is deterministic.
\end{enumerate}
\end{theorem}

\paragraph*{Structural Equivalence} the calculus is endowed with a structural equivalence, noted $\eqstruct$, whose property is to be a strong bisimulation with respect to $\togens$. It is the least equivalence relation closed by weak contexts defined in \reftab{explicit-rewritingrules}.

\begin{atend}
\subsection{Structural Equivalence}
\label{ssect:app-struct-eq}
The aim is to prove the strong bisimulation of structural equivalence, whose proof
relies on the next lemma.
\begin{lemma}
\label{l:nonstrict_cbv_eqstruct_preserves_shapes}
The equivalence relation $\tostruct$ preserves the ``shapes'' of $\sctxp{\fire}$ and $\evctxp{\var}$. Formally:
\begin{enumerate}
  \item \label{p:nonstrict_cbv_eqstruct_preserves_shapes-one}If $\sctxp{\fire} \eqstruct \tm$, then $\tm$ is of the form $\sctxtwop{\firetwo}$.
  \item \label{p:nonstrict_cbv_eqstruct_preserves_shapes-two}If $\evctxp{\var} \eqstruct \tm$, with $\var$ not bound by $\evctx$,
  then $\tm$ is of the form $\evctxtwop{\var}$, with $\var$ not bound by $\evctxtwo$.
\end{enumerate}
\end{lemma}
\begin{IEEEproof}
 \begin{enumerate}
  \item By induction on $\sctx$.
  \item By induction on $\evctx$.
 \end{enumerate}

\end{IEEEproof}

Now, we are ready for the bisimulation property.
\end{atend}

\begin{proposition}[$\tostruct$ is a Strong Bisimulation wrt $\togens$]
	\label{prop:GenStrongBisim}
	\label{prop:strong-bis}
	Let $\mathtt{x}\in\set{\ssym\msym,\ssym\esym}$. Then, $\tm\eqstruct\tmtwo$ and $\tm\togenx\tmp$ implies that there exists $\tmtwop$ such that $\tmtwo\togenx\tmtwop$ and $\tmp\eqstruct\tmtwop$.
\end{proposition}
\begin{proofatend}

Let $\tostructsym$ be the symmetric closure of the
union of the axioms defining $\eqstruct$, \ie\ of $\tostructcom \cup \tostructapl \cup \tostructapr \cup \tostructes$.
Note that $\eqstruct$ is the reflexive--transitive closure of $\tostructsym$.
The proof is in two parts:
\begin{itemize}
\item[] (I) Prove the property holds for $\tostructsym$, \ie\
            if $\tm\towhl_{a}\tmtwo$ and $\tm\tostructsym\tmthree$, there exists $\tmfour$ s.t. $\tmthree\towhl_{a}\tmfour$ and $\tmtwo\eqstruct\tmfour$.
\item[] (II) Prove the property holds for $\eqstruct$ (\ie\ for many steps of $\tostructsym$) by resorting to (I).
\end{itemize}

The proof of (II) is immediate by induction on the number of $\tostructsym$ steps.
The proof of (I) goes by induction on the rewriting step $\togen$ (that, since $\togen$ is closed by evaluation contexts, becomes a proof by induction on the evaluation context $\evctx$). In principle, we should always consider the two directions of $\tostructsym$. Most of the time, however, one direction is obtained by simply reading the diagram of the other direction bottom-up, instead than top-down; these cases are simply omitted, we distinguish the two directions only when it is relevant.

The proof of the strong bisimulation property is by induction on $\togen$.

\begin{enumerate}
\item \casealt{Base case 1: multiplicative root step $
            \tm = \sctxp{\l\var.\tmp}\sctxtwop{\fire}
            \rtodbv
            \tmtwo = \sctxp{\tmp\esub{\var}{\sctxtwop{\fire}}}
            $}
    The nontrivial cases are when the $\tostructsym$ step overlaps
    the pattern of the $\dbv$-redex. Note that by \reflemma{nonstrict_cbv_eqstruct_preserves_shapes}.\ref{p:nonstrict_cbv_eqstruct_preserves_shapes-one},
    if the $\tostructsym$ is internal to $\sctxtwop{\fire}$,
    the proof is direct, since the $\dbv$-redex is preserved. More precisely,
    if $\sctxtwop{\fire} \tostructsym \sctxthreep{\firetwo}$, we have:

          $$
          \commutesdbvEEABCD{\tostructsym}{\tostructsym}{
            \sctxp{\l\var.\tmp}\,\sctxtwop{\fire}
          }{
            \sctxp{\tmp\esub{\var}{\sctxtwop{\fire}}}
          }{
            \sctxp{\l\var.\tmp}\,\sctxthreep{\firetwo}
          }{
            \sctxp{\tmp\esub{\var}{\sctxthreep{\firetwo}}}
          }
          $$
    
    Consider the remaining possibilities for $\tostructsym$:

    \begin{enumerate}

    \item \caselight{Commutation of independent substitutions $\tostructcom$}.
        The commutation of substitutions must be in $\sctx$,
        \ie\ $\sctx$ must be of the form $\sctxOnep{\sctxTwo\esub{\vartwo}{\tmtwop}\esub{\varthree}{\tmthreep}}$
        with $\varthree \not\in \fv{\tmtwop}$.
        Let $\sctxal := \sctxOnep{\sctxTwo\esub{\varthree}{\tmthreep}\esub{\vartwo}{\tmtwop}}$.
        Then:
        $$
        \commutesdbvEEABCD{\tostructcom}{\tostructcom}{
            \sctxp{\l\var.\tmp}\,\sctxtwop{\fire}
        }{
            \sctxp{\tmp\esub{\var}{\sctxtwop{\fire}}}
        }{
            \sctxpal{\l\var.\tmp}\,\sctxtwop{\fire}
        }{
            \sctxpal{\tmp\esub{\var}{\sctxtwop{\fire}}}
        }
        $$

    \item \label{p:str-bis-nscbv-base-mul-apl}\caselight{Commutation with the left of an application $\tostructapl$}.
        The diagram is:
        
        $$\hspace{-1.1cm}
          \commutesdbvEEABCD{\tostructap}{=}{
              \sctxp{\l\var.\tmp}\esub{\vartwo}{\tmfive}\,\sctxtwop\fire
          }{
              \sctxp{\tmp\esub\var{\sctxtwop\fire}}\esub\vartwo\tmfive
          }{
              (\sctxp{\l\var.\tmp}\,\sctxtwop{\fire})\esub\vartwo\tmfive
          }{
              \sctxp{\tmp\esub\var{\sctxtwop\fire}}\esub\vartwo\tmfive
          }
          $$
          
    \item \label{p:str-bis-nscbv-base-mul-apr}\caselight{Commutation with the right of an application $\tostructapr$}.
        The diagram is:        
        
          $$\hspace{-1.0cm}
            \begin{tikzpicture}[ocenter]
             \node (s) {\normalsize\ensuremath{ \sctxp{\l\var.\tmp}\,\sctxtwop\fire\esub{\vartwo}{\tmfive} }};
             \node at (s.center)  [right=1.7*\nodeHorDist](s1){\normalsize\ensuremath{\sctxp{\tmp\esub\var{\sctxtwop\fire\esub\vartwo\tmfive}}}};
             \node at (s1.center) [below=\nodeVerDist](t) {\normalsize\ensuremath{ 
             \sctxp{\tmp\esub\var{\sctxtwop\fire}\esub\vartwo\tmfive}
             }};
             \node at (s.center)  [below=2*\nodeVerDist](s3) {\normalsize\ensuremath{ (\sctxp{\l\var.\tmp}\,\sctxtwop{\fire})\esub\vartwo\tmfive }};
             \node at (s3.center) [right=1.7*\nodeHorDist](s4) {\normalsize\ensuremath{ \sctxp{\tmp\esub\var{\sctxtwop\fire}}\esub\vartwo\tmfive }};
             \node at (s.center)  [below=\nodeVerDist](eq1){\normalsize\ensuremath{\tostructapl}};
             \node at (s1.center) [below=\nodeVerDist/2](eq2){\normalsize\ensuremath{\tostructes}};
             \node at (s1.center) [below=3*\nodeVerDist/2](eq2){\normalsize\ensuremath{\tostructcom^*}};
             \draw[-o] (s) to node {$\lsvsym$} (s1);
             \draw[-o, dashed] (s3) to node {$\lsvsym$} (s4);
            \end{tikzpicture} 
            $$

    \item \caselight{Composition of substitutions $\tostructes$}.
        The composition of substitutions must be in $\sctx$,
        \ie\ $\sctx$ must be of the form $\sctxOnep{\sctxTwo\esub{\vartwo}{\tmtwop}\esub{\varthree}{\tmthreep}}$
        with $\varthree \not\in \fv{\sctxTwop{\l\var.\tmp}}$.
        Let $\sctxal := \sctxOnep{\sctxTwo\esub{\vartwo}{\tmtwop\esub{\varthree}{\tmthreep}}}$.
        Then:
        $$
        \commutesdbvEEABCD{\tostructes}{\tostructes}{
            \sctxp{\l\var.\tmp}\,\sctxtwop{\fire}
        }{
            \sctxp{\tmp\esub{\var}{\sctxtwop{\fire}}}
        }{
            \sctxpal{\l\var.\tmp}\,\sctxtwop{\fire}
        }{
            \sctxpal{\tmp\esub{\var}{\sctxtwop{\fire}}}
        }
        $$
    \end{enumerate}

  \item \casealt{Base case 2: exponential root step $
            \tm = \evctxp{\var}\esub{\var}{\sctxp{\fire}}
            \rtolsv
            \tmtwo = \sctxp{\evctxp{\fire}\esub{\var}{\fire}}
            $}
    Consider first the case when the $\tostructsym$-redex is internal
    to $\evctxp{\var}$. By \reflemma{nonstrict_cbv_eqstruct_preserves_shapes}.\ref{p:nonstrict_cbv_eqstruct_preserves_shapes-two}
    we know $\tostructsym$ preserves the shape of $\evctxp{\var}$,
    \ie\ $\evctxp{\var} \tostructsym \evctxpal{\var}$. Then:
    $$
    \commuteslsvEEABCD{\tostructsym}{\eqstruct}{
        \evctxp{\var}\esub{\var}{\sctxp{\fire}}
    }{
        \sctxp{\evctxp{\fire}\esub{\var}{\fire}}
    }{
        \evctxpal{\var}\esub{\var}{\sctxp{\fire}}
    }{
        \sctxp{\evctxpal{\fire}\esub{\var}{\fire}}
    }
    $$
    If the $\tostructsym$-redex is internal to one of the substitutions in $\sctx$, the proof is similarly straightforward.
    Note that the $\tostructsym$-redex has always a substitution
    at the root. The remaining possibilities are that such a 
    substitution is in $\sctx$ and it interact with constructors outside $\sctx$, or that it is precisely $\esub{\var}{\sctxp{\fire}}$.
    Axiom by axiom:
    
    \begin{enumerate}

    \item \caselight{Commutation of independent substitutions $\tostructcom$}.
        The case where both commuted substitutions belong to $\sctx$ has already been treated. The remaining possibility is that 
        $\evctx = \evctxtwo\esub{\vartwo}{\tmp}$ and $\esub{\var}{\sctxp{\fire}}$ commutes with $\esub{\vartwo}{\tmp}$ (which implies $\var \not\in \fv{\tmp}$). Then:
        $$\hspace{-1.0cm}
        \commuteslsvEEABCD{\tostructcom}{\tostructcom^*}{
            \evctxtwop{\var}\esub{\vartwo}{\tmp}\esub{\var}{\sctxp{\fire}}
        }{
            \sctxp{\evctxtwop{\fire}\esub{\vartwo}{\tmp}\esub{\var}{\fire}}
        }{
            \evctxtwop{\var}\esub{\var}{\sctxp{\fire}}\esub{\vartwo}{\tmp}
        }{
            \sctxp{\evctxtwop{\fire}\esub{\var}{\fire}}\esub{\vartwo}{\tmp}
        }
        $$

    \item \label{p:str-bis-nscbv-base-exp-apl} \caselight{Commutation with the left of an application $\tostructapl$}.
        The only possibility is that the substitution $\esub{\var}{\sctxp{\fire}}$
        is commuted with the outermost application in $\evctxp{\var}$, \ie\ $\evctx = \evctxtwo\sctxtwop\firetwo$. Then,
        
            $$\hspace{-1.3cm}
            \begin{tikzpicture}[ocenter]
             \node (s) {\normalsize\ensuremath{ (\evctxtwop{\var}\,\sctxtwop\firetwo)\esub{\var}{\sctxp{\fire}} }};
             \node at (s.center)  [right=1.7*\nodeHorDist](s1){\normalsize\ensuremath{\sctxp{(\evctxtwop{\fire}\,\sctxtwop\firetwo)\esub{\var}{\fire}}}};
             \node at (s1.center) [below=\nodeVerDist](t) {\normalsize\ensuremath{ 
             \sctxp{\evctxtwop{\fire}\esub{\var}{\fire} \sctxtwop\firetwo}
             }};
             \node at (s.center)  [below=2*\nodeVerDist](s3) {\normalsize\ensuremath{ \evctxtwop{\var}\esub{\var}{\sctxp{\fire}}\sctxtwop\firetwo }};
             \node at (s3.center) [right=1.7*\nodeHorDist](s4) {\normalsize\ensuremath{ \sctxp{\evctxtwop{\fire}\esub{\var}{\fire}}\sctxtwop\firetwo }};
             \node at (s.center)  [below=\nodeVerDist](eq1){\normalsize\ensuremath{\tostructapl}};
             \node at (s1.center) [below=\nodeVerDist/2](eq2){\normalsize\ensuremath{\tostructapl}};
             \node at (s1.center) [below=3*\nodeVerDist/2](eq2){\normalsize\ensuremath{\tostructapl^*}};
             \draw[-o] (s) to node {$\lsvsym$} (s1);
             \draw[-o, dashed] (s3) to node {$\lsvsym$} (s4);
            \end{tikzpicture} 
            $$
         The $\tostructapl^*$ step is justified by the fact that in the source term $(\evctxtwop{\var}\,\sctxtwop\firetwo)\esub{\var}{\sctxp{\fire}}$ the context $\sctx$ is only around $\fire$, and so it cannot capture variables in $\sctxtwop\firetwo$.

    \item \label{p:str-bis-nscbv-base-exp-apr}\caselight{Commutation with the right of an application $\tostructapr$}.
       similarly to the previous case
               $$\hspace{-0.5cm}
            \begin{tikzpicture}[ocenter]
             \node (s) {\normalsize\ensuremath{ (\tmp\,\evctxtwop{\var})\esub{\var}{\sctxp{\fire}} }};
             \node at (s.center)  [right=1.7*\nodeHorDist](s1){\normalsize\ensuremath{ \sctxp{(\tmp\,\evctxtwop{\fire})\esub{\var}{\fire}} }};
             \node at (s1.center) [below=\nodeVerDist](t) {\normalsize\ensuremath{ 
             \sctxp{\tmp\evctxtwop{\fire}\esub{\var}{\fire}}
             }};
             \node at (s.center)  [below=2*\nodeVerDist](s3) {\normalsize\ensuremath{ \tmp\evctxtwop{\var}\esub{\var}{\sctxp{\fire}} }};
             \node at (s3.center) [right=1.7*\nodeHorDist](s4) {\normalsize\ensuremath{ \tmp\sctxp{\evctxtwop{\fire}\esub{\var}{\fire}} }};
             \node at (s.center)  [below=\nodeVerDist](eq1){\normalsize\ensuremath{\tostructapr}};
             \node at (s1.center) [below=\nodeVerDist/2](eq2){\normalsize\ensuremath{\tostructapr}};
             \node at (s1.center) [below=3*\nodeVerDist/2](eq2){\normalsize\ensuremath{\tostructapr^*}};
             \draw[-o] (s) to node {$\lsvsym$} (s1);
             \draw[-o, dashed] (s3) to node {$\lsvsym$} (s4);
            \end{tikzpicture} 
            $$

            The $\tostructapr^*$ step is justified by the fact that in the source term $(\evctxtwop{\var}\,\sctxtwop\firetwo)\esub{\var}{\sctxp{\fire}}$ the context $\sctx$ is only around $\fire$, and so it cannot capture variables in $\tmp$.
    \item \caselight{Composition of substitutions $\tostructes$}.
       The only possible case is that $\esub{\var}{\sctxp{\fire}}$ is
        the outermost substitution composed by $\tostructes$.
        This is not possible if the rule is applied from left to right,
        since it would imply that
        $\evctxp\var = \evctxtwop\var\esub{\vartwo}{\tmp}$ with
        $\var \not\in \evctxtwop{\var}$, which is a contradiction.

        Finally, if the $\tostructes$ rule is applied from right to left,
        $\sctx$ is of the form $\sctxtwo\esub{\vartwo}{\tmp}$ and:
        $$\hspace{-1.0cm}
        \commuteslsvEEABCD{\tostructes}{=}{
            \evctxp{\var}\esub{\var}{\sctxtwop{\fire}\esub{\vartwo}{\tmp}}
        }{
            \sctxtwop{\evctxp{\fire}\esub{\var}{\fire}}\esub{\vartwo}{\tmp}
        }{
            \evctxp{\var}\esub{\var}{\sctxtwop{\fire}}\esub{\vartwo}{\tmp}
        }{
            \sctxtwop{\evctxp{\var}\esub{\var}{\fire}}\esub{\vartwo}{\tmp}
        }
        $$
    \end{enumerate}
  
\item \casealt{Inductive case 1: left of an application $\evctx = \evctxtwo\sctxp\fire$}
  The situation is:
  $$\tm = \evctxtwop{\tmp}\,\sctxp\fire \towhlcek \tmthree\,\sctxp\fire = \tmtwo$$
  for some $\tmthree$. If the $\tostructsym$ step is internal to $\evctxtwop{\tmp}$, the result follows
  by \ih, and if it is internal to $\sctxp{\fire}$, it is straightforward to close
  the diagram by resorting to the fact that $\eqstruct$ preserves the shape of
  $\sctxp{\fire}$ (\reflemma{nonstrict_cbv_eqstruct_preserves_shapes}).
  The nontrivial case is when the $\tostructsym$ step overlaps $\evctxtwop{\tmp}$ and $\sctxp\fire$. There are two cases:
  \begin{enumerate}
  \item \caselight{The substitution comes from $\tmp$.}
        That is, $\evctxtwo = \ctxhole$ and $\tmp$ has a substitution at its root.
        Then $\tmp$ must be a $\rtolsv$-redex
        $\tmp = \evctxthreep{\var}\esub{\var}{\sctxp{\fire}}$. The diagram then is the same as in case \ref{p:str-bis-nscbv-base-exp-apl}, reading it bottom-up.
 
        
  \item \label{p:str-bis-nscbv-ind1-b}\caselight{The substitution comes from $\evctxtwo$.}
        That is: $\evctxtwo = \evctxthree\esub{\var}{\tmfour}$ and the rewriting step is internal to $\evctxthreep\tmp$, reducing it to $\tmthreep$, \ie\ $\tmthree = \tmthreep\esub{\var}{\tmfour}$. The proof is then straightforward:
        $$
        \commutesredEEABCD{\tostructapl}{\tostructapl}{
            \evctxthreep{\tmp}\esub{\var}{\tmfour}\,\sctxp\fire
        }{
            \tmthreep\esub{\var}{\tmfour}\,\sctxp\fire
        }{
            (\evctxthreep{\tmp}\,\sctxp\fire)\esub{\var}{\tmfour}
        }{
            (\tmthreep,\sctxp\fire)\esub{\var}{\tmfour}
        }
        $$
 
 \item \label{p:str-bis-nscbv-ind1-c}\caselight{The substitution comes from $\sctx$.}
        That is: $\sctx = \sctxtwo\esub{\var}{\tmfour}$. Then
               $$
        \commutesredEEABCD{\tostructapr}{\tostructapr}{
            \evctxtwop{\tmp}\,\sctxtwop\fire\esub{\var}{\tmfour}
        }{
            \tmthree\,\sctxtwop\fire\esub{\var}{\tmfour}
        }{
            (\evctxtwop{\tmp}\,\sctxtwop\fire)\esub{\var}{\tmfour}
        }{
            (\tmthree,\sctxtwop\fire)\esub{\var}{\tmfour}
        }
        $$

  \end{enumerate}

\item \casealt{Inductive case 2: right of an application $\evctx = \tmfive\evctxtwo$}
  The situation is:
  $$\tm = \tmfive\,\evctxtwop{\tmp} \towhlcek \tmfive\,\tmfour = \tmtwo$$
  for some $\tmfour$. Reasoning as in the previous case (\caselight{left of an application}),
  if the $\tostructsym$ step is internal to $\evctxtwop{\tmp}$, the result follows
  by \ih, and it is immediate also if it is internal to $\tmfive$.

  The remaining possibility is that the $\tostructsym$ step overlaps
  with $\tmfive$ or $\evctxtwop{\tmp}$. As in the previous case,
  this is only be possible because of a \caselight{commutation with application} rule. Cases:
  \begin{enumerate}
  \item \caselight{The substitution comes from $\tmp$.}
  That is, $\evctxtwo = \ctxhole$ and $\tmp$ is a $\rtolsv$-redex $\tmp = \evctxthreep{\vartwo}\esub{\vartwo}{\sctxtwop{\fire}}$. The diagram then is the same as in case \ref{p:str-bis-nscbv-base-exp-apr}, reading it bottom-up.
  
  \item \label{p:str-bis-nscbv-ind2-b}\caselight{The substitution comes from $\evctxtwo$.}
        That is, $\evctxtwo = \evctxthree\esub{\var}{\tmthreep}$.        
        This case is then straightforward:
        $$
        \commutesredEEABCD{\tostructapr}{\tostructapr}{
            \tmfive\,\evctxthreep{\tmp}\esub{\var}{\tmthreep}
        }{
            \tmfive\,\tmfour\esub{\var}{\tmthreep}
        }{
            (\tmfive\,\evctxthreep{\tmp})\esub{\var}{\tmthreep}
        }{
            (\tmfive\,\tmfour)\esub{\var}{\tmthreep}
        }
        $$
  \item \label{p:str-bis-nscbv-ind2-c}\caselight{The substitution comes from $\tmfive$.}
        That is, $\tmfive = \tmfivep\esub{\var}{\tmthreep}$.        
        This case is  straightforward:
        $$
        \commutesredEEABCD{\tostructapl}{\tostructapl}{
            \tmfivep\esub{\var}{\tmthreep}\,\evctxtwop{\tmp}
        }{
            \tmfivep\esub{\var}{\tmthreep}\,\tmfour
        }{
            (\tmfivep\,\evctxtwop{\tmp})\esub{\var}{\tmthreep}
        }{
            (\tmfivep\,\tmfour)\esub{\var}{\tmthreep}
        }
        $$
  \end{enumerate}

\item \casealt{Inductive case 3: left of a substitution $\evctx = \evctxtwo\esub{\var}{\tmfive}$}
  The situation is:
  $$\tm = \evctxtwop{\tmp}\esub{\var}{\tmfive} \towhlcek \tmfour\esub{\var}{\tmfive} = \tmtwo$$
  If the $\tostructsym$ step is internal to $\evctxtwop{\tmp}$,
  the result follows by \ih. If it is internal to $\tmfive$, the steps are
  orthogonal, which makes the diagram trivial.  
  The remaining possibility is that the substitution $\esub{\var}{\tmfive}$
  is involved in the $\tostructsym$ redex.
  By case analysis on the kind of the step $\eqstruct$:
  \begin{enumerate}
   \item \caselight{Commutation of independent substitutions $\tostructcom$}.
	Since $\evctxtwop{\tmp}$ must have a substitution at the root,
    there are two possibilities:
    \begin{enumerate}
    \item \caselight{The substitution comes from $\tmp$.}
          That is, $\evctxtwo = \ctxhole$ and $\tmp$ is a $\rtolsv$-redex
          $\tmp = \evctxthreep{\vartwo}\esub{\vartwo}{\sctxp{\fire}}$,
          with $\var \not\in \fv{\sctxp{\fire}}$. Then:
          $$\hspace{-1.5cm}
          \commuteslsvEEABCD{\tostructcom}{\tostructcom^*}{
			\evctxthreep{\vartwo}\esub{\vartwo}{\sctxp{\fire}}\esub{\var}{\tmfive}
          }{
			\sctxp{\evctxthreep{\fire}\esub{\vartwo}{\fire}}\esub{\var}{\tmfive}
          }{
			\evctxthreep{\vartwo}\esub{\var}{\tmfive}\esub{\vartwo}{\sctxp{\fire}}
          }{
			\sctxp{\evctxthreep{\fire}\esub{\var}{\tmfive}\esub{\vartwo}{\fire}}
          }
          $$
    \item \caselight{The substitution comes from $\evctxtwo$.}
		That is, $\evctxtwo = \evctxthree\esub{\vartwo}{\tmthreep}$
		with $\var \not\in \fv{\tmthreep}$. This case is
		direct:
          $$\hspace{-1.0cm}
          \commuteslsvEEABCD{\tostructcom}{\tostructcom}{
			\evctxthreep{\tmp}\esub{\vartwo}{\tmthreep}\esub{\var}{\tmfive}
          }{
			\evctxthreep{\tmtwop}\esub{\vartwo}{\tmthreep}\esub{\var}{\tmfive}
          }{
			\evctxthreep{\tmp}\esub{\var}{\tmfive}\esub{\vartwo}{\tmthreep}
          }{
			\evctxthreep{\tmtwop}\esub{\var}{\tmfive}\esub{\vartwo}{\tmthreep}
          }
          $$
    \end{enumerate}
    
  \item \caselight{Commutation with application $\tostructap$}.
    $\evctxtwop{\tmp}$ must be an application. This allows for three
    possibilities:
    \begin{enumerate}
    \item \caselight{The application comes from $\tmp$.}
          That is, $\evctxtwo = \ctxhole$ and $\tmp$ is a $\rtodbv$-redex
          $\tmp = \sctxp{\l\vartwo.\tmpp}\,\sctxtwop{\fire}$. two sub-cases, whether $\esub{\var}{\tmfive}$ commutes on the left or on the right of the application. The left case is case \ref{p:str-bis-nscbv-base-mul-apl} (read bottom-up), while the right case is case \ref{p:str-bis-nscbv-base-mul-apr} (again bottom-up).

    \item \caselight{The application comes from $\evctxtwo = \evctxthree\,\sctxp\tmthreep$.}
        There are two sub-cases, whether $\esub{\var}{\tmfive}$ commutes on the left or on the right of the application. The left case is case \ref{p:str-bis-nscbv-ind1-b} (read bottom-up), while the right case is case \ref{p:str-bis-nscbv-ind1-c} (again bottom-up).

    \item \caselight{The application comes from $\evctxtwo = \tmfive\,\evctxthree$.}         
         Similarly to the previous case, it reduces to cases \ref{p:str-bis-nscbv-ind2-b} and \ref{p:str-bis-nscbv-ind2-c}.
    \end{enumerate}

  \item \caselight{Composition of substitutions $\tostructes$}.
	Two sub-cases:
    \begin{enumerate}
    \item \caselight{The substitution comes from $\tmp$.}
          That is, $\evctxtwo = \ctxhole$ and $\tmp$ is a $\rtolsv$-redex
          $\tmp = \evctxthreep{\vartwo}\esub{\vartwo}{\sctxp{\fire}}$,
          with $\var \not\in \fv{\evctxthreep{\vartwo}}$. Then:
          $$\hspace{-1.5cm}
          \commuteslsvEEABCD{\tostructes}{=}{
			\evctxthreep{\vartwo}\esub{\vartwo}{\sctxp{\fire}}\esub{\var}{\tmfive}
          }{
			\sctxp{\evctxthreep{\fire}\esub{\vartwo}{\fire}}\esub{\var}{\tmfive}
          }{
			\evctxthreep{\vartwo}\esub{\vartwo}{\sctxp{\fire}\esub{\var}{\tmfive}}
          }{
			\sctxp{\evctxthreep{\fire}\esub{\vartwo}{\fire}}\esub{\var}{\tmfive}
          }
          $$
    \item \caselight{The substitution comes from $\evctxtwo$.}
		That is, $\evctxtwo = \evctxthree\esub{\vartwo}{\tmthreep}$
		with $\var \not\in \fv{\evctxthreep{\tmp}}$. Then:
          $$\hspace{-1.5cm}
          \commutesredEEABCD{\tostructes}{\tostructes}{
			\evctxthreep{\tmp}\esub{\vartwo}{\tmthreep}\esub{\var}{\tmfive}
          }{
			\evctxthreep{\tmtwop}\esub{\vartwo}{\tmthreep}\esub{\var}{\tmfive}
          }{
			\evctxthreep{\tmp}\esub{\vartwo}{\tmthreep\esub{\var}{\tmfive}}
          }{
			\evctxthreep{\tmtwop}\esub{\vartwo}{\tmthreep\esub{\var}{\tmfive}}
          }
          $$
    
	\end{enumerate}\qedhere
  \end{enumerate}
  
\end{enumerate}

\end{proofatend}

\begin{atend}
A final lemma about the $\tostruct$ relation will be useful later:

\begin{lemma}[ES Commute with Evaluation Contexts via $\tostruct$] 
	\label{l:ev-comm-struct}
	Let $\tctx$ be a shallow context s.t. $\var\notin\fv\tctx$ and $\tctx$ doe not capture the variables in $\fv\tmtwo$. Then $\tctxp{\tm\esub{\var}{\tmtwo}} \eqstruct \tctxp{\tm}\esub{\var}{\tmtwo}$.
\end{lemma}

\begin{IEEEproof}
  by induction on $\tctx$.
  \begin{enumerate}
   \item \emph{Empty Context $\tctx = \ctxhole$}. Then $\tctxp{\tm}\esub{\var}{\tmtwo} = \tm\esub{\var}{\tmtwo} = \tctxp{\tm\esub{\var}{\tmtwo}}$.
   \item \emph{Application Left $\tctx = \tctxtwo\tmthree$}. Then
   
   \begin{center}$\begin{array}{lll}
    \tctxp{\tm\esub{\var}{\tmtwo}} & = \\
    \tctxtwop{\tm\esub{\var}{\tmtwo}}\tmthree & = & \mbox{(by \ih)}\\
    \tctxtwop{\tm}\esub{\var}{\tmtwo}\tmthree & \tostructapl \\ 
    (\tctxtwop{\tm}\tmthree) \esub{\var}{\tmtwo} & =\\
    \tctxp{\tm}\esub{\var}{\tmtwo}    
   \end{array}$\end{center}
   
   \item \emph{Application Right $\evctx = \tmthree \evctxtwo$}. Then
   
   \begin{center}$\begin{array}{lll}
    \tctxp{\tm\esub{\var}{\tmtwo}} & = \\
    \tmthree\tctxtwop{\tm\esub{\var}{\tmtwo}} & = & \mbox{(by \ih)}\\
    \tmthree\tctxtwop{\tm}\esub{\var}{\tmtwo} & \tostructapr \\
    (\tmthree\tctxtwop{\tm}) \esub{\var}{\tmtwo} & =\\
    \tctxp{\tm}\esub{\var}{\tmtwo}    
   \end{array}$\end{center}
   
   \item \emph{Substitution $\evctx = \evctxtwo\esub\vartwo\tmthree$}. Then
   
   \begin{center}$\begin{array}{lll}
    \tctxp{\tm\esub{\var}{\tmtwo}} & =\\
    \tctxtwop{\tm\esub{\var}{\tmtwo}}\esub\vartwo\tmthree & = & \mbox{(by \ih)}\\
    \tctxtwop{\tm}\esub{\var}{\tmtwo}\esub\vartwo\tmthree & \tostructcom \\
    \tctxtwop{\tm}\esub\vartwo\tmthree \esub{\var}{\tmtwo} & = \\
    \tctxp{\tm}\esub{\var}{\tmtwo}    
   \end{array}$\end{center}
   Note that $\tostructcom$ can be applied because of the hypotheses $\var\notin\fv\tctx$ and $\tctx$ doe not capture the variables in $\fv\tmtwo$.
  \end{enumerate}
\end{IEEEproof}
\end{atend}

\paragraph*{Size Explosion, Again} coming back to the size explosion example, the idea is that---to circumvent it---$\tm_n$ should better $\togenm$-evaluate to:
$$ \tmfour_n \defeq (\var_{0}\var_{0})\esub{\var_{0}}{\var_{1}^2}\esub{\var_{1}}{\var_{2}^2}\ldots\esub{\var_{n-1}}{\var_{n}^2}\esub{\var_{n}}\gconst$$
which is an alternative, compact representation of $\gconst^{2^n}$, of size linear in $n$, and with just one occurrence of $\gconst$.
Without \symbs, ES are enough to circumvent size explosion \cite{DBLP:conf/fpca/BlellochG95,DBLP:conf/birthday/SandsGM02,DBLP:journals/tcs/LagoM08}. In our case however they fail. The evaluation we just defined indeed does not stop on the desired compact representation, and in fact a linear number of steps (namely $3n$) may still produce an exponential output (in a substitution context).

\begin{proposition}[Size Explosion in the \ShFC]
	$\tm_n \gconst (\togensm\togense^2)^n \sctxp{\gconst^{2^n}}$.
\end{proposition}

\begin{IEEEproof}
	by induction on $n$. Let $\gconsttwo \defeq \gconst^2 = \gconst \gconst$. Cases:
$$\begin{array}{llllll}
	\tm_1 & = & (\la{\var_{1}}(\var_{1}\var_{1})) \gconst &	\togensm\\
	&   & (\var_{1}\var_{1})\esub{\var_{1}}\gconst &\togense \\
	&   & (\var_{1}\gconst)\esub{\var_{1}}\gconst &\togense \\
	&   & (\gconst\gconst)\esub{\var_{1}}\gconst &=&\gconst^2\esub{\var_{1}}\gconst	
\end{array}$$
$$\begin{array}{llllll}
	\tm_{n+1} & = & (\la{\var_{n+1}}(\tm_{n}(\var_{n+1}\var_{n+1}))) \gconst & \togensm\togense^2 \\
	& & (\tm_n \gconst^2)\esub{\var_{1}}\gconst = \sctxp{\tm_n \gconsttwo} & (\togensm\togense^2)^n\mbox{ (\ih)}\\
	& & \sctxtwop{\gconsttwo^{2^n}}  =  \sctxtwop{\gconst^{2^{n+1}}} & \hfill\qedhere
\end{array}$$
\end{IEEEproof}

Before introducing useful evaluation---that will liberate us from size explosion---we are going to fully set up the architecture of the problem, by explaining 1) how ES implement a calculus, 2) how an abstract machine implements a calculus with ES, and 3) how to define an abstract machine for the inefficient \ShFC. Only by then (\refsect{useful}) we will start optimising the framework, first with useful sharing and then by eliminating renaming chains.

\section{Two Levels Implementation}
Here we explain how the the small-step strategy $\tof$ of the \FC\ is implemented by a micro-step strategy $\togen$. We are looking for an appropriate strategy $\calculus$ with ES which is polynomially related to both $\tof$ and an abstract machine. Then we need two theorems:
\begin{enumerate}
\item 
  \emph{High-Level Implementation}: $\tof$ terminates iff
  $\calculus$ terminates. Moreover, $\tof$ is implemented by $\calculus$
  with only a polynomial overhead. Namely, $\tm\calculus^k\tmtwo$ iff
  $\tm\tof^h\unf{\tmtwo}$ with $k$ polynomial in $h$;
\item 
  \emph{Low-Level Implementation}: $\calculus$ is implemented on an abstract machine with an overhead in time which is polynomial in both $k$ and the size of $\tm$.
\end{enumerate}
We will actually be more accurate, giving linear or quadratic bounds, but this is the general setting.

\subsection{High-Level Implementation}
\label{ssect:high-level-impl}
\begin{atend}\section{Proofs Omitted From \refssect{high-level-impl}\\ (High-Level Implementation)}
\label{sect:app-high-level-impl}
First, the High-Level Implementation Theorem.
\end{atend}
First, terminology and notations. \emph{Derivations} $\deriv,\derivtwo,\ldots$ are sequences of rewriting steps. With $\size\deriv$, $\sizem\deriv$, and $\sizee\deriv$ we denote respectively the length, the number of multiplicative, and exponential steps of $\deriv$. 

\begin{definition}\label{def:hlis}
Let $\tof$ be a deterministic strategy on \FC-terms and $\calculus$ a
deterministic strategy for terms with ES. The pair $(\tof,\calculus)$ is a
\deff{high-level implementation system} if whenever $\tm$ is a
$\l$-term and $\deriv:\tm\calculus^*\tmtwo$ then:
\begin{enumerate}
  \item 
    \emph{Normal Form}: if $\tmtwo$ is a $\calculus$-normal form then
    $\unf{\tmtwo}$ is a $\tof$-normal form.
  \item 
    \emph{Projection}: $\unf{\deriv}:\unf{\tm}\tof^*\unf{\tmtwo}$ and
    $\size{\unf{\deriv}}=\sizem\deriv$.
    \end{enumerate}
Moreover, it is 
\begin{enumerate}
	\item \emph{locally bounded}: if the length of a sequence of substitution $\esym$-steps from $\tmtwo$ is linear in the number $\sizem\deriv$ of $\msym$-steps in $\deriv$;
	\item \emph{globally bounded}: if $\sizee\deriv$ is linear in $\sizem\deriv$.
\end{enumerate}
\end{definition}

The normal form and projection properties address the
\emph{qualitative} part, \ie\
the part about termination. The normal form property guarantees that
$\calculus$ does not stop prematurely, so that when $\calculus$ terminates
$\tof$ cannot keep going. The projection property guarantees that
termination of $\tof$ implies termination of $\calculus$. The two
properties actually state a stronger fact: \emph{$\tof$ steps can
  be identified with the $\togenm$-steps of the $\calculus$ strategy}.

The local and global bounds allow to bound the overhead introduced by the \ShFC{} wrt the \FC, because by relating $\togenm$ and $\togene$ steps, they relate $\size\deriv$ and $\size{\unf\deriv}$, since $\tof$ and $\togenm$ steps can be identified.

The high-level part can now be proved abstractly.
\begin{theorem}[High-Level Implementation]\label{thm:hlithm}
  Let $\tm$ be an ordinary $\l$-term and $(\tof,\calculus)$ a
  high-level implementation system.
  \begin{enumerate}
	  \item \emph{Normalisation}: $\tm$ is $\tof$-normalising iff it is $\calculus$-normalising.
	  \item \emph{Projection}: if $\deriv:\tm\calculus^*\tmtwo$ then
    $\unf{\deriv}:\tm\tof^*\unf{\tmtwo}$.
  \end{enumerate}
  Moreover, the overhead of $\calculus$ is, depending on the system:
  \begin{enumerate}
	\item \emph{locally bounded}: quadratic, \ie\ $\size\deriv=O(\size{\unf{\deriv}}^2)$.
	\item \emph{globally bounded}: linear, \ie\ $\size\deriv=O(\size{\unf{\deriv}})$.
\end{enumerate}    
\end{theorem}

\begin{proofatend}
the proof is a minimal variation over the proof of Theorem 4.2, page 4, in \cite{DBLP:conf/csl/AccattoliL14}. Essentially we merged the trace and syntactic bound properties of that statement into our locally bound property. Note that for the global bound there is nothing to prove, it follows from the the hypothesis itself and projection.
\end{proofatend}
\noindent For the low-level part, in contrast to \cite{DBLP:conf/csl/AccattoliL14}, we rely on abstract machines, introduced in the next section.

Let us see our framework at work. We have the following result: 

\begin{atend}

Now, we prove that $(\tof,\togens)$ is a high-level implementation system, \ie\ \refthm{explicithlis}.

The normal form property required for high-level implementation
system has already been proved (\refthm{ES-determ-and-nfs}). It only remains to prove the projection property.

\begin{lemma}[Projection of a Rewriting Step]
	\label{l:unfolding-steps} 
	Let $\tm = \evctxp\tmtwo$ and $\evctx$ be an evaluation context. 
	\begin{enumerate}
		\item \label{p:unfolding-steps-two} \emph{Multiplicative Projection}: if $\tm\togensm\tmthree$ then $\unf\tm \tof \unf\tmthree$. More precisely, if $\evctxp\tmtwo \togensm \evctxp\tmthree$ with $\tmtwo\rtodbv\tmthree$ then $\unf\evctx{\ctxholep{\relunf\tmtwo\evctx}} \tof \unf\evctx{\ctxholep{\relunf\tmthree\evctx}}$ with $\relunf\tmtwo\evctx \rtof \relunf\tmthree\evctx$;
		\item \label{p:unfolding-steps-three} \emph{Exponential Projection}: if $\tm\togense\tmthree$ then $\unf\tm = \unf\tmthree$;
	\end{enumerate}
\end{lemma}

\myproof{
\begin{IEEEproof}
	\hfill
	\begin{enumerate}
		\item By induction on $\evctx$. Cases:
		
		\begin{enumerate}
			\item \emph{Empty Context $\evctx = \ctxhole$}. Let $\tm = \sctxp{\l\var.\tmfour}\sctxtwop\fire \rtodbv \sctxp{\tmfour\esub\var{\sctxtwop\fire}} = \tmthree$. By induction on $\sctx$. Two cases:
				\begin{enumerate}
					\item \emph{Empty context $\sctx = \ctxhole$}. Then\\ $\tm = \l\var.\tmfour\sctxtwop\fire \rtodbv \tmfour\esub\var{\sctxtwop\fire} = \tmthree$
					\[\begin{array}{ll}
						\unf\tm & =\\
                                                \unf{((\l\var.\tmfour)\sctxtwop\fire)} & = \\
						(\l\var.\unf{\tmfour})\unf{\sctxtwop\fire} & \tof \\
						\unf{\tmfour}\isub\var{\unf{\sctxtwop\fire}} & =\\
						\unf{\tmfour\esub\var{\sctxtwop\fire}} & =\\
                                                \unf\tmthree
					\end{array}\]
			
					\item \emph{Substitution $\sctx = \sctxtwo \esub\vartwo\tmfive$}. Then\\ \mbox{$\tm = \sctxp{\l\var.\tmfour}\esub\vartwo\tmfive\sctxtwop\fire \rtodbv$} \mbox{$\sctxp{\tmfour\esub\var{\sctxtwop\fire}}\esub\vartwo\tmfive = \tmthree$}. We have
					\[\begin{array}{lll}
						\unf\tm & =\\
                                                \unf{(\sctxp{\l\var.\tmfour}\esub\vartwo\tmfive\sctxtwop\fire)} & =	\\
						\unf{\sctxp{\l\var.\tmfour}\esub\vartwo\tmfive}\unf{\sctxtwop\fire} & = \\
						\unf{\sctxp{\l\var.\tmfour}}\isub\vartwo{\unf\tmfive}\unf{\sctxtwop\fire} & =\\
						\unf{\sctxp{\l\var.\tmfour}}\unf{\sctxtwop\fire} \isub\vartwo{\unf\tmfive} & = \\
						\unf{(\sctxp{\l\var.\tmfour}\sctxtwop\fire)} \isub\vartwo{\unf\tmfive} & \rtof & \mbox{(\ih}\\
                                               && \mbox{and \reflemma{tof-and-subs})} \\
						\unf{\sctxp{\tmfour\esub\var{\sctxtwop\fire}}} \isub\vartwo{\unf\tmfive} & = \\
						\unf{\sctxp{\tmfour\esub\var{\sctxtwop\fire}} \esub\vartwo{\tmfive}} & =\\
						\unf\tmthree
					\end{array}\]
				\end{enumerate}
		
			\item \emph{Application Left $\evctx = \evctxtwo\sctxp\fire$}. Then\\ $\evctxtwop\tmtwo\sctxp\fire \togensm \evctxtwop\tmthree\sctxp\fire$ with $\tmtwo\rtodbv \tmthree$. We have:
			\[\begin{array}{lll}
				\unf{\evctxp\tmtwo} & =\\
                                \unf{(\evctxtwop\tmtwo\sctxp\fire)} & =\\
				\unf{\evctxtwop\tmtwo} \unf{\sctxp\fire} & \tof & \mbox{(\ih)} \\
				\unf{\evctxtwop\tmthree} \unf{\sctxp\fire} & =	 \\ 
				\unf{(\evctxtwop\tmthree{\sctxp\fire})} & =\\
				\unf{\evctxp\tmthree}
			\end{array}\]
		Actually, the $\tof$ step is justified by the \ih\ and the fact that $\ctxhole \unf{\sctxp\fire}$ is an evaluation context because $\unf{\sctxp\fire}$ is a \fireball\ (by \reflemma{es-answers-are-normal}). The \ih\ also gives $\relunf\tmtwo\evctxtwo \rtof \relunf\tmthree\evctxtwo$. To conclude note that $\relunf\tmtwo\evctx = \relunf\tmtwo{\evctxtwo\fire} = \relunf\tmtwo\evctxtwo \rtof \relunf\tmthree\evctxtwo = \relunf\tmthree{\evctxtwo\fire} = \relunf\tmthree\evctx$.

			\item \emph{Application Right $\evctx = \tmthree \evctxtwo$}. Follows from the \ih, along the lines of the previous case.
			
			\item \emph{Substitution $\evctx = \evctxtwo\esub\var\tmfour$}. Then $\evctxtwop\tmtwo\esub\var\tmfour \togensm \evctxtwop\tmthree\esub\var\tmfour$ with $\tmtwo\rtodbv \tmthree$. We have:
			\[\begin{array}{lll}
				\unf{\evctxp\tmtwo} & =\\
                                \unf{\evctxtwop\tmtwo\esub\var\tmfour} & =\\
				\unf{\evctxtwop\tmtwo} \isub\var{\unf\tmfour} 	& \tof & \mbox{(\ih\ and \reflemma{tof-and-subs})} \\
				\unf{\evctxtwop\tmthree} \isub\var{\unf\tmfour}  & = \\
				\unf{\evctxtwop\tmthree\esub\var\tmfour} & =\\
				\unf{\evctxp\tmthree}
			\end{array}\]
			The \ih\ also gives $\relunf\tmtwo\evctxtwo \rtof \relunf\tmthree\evctxtwo$. To conclude note that 
			\[\begin{array}{lll}
				\relunf\tmtwo\evctx & =\\
                                \relunf\tmtwo{\evctxtwo\esub\var\tmfour} & =\\
				\relunf\tmtwo\evctxtwo \isub\var{\unf\tmfour} 	& \rtof & \mbox{(\reflemma{tof-and-subs})} \\
				\relunf\tmthree\evctxtwo \isub\var{\unf\tmfour} & =	 \\ 
				\relunf\tmthree{\evctxtwo\esub\var\tmfour} & =\\
				\relunf\tmthree\evctx
			\end{array}\]
		\end{enumerate}
		
		\item We prove that if $\tm\rtolsv\tmthree$ then $\unf\tm=\unf\tmthree$ for any evaluation context $\evctx$. From \reflemmap{relunf-properties}{two} the statement follows. By induction on $\evctx$. We have $\tm = \evctxtwop\var\esub\var{\sctxp{\fire}} \rtolsv  \sctxp{\evctxtwop\fire\esub\var\fire} = \tmthree$. By induction on $\sctx$. Two cases:
				\begin{enumerate}
					\item \emph{Empty context $\sctx = \ctxhole$}. Then $\tm = \evctxtwop\var\esub\var{\fire} \rtolsv  \evctxtwop\fire\esub\var\fire = \tmthree$
					\[\begin{array}{lll}
						\unf\tm & =\\
                                                \unf{\evctxtwop\var\esub\var\fire}  &=	\\
						\unf{\evctxtwop\var}\isub\var{\unf\fire} & =\\
						\unf{\evctxtwo}\ctxholep{\relunf\var\evctxtwo}\isub\var{\unf\fire} &=& \mbox{(by \reflemmap{relunf-properties}{four})}\\
						\unf{\evctxtwo}\ctxholep{\var}\isub\var{\unf\fire} &=\\
						\unf{\evctxtwo}\ctxholep{\unf\fire}\isub\var{\unf\fire} &=\\
						\unf{\evctxtwo}\ctxholep{\relunf\fire\evctxtwo}\isub\var{\unf\fire} &=\\
						\unf{\evctxtwop\fire}\isub\var{\unf\fire} &=\\
						\unf{\evctxtwop\fire\esub\var\fire} \\
						\unf\tmthree
					\end{array}\]
			
					\item \emph{Substitution $\sctx = \sctxtwo \esub\vartwo\tmfive$}. Then $\tm = \evctxtwop\var\esub\var{\sctxp{\fire}\esub\vartwo\tmfive} \rtolsv  \sctxp{\evctxtwop\fire\esub\var\fire}\esub\vartwo\tmfive = \tmthree$. 
					\[\begin{array}{lll}
						\unf\tm & =\\
                                                \unf{\evctxtwop\var\esub\var{\sctxp{\fire}\esub\vartwo\tmfive}} &=	\\
						\unf{\evctxtwop\var}\isub\var{\unf{\sctxp\fire\esub\vartwo\tmfive}} &= \\
						\unf{\evctxtwop\var}\isub\var{\unf{\sctxp\fire}\isub\vartwo{\unf\tmfive}} &=\\
						\unf{\evctxtwop\var}\isub\var{\unf{\sctxp\fire}}\isub\vartwo{\unf\tmfive} &=\\
						\unf{\evctxtwop\var\esub\var{\sctxp\fire}}\isub\vartwo{\unf\tmfive} & = & \mbox{(by \ih)} \\
						\unf{\sctxp{\evctxtwop\fire\esub\var\fire}}\isub\vartwo{\unf\tmfive} &=\\
						\unf{\sctxp{\evctxtwop\fire\esub\var\fire}\esub\vartwo\tmfive}&= \\
						\unf\tmthree
					\end{array}\]
				\end{enumerate}
	\end{enumerate}
\end{IEEEproof}
}
 
 To prove that $(\tof,\togens)$ is a high-level implementation system we only have to put together the various results.
\end{atend}

\begin{theorem}\label{thm:explicithlis}
	$(\tof,\togens)$ is a high-level implementation system.
\end{theorem}
\begin{proofatend}
immediate from \refcoro{es-closed-determ} and
\reflemma{unfolding-steps}.
\end{proofatend}

Note the absence of complexity bounds. In fact, $(\tof,\togens)$ is not even locally bounded. Let $\tm^n$ here be defined by $\tm^1 = \tm$ and $\tm^{n+1} = \tm^{n} \tm$, and $\tmtwo_n \defeq (\la\var\var^n) \gconst$. Then $\deriv:\tmtwo_n \togenm\togene^n \gconst^n\esub\var\gconst$ is a counter-example to local boundedness. Moreover, the \ShFC\ also suffers of size explosion, \ie\ implementing a single step may take exponential time. In \refsect{useful} the introduction of useful sharing will solve these issues.

\subsection{Low-Level Implementation: Abstract Machines}
\label{ssect:low-level-impl}
\begin{atend}\section{Proofs Omitted From \refssect{low-level-impl}\\ (Low-Level Implementation: Abstract Machines)}

\end{atend}

\paragraph*{Introducing Distilleries} an abstract machine $\mach$ is meant to implement a strategy $\calculus$ via a \emph{distillation}, \ie\ a decoding function $\decodefun$. A machine has a state $\state$, given by a \emph{code} $\code$, \ie\ a $\l$-term $\tm$ without ES and not considered up to $\alpha$-equivalence, and some data-structures like stacks, dumps, environments, and eventually heaps. The data-structures are used to implement the search of the next $\calculus$-redex and some form of parsimonious substitution, and they distill to evaluation contexts for $\calculus$. Every state $\state$ decodes to a term $\decode\state$, having the shape $\evctxp\tm$, where $\tm$ is a $\l$-term and $\evctx$ is some kind of evaluation context for $\calculus$. 

A machine computes using transitions, whose union is noted $\tomach$, of two types. The \emph{principal} one, noted $\tomachp$, corresponds to the firing of a rule defining $\calculus$. In doing so, the machine can differ from the calculus implemented by a transformation of the evaluation context to an equivalent one, up to a structural congruence $\eqstruct$. The \emph{commutative} transitions, noted $\tomachc$, implement the search for the next redex to be fired by rearranging the data-structures to single out a new evaluation context, and they are invisible on the calculus. The names reflect a proof-theoretical view, as machine transitions can be seen as cut-elimination steps \cite{DBLP:journals/toplas/AriolaBS09,DBLP:conf/icfp/AccattoliBM14}. Garbage collection is here simply ignored, as in the LSC it can always be postponed.

To preserve correctness, structural congruence $\eqstruct$ is required to commute with evaluation $\calculus$, \ie\ to satisfy
\begin{center}
\vspace{-0.5cm}
\begin{tabular}{c@{}c@{}c@{}c@{}c@{}c@{}c}\hspace{-0.55cm}
 $\Bigg($\begin{tikzpicture}[ocenter]
  \node (s) {\footnotesize$\tm$};
  \node at (s.center)  [below =0.8*\nodeVerDist](s2) {\footnotesize$\tmtwo$};
  \node at (s.center) [right= 0.4*\nodeHorDist](t) {\footnotesize$\tmfour$};

  \node at (s.center)[anchor = center, below=0.3*\nodeVerDist](eq1){\footnotesize$\tostruct$};
    \draw[-o] (s) to  (t);
\end{tikzpicture}

&  {\footnotesize $\Rightarrow \exists \tmfive$ s.t.} & 
 \begin{tikzpicture}[ocenter]
  \node (s) {\footnotesize$\tm$};
  \node at (s.center)  [below =0.8*\nodeVerDist](s2) {\footnotesize$\tmtwo$};
  \node at (s.center) [right= 0.4*\nodeHorDist](t) {\footnotesize$\tmfour$};
  \node at (s2-|t) [](s1){\footnotesize$\tmfive$};

  \node at (s.center)[anchor = center, below=0.3*\nodeVerDist](eq1){\footnotesize$\tostruct$};
  \node at (t.center)[anchor = center, below=0.3*\nodeVerDist](eq2){\footnotesize$\tostruct$};
    \draw[-o] (s) to  (t);
	\draw[-o] (s2) to  (s1);
\end{tikzpicture}$\Bigg)$ 
&
$\wedge$
&
 $\Bigg($\begin{tikzpicture}[ocenter]
  \node (s) {\footnotesize$\tm$};
  \node at (s.center)  [below =0.8*\nodeVerDist](s2) {\footnotesize$\tmtwo$};
  \node at (s.center) [right= 0.4*\nodeHorDist](t) {};
  \node at (s2-|t) [](s1){\footnotesize$\tmfive$};

  \node at (s.center)[anchor = center, below=0.3*\nodeVerDist](eq1){\footnotesize$\tostruct$};
	\draw[-o] (s2) to  (s1);
\end{tikzpicture}

&  {\footnotesize $\Rightarrow \exists \tmfour$ s.t.} & 
 \begin{tikzpicture}[ocenter]
  \node (s) {\footnotesize$\tm$};
  \node at (s.center)  [below =0.8*\nodeVerDist](s2) {\footnotesize$\tmtwo$};
  \node at (s.center) [right= 0.4*\nodeHorDist](t) {\footnotesize$\tmfour$};
  \node at (s2-|t) [](s1){\footnotesize$\tmfive$};

  \node at (s.center)[anchor = center, below=0.3*\nodeVerDist](eq1){\footnotesize$\tostruct$};
  \node at (t.center)[anchor = center, below=0.3*\nodeVerDist](eq2){\footnotesize$\tostruct$};
    \draw[-o] (s) to  (t);
	\draw[-o] (s2) to  (s1);
\end{tikzpicture}$\Bigg)$
\end{tabular}
\end{center}
for each of the rules of $\calculus$, preserving the kind of rule. In fact, this means that $\eqstruct$ is a \emph{strong} bisimulation (\ie\ \emph{one} step to \emph{one} step) with respect to $\calculus$. Strong bisimulations formalise transformations which are transparent with respect to the behaviour, even at the level of complexity, because they can be retarded without affecting the length of evaluation:

\begin{lemma}[$\tostruct$ Postponement]
	\label{l:postponement}
	If $\eqstruct$ is a strong bisimulation and $\tm\mathrel{(\to\cup\eqstruct)^*}\tmtwo$ then $\tm\mathrel{\to^*\eqstruct}\tmtwo$ and the number and kind of steps of $\calculus$ in the two reduction sequences is the same.
\end{lemma}
\begin{proofatend}
straightforward induction on the length of $\tm\mathrel{(\to\cup\eqstruct)^*}\tmtwo$, using the strong bisimulation property.
\end{proofatend}

We can finally introduce distilleries, \ie\ systems where a strategy $\calculus$ simulates a machine $\mach$ up to structural equivalence $\eqstruct$ (via the decoding $\decodefun$).

\begin{definition}
A \emph{distillery} $\distil = (\mach, \calculus, \tostruct,\decodefun)$ is given by:
\begin{enumerate}
	\item An \emph{abstract machine} $\mach$, given by
	\begin{enumerate}
		\item a deterministic labeled \emph{transition} system $\tomach$ on states $\state$;
		\item a distinguished class of states deemed \emph{initial}, in bijection with closed $\l$-terms and from which one obtains the \emph{reachable} states by applying $\tomach^\ast$;
		\item a partition of the labels of the transition system $\tomach$ as:
		\begin{itemize}			
			\item \emph{principal} transitions, noted $\tomachp$,
			\item \emph{commutative} transitions, noted $\tomacha$;
		\end{itemize}
		
	\end{enumerate}
	
		\item a deterministic \emph{strategy} $\calculus$;

	 \item a \emph{structural equivalence} $\eqstruct$ on terms s.t. it is a strong bisimulation with respect to $\calculus$;

	\item a \emph{distillation} $\decodefun$, \ie\ a decoding function from states to terms, s.t. on reachable states:
	\begin{itemize}
		\item \emph{Principal}: $\state\tomachp\statetwo$ implies $\decode\state\calculus\eqstruct\decode\statetwo$,
		\item \emph{Commutative}: $\state\tomacha\statetwo$ implies $\decode\state\eqstruct\decode\statetwo$.		
	\end{itemize}
\end{enumerate}
\end{definition}

We will soon prove that a distillery implies a simulation theorem, but we want a stronger form of relationship. Additional hypothesis are required to obtain the converse simulation, handle explicit substitution, and talk about complexity bounds. 

Some terminology first. An \emph{execution} $\exec$ is a sequence of transition from an initial state. With $\size\exec$, $\sizep\exec$, and $\sizecom\exec$ we denote respectively the length, the number of principal, and commutative transitions of $\exec$. The \emph{size} of a term is noted $\size\tm$.

\begin{definition}[Distillation Qualities]\label{def:distillationqual}
A distillery is 
\begin{itemize}
	\item \emph{Reflective} when on reachable states:
\begin{itemize}
	\item \emph{Termination}: 	$\tomacha$ \emph{terminates};
	
	\item \emph{Progress}: if $\decode\state$ reduces then $\state$ is not final.
\end{itemize}

	\item \emph{Explicit} when 
	\begin{itemize}
		\item \emph{Partition}: principal transitions are partitioned into \emph{multiplicative} $\tomachm$ and \emph{exponential} $\tomache$, like for the strategy $\togen$.
		\item \emph{Explicit decoding}: the partition is preserved by the decoding, \ie
		\begin{itemize}
			\item \emph{Multiplicative}: $\state\tomachm\statetwo$ implies $\decode\state\togenm\eqstruct\decode\statetwo$;
			\item \emph{Exponential}: $\state\tomache\statetwo$ implies $\decode\state\togene\eqstruct\decode\statetwo$;
		\end{itemize}
	\end{itemize}

	\item \emph{Bilinear} when it is reflective and 
\begin{itemize}
	\item \emph{Execution Length}: given an execution $\exec$ from an initial term $\tm$, the number of commutative steps $\sizecom\exec$ is linear in both $\size\tm$ and $\sizep\exec$ (with a slightly stronger dependency on $\size\tm$, due to the time needed to recognise a normal form), \ie\ if $ \sizecom\exec= O((1+\sizep\exec)\cdot\size{\tm})$.
	\item \emph{Commutative}: $\tomachc$ is implementable on RAM in a constant number of steps;
	\item \emph{Principal}: $\tomachp$ is implementable on RAM in $O(\size\tm)$ steps.
\end{itemize}
\end{itemize}
\end{definition}

A reflective distillery is enough to obtain a bisimulation between the strategy $\togen$ and the machine $\mach$, that is strong up to structural equivalence $\eqstruct$. With $\sizem\exec$ and  $\sizee\exec$ we denote respectively the number of multiplicative and exponential transitions of $\exec$.
\begin{table*}
\caption{\GLAM: data-structures, decoding and transitions}
\label{tab:explicit-transitions}
\label{tab:transitions}

\centering
$
\begin{array}{c}
\begin{array}{c|c}
\begin{array}{rclllrcllllllll}
        \stackitem & \grameq & \code \mid \pair{\code}{\stack}&&
	\genv,\genvtwo		& \grameq & \stempty \mid \esub\var{\code}\cons\genv \\
	\stack,\stacktwo 	& \grameq & \stempty \mid \stackitem \cons \stack&&

	\state,\statetwo		& \grameq & \glamst\dump\code\stack\genv\\
	\dump, \dumptwo	& \grameq & \stempty \mid \dump \cons \dentry\code\stack \\
\end{array} &
\begin{array}{rclllrcllllllll}
	\decode{\stempty}	& \defeq & \ctxhole &&	
        \decode{ \esub\var\code \cons \genv} 				& \defeq & \decgenvp{\ctxhole\esub\var\code} \\
	\decode{ \stackitem \cons \stack} 			& \defeq & \decstackp{\ctxhole\decode{\stackitem}} &&
	\decode{\evctx_{\state}}				& \defeq & \decgenvp{\decdump\ctxholep{\decstack}} \\
        \decode{\pair{\tm}{\stack}} & \defeq & \decstackp{\tm} &&
	\decode{\state}				& \defeq & \decodep{\evctx_{\state}}{\tm} \\
        \decode{\dump\cons\dentry\code\stack} & \defeq & \decdumpp{\decstackp{\code\ctxholep}} &&\multicolumn{3}{l}{\mbox{where $\state = \glamst\dump\code\stack\genv$}}\\
\end{array}
\end{array}
\\\\\hline\\
  	{\setlength{\arraycolsep}{1em}
  	\begin{array}{c|c|c|ccc|c|c|c}
		\dump & \code\codetwo & \stack & \genv
	  	&\tomachcone&
	  	\dump\cons\dentry\code\stack & \codetwo & \stempty &\genv
	  	\\
	  	
		\dump & \l\var.\code & \codetwo\cons\stack & \genv
	  	&\tomachsm &
		\dump & \code & \stack & \esub\var\codetwo\genv
	  	\\
	  	
		\dump\cons(\code,\stack) & \const & \stacktwo & \genv
		& \tomachcthree &
		\dump & \code & \pair{\const}{\stacktwo}\cons\stack & \genv
		\\

		\dump\cons\dentry\code\stack & \l\var.\codetwo & \stempty & \genv
		& \tomachctwo &
		\dump & \code & \l\var.\codetwo\cons\stack & \genv
		\\

		\dump & \var & \stack & \genv_1\esub\var\codetwo\genv_2
		& \tomachse &
		\dump & \rename{\codetwo} & \stack & \genv_1\esub\var\codetwo\genv_2
  	\end{array}}
\end{array}$

\begin{minipage}{14cm}
~\\where $\rename{\codetwo}$ is any code $\alpha$-equivalent to $\codetwo$ that preserves well-naming of the machine, i.e. such that any bound name in $\rename{\codetwo}$ is fresh with respect to those in $\dump$, $\stack$ and $\genv_1\esub\var\codetwo\genv_2$.
\end{minipage}
  \end{table*}

\begin{theorem}[Correctness and Completeness]
	\label{tm:GenSim} 
	Let $\distil$ be a reflective distillery and $\state$ an initial state.
	\begin{enumerate}
		\item \emph{Strong Simulation}: for every execution $\exec:\state\tomach^*\statetwo$ there is a derivation $\deriv:\decode\state\togen^*\tostruct\decode\statetwo$ s.t. $\sizep\exec=\size\deriv$.
		\item \label{p:GenSim-two}\emph{Reverse Strong Simulation}: for every derivation $\deriv:\decode\state\togen^*\tm$ there is an execution $\exec:\state\tomach^*\statetwo$ s.t. $\tm\tostruct\decode\statetwo$ and $\sizep\exec=\size\deriv$.
	\end{enumerate}
	Moreover, if $\distil$ is explicit then $\sizem\exec=\sizem\deriv$ and $\sizee\exec=\sizee\deriv$.
\end{theorem}
\begin{proofatend}
the proof can be found in~\cite{DBLP:conf/icfp/AccattoliBM14}
(Theorems 4.2 and 4.4) up to trivial modifications due to minor changes
in the definition of distilleries and their properties.
\end{proofatend}

Bilinearity, instead, is crucial for the low-level theorem. 

\begin{theorem}[Low-Level Implementation Theorem]\label{th:lowlevelimplth}
Let $\calculus$ be a strategy on terms with ES s.t. there exists a bilinear distillery $\distil = (\mach, \calculus, \tostruct,\decodefun)$. Then a $\togen$-derivation $\deriv$ is implementable on RAM machines in $O((1+\size\deriv)\cdot \size\tm)$ steps, \ie\ bilinear in the size of the initial term $\tm$ and the length of the derivation $\size\deriv$.
\end{theorem}

\begin{IEEEproof}
given $\deriv:\tm\calculus^n\tmtwo$ by \reftm{GenSim}.\refpointmute{GenSim-two} there is an execution $\exec:\state\tomach^*\statetwo$ s.t. $\tmtwo\tostruct\decode\statetwo$ and $\sizep\exec=\size\deriv$. The number of RAM steps to implement $\exec$ is the sum of the number for the commutative and the principal transitions. By bilinearity, $ \sizecom\exec =  O((1+\sizep\exec)\cdot\size{\tm})$ and so all the commutative transitions in $\exec$ require $O((1+\sizep\exec)\cdot\size{\tm})$ steps, because a single one takes a constant number of steps. Again by bilinearity, each principal one requires $O(\size{\tm})$ steps, and so all the principal transitions together require $O(\sizep\exec\cdot\size{\tm})$ steps.
\end{IEEEproof}

We will discuss three distilleries, summarised in \reftab{pre-useful-rewritingrules} (page \pageref{tab:pre-useful-rewritingrules}), and two of them will be bilinear. The machines will be sophisticated, so that we will first present a machine for the inefficient \ShFC\ (\refsect{glam}, called \GLAM), that we will later refine with useful sharing (\refsect{glamour}, \glamour) and with renaming chains elimination (\refsect{lazyglamour}, \Fglamour).

Let us point out an apparent discrepancy with the literature. For the simpler case without \symbs, the number of commutative steps of the abstract machine studied in \cite{DBLP:conf/birthday/SandsGM02} is truly linear (and not bilinear), \ie\ it does not dependent on the size of the initial term. Three remarks:
\begin{enumerate}
\item \emph{Complete Evaluation}: it is true only for evaluation to normal form, while our low-level theorem is also valid for both any prefix of the evaluation and diverging evaluations. 
\item \emph{Normal Form Recognition}: it relies on the fact that closed normal forms (\ie\ values) can be recognised in constant time, by simply checking the topmost constructor. With \symbs checking if a term is normal requires time linear in its size, and so linearity is simply not possible.
\item \emph{Asymptotically Irrelevant}:  the dependency from the initial term disappears from the number of commutative transitions but still affects the cost of the principal ones, because every exponentials transition copies a subterm of the initial term, and thus it takes $O(\size\tm)$ time.
\end{enumerate}

\begin{table*}
\caption{Context and Relative Unfolding}
\label{tab:unfolding}
\centering
\begin{tabular}{c@{\sep\sep}c@{\sep\sep}c}

Context Unfolding & Relative Unfolding & Relative Context Unfolding\\
$\begin{array}{lllllllllllll}
	\unf\ctxhole 				& \defeq & \ctxhole\\
	\unf{(\tm \tctx)} 		& \defeq & \unf\tm \unf{\tctx}\\
	\unf{(\tctx \tm)} 		& \defeq & \unf{\tctx} \unf\tm\\
	\unf{\tctx \esub\var\tm} 		& \defeq & \unf\tctx \isub\var{\unf\tm}
\end{array}$

&

$\begin{array}{lllllllllllll}
	\relunf\tm\ctxhole 				& \defeq & \unf\tm\\
	\relunf\tm{\tmtwo \tctx} 		& \defeq & \relunf\tm \tctx\\
	\relunf\tm{\tctx \tmtwo} 		& \defeq & \relunf\tm\tctx\\
	\relunf\tm{\tctx \esub\var\tmtwo} 		& \defeq & \relunf{\tm}{\tctx} \isub\var{\unf\tmtwo}
\end{array}$ &
$\begin{array}{lllllllllllll}
	\relunf\tctxtwo\ctxhole 				& \defeq & \unf\tctxtwo\\
	\relunf\tctxtwo{\tmtwo \tctx} 		& \defeq & \relunf\tctxtwo \tctx\\
	\relunf\tctxtwo{\tctx \tmtwo} 		& \defeq & \relunf\tctxtwo\tctx\\
	\relunf\tctxtwo{\tctx \esub\var\tmtwo} 		& \defeq & \relunf{\tctxtwo}{\tctx} \isub\var{\unf\tmtwo}
\end{array}$
\end{tabular}
\end{table*}

\section{An Inefficient Distillery: the \GLAM\ Machine}\label{sect:glam}
\begin{atend}\section{Proofs Omitted From \refsect{glam}\\ (An Inefficient Distillery: the \GLAM\ Machine)}\end{atend}
\begin{atend}
The aim of this section is to prove \refth{glam-refl-distillation}, \ie\ that 
$(\GLAM,\togens, \eqstruct, \decodefun)$ is a reflective explicit distillery.

\end{atend}

In this section we introduce the \GLAM\ machine and show that it distills to the \ShFC. The distillery is inefficient, because \ShFC\ suffers of size explosion, but it is a good case study to present distilleries before the optimisations. Moreover, it allows to show an unexpected fact: while adding useful sharing to the calculus will be a quite tricky and technical affair (\refsect{useful}), adding usefulness to the \GLAM\ will be surprisingly simple (\refsect{glamour}), and yet tests of usefulness will only require constant time.

The machine of this section is the Global LAM (\GLAM). The name is due to a similar machine, based on \emph{local} environments, introduced in \cite{DBLP:conf/icfp/AccattoliBM14} and called LAM---standing for Leroy Abstract Machine. The \GLAM\ differs from the LAM in two respects: 1) it uses \emph{global} rather than local environments, and 2) it has an additional rule to handle constructors. 

\paragraph*{Data-Structures} at the machine level, \emph{terms} are replaced by \emph{codes}, \ie\ terms not considered up to $\alpha$-equivalence. To distinguish codes from terms, we over-line codes like in $\code$.

States (noted $\state,\statetwo,\ldots$) of the abstract machine are made out of a \emph{context dump} $\dump$, a \emph{code} $\tm$, an \emph{argument stack} $\stack$, and a global environment $\genv$, defined by the grammars in \reftab{transitions}. To save space, sometimes we write $\esub\var{\code}\genv$ for $\esub\var{\code}\cons\genv$. Note that stacks may contain pairs $\pair{\code}{\stack}$ of a code and a stack, used to code the application of $\code$ to the stack $\stack$. We choose this representation to implement commutative rules in constant time.

\paragraph*{The Machine} the machine transitions are given in \reftab{explicit-transitions}. Note that the multiplicative one $\tomachsm$ puts a new entry in the environment, while the exponential one $\tomachse$ performs a clashing-avoiding substitution from the environment. The idea is that the principal transitions $\tomachsm$ and $\tomachse$ implement $\togensm$ and $\togense$ while the commutative transitions $\tomachcone$, $\tomachctwo$, and $\tomachcthree$ locate and
expose the next redex following a right-to-left strategy.

The commutative rule $\tomachcone$ forces evaluation to be right-to-left on applications: the machine processes first the argument $\codetwo$, saving the left sub term $\code$ on the dump together with its current stack $\stack$.
The role of $\tomachctwo$ and $\tomachcthree$ is to backtrack to the saved sub-term. Indeed, when the argument, \ie\ the current code, is finally put in normal form, encoded by
a \emph{stack item} $\stackitem$, the stack item is pushed on the stack, and the machine backtracks to the pair on the dump.

\paragraph*{The Distillery} machines start an execution on \emph{initial states} defined as $ \glamst\stempty\code\stempty\stempty$, \ie\ obtained by taking the term, seen now as the code $\code$, and setting to
$\stempty$ the other machine components. A state represents a term---given by the code---and an evaluation context, that for the \GLAM\ is obtained by decoding $\dump$, $\stack$, and $\genv$. The decoding $\decodefun$ (or distillation) function is defined in \reftab{explicit-transitions}.
Note that stacks are decoded to contest in postfix notation for plugging. To improve readability, when we decode machines, we will denote $\wctxp\tm$ with $\ctxholep\tm\wctx$, if the component
occurs on the right of $\tm$ in the machine representation.

A machine state is \emph{closed} when all free variables in any component of the state are bound
in $\genv$ or, equivalently, when $\decode{\state}$ is closed in the usual sense.
It is \emph{well-named} when all variables bound in the state are
distinct. We require well-namedness as a machine invariant to allow
every environment entry $\esub\var{\code}$ to be global (i.e. to bind $\var$
everywhere in the machine state). From now on, the initial state associated to a term $\tm$ has as code the term obtained $\alpha$-converting $\tm$ to make it well-named.

For every machine  we will have invariants, in order to prove the properties of a distillery. They are always proved by induction over the length of the execution, by a simple inspection of the transitions. For the \GLAM:

\begin{lemma}[GLAM Invariants]\label{l:GLAMInvariants}
	Let $\state = (\dump,\tmtwo,\stack,\genv)$ be a state reachable from an initial code $\code$. Then:
	\begin{enumerate}
		\item \label{p:GLAMInvariants-one}\emph{Closure}: $\state$ is closed and well-named;
		\item \label{p:GLAMInvariants-two}\emph{Value}: values in components of $\state$ are sub-terms of $\code$;
		\item \label{p:GLAMInvariants-three}\emph{\Fireball}: every term in $\stack$, in $\genv$, and in every stack in $\dump$ is a \fireball;
		\item \label{p:GLAMInvariants-four}\emph{Contextual Decoding}: $\decgenv$, $\decdump$, $\decstack$, and $\evctx_\state$ are evaluation contexts;
	\end{enumerate}
\end{lemma}

\begin{proofatend}
by induction over the length of the execution.
The base case holds because $\code$ is initial.
The inductive step is by cases over the kind of transition.
All the verifications are trivial inspections of the transition.
\end{proofatend}

\begin{atend}
The first step to prove \refth{glam-refl-distillation} is the distillation
property. Note from the statement that the distillation is explicit
(see \refdef{distillationqual}).

\begin{lemma}[Explicit Distillation]\label{l:glam-distillation}
	Let $\state$ be a reachable state. Then:
		\begin{enumerate}
		\item \emph{Commutative}: If $\state\tomachcp{_{1,2,3}}\statetwo$ then $\decode\state=\decode\statetwo$;
		\item \emph{Multiplicative}: If $\state\tomachsm\statetwo$ then $\decode\state\togensm\eqstruct\decode\statetwo$;
		\item \emph{Exponential}: If $\state\tomachse\statetwo$ then $\decode\state\togense\decode\statetwo$.
	\end{enumerate}
\end{lemma}

\begin{IEEEproof}~

	\begin{itemize}
		\item Case $\tomachcone$:
		\[\begin{array}{lll}
        		\decode{(\dump,\code\codetwo,\stack,\genv)} & =\\
                        \decgenvp{\decdump\ctxholep{\decstackp{\code\codetwo}}} & = \\
                        \decgenvp{\decdump\ctxholep{\decstackp{\code\ctxhole}}\ctxholep\codetwo} & = \\
                        \decgenvp{\decode{\dentry\code\stack \cons\dump}\ctxholep{{\ctxholep{\codetwo}}}} & = \\
                        \decgenvp{\decode{\dentry\code\stack \cons\dump}\ctxholep{\decode{\stempty}{\ctxholep{\codetwo}}}} & =\\
                        \decode{(\dentry\code\stack \cons\dump,\codetwo,\stempty,\genv)}
    		\end{array}\]		

\item Case $\tomachsm$:
		\[\begin{array}{lll}
        		\decode{(\dump,\l\var.\code,\codetwo\cons\stack,\genv)} & = \\
                         \decgenvp{\decdump\ctxholep{\decode{\codetwo\cons\stack}\ctxholep{\l\var.\code}}} & =\\
                         \decgenvp{\decdump\ctxholep{\decode{\stack}\ctxholep{(\l\var.\code) \codetwo}}} &\togensm & \mbox{ (by \reflemma{GLAMInvariants}.\ref{p:GLAMInvariants-three},\ref{p:GLAMInvariants-four}})\\
			\decgenvp{\decdump\ctxholep{\decode{\stack}\ctxholep{\code \esub\var\codetwo}}} & \tostruct & \mbox{ (by \reflemma{ev-comm-struct})}\\
			\decgenvp{\decdump\ctxholep{\decstackp{\tm}}\esub\var\codetwo} & =\\
                        \decgenvpx{\esub\var\codetwo\cons\genv}{\decdump\ctxholep{\decstackp{\tm}}} & =\\
                        \decode{(\dump,\code,\stack,\esub\var\codetwo\cons\genv)}
    		\end{array}\]

Note that the multiplicative step is justified by points \ref{p:GLAMInvariants-three} and \ref{p:GLAMInvariants-four} of \reflemma{GLAMInvariants}, for which $\codetwo$ is a \fireball{} and $\decgenvp{\decdump\ctxholep\decstack}$ is an evaluation context. Moreover, the $\tostruct$ step holds because by \reflemma{GLAMInvariants}.\ref{p:GLAMInvariants-one} (well-namedness) $\var$ occurs only in $\code$ and so by \reflemma{ev-comm-struct} the substitution $\esub\var\codetwo$ commutes with the environment $\decdump\ctxholep{\decstack}$.
	  	
\item Case $\tomachctwo$:
		\[\begin{array}{lll}
        		\decode{(\dentry\code\stack \cons\dump,\l\var.\codetwo,\stempty,\genv)} & =\\
                        \decgenvp{\decode{\dentry\code\stack \cons\dump}\ctxholep{\decode\stempty\ctxholep{\l\var.\codetwo}}} & =\\
                         \decgenvp{\decdump\ctxholep{\decstackp{\code\l\var.\codetwo}}} & =\\
                         \decgenvp{\decdump\ctxholep{\decode{\l\var.\codetwo\cons\stack}\ctxholep\code}} & =\\
                         \decode{(\dump,\code,\l\var.\codetwo\cons\stack,\genv)}
    		\end{array}\]		
		
\item Case $\tomachcthree$:
		\[\begin{array}{lll}
        		\decode{((\code,\stack)\cons\dump,\const,\stacktwo,\genv)} & =\\
                        \decgenvp{\decode{(\code,\stack)\cons\dump}\ctxholep{\decode{\stacktwo}\ctxholep{\const}}} & =\\
                        \decgenvp{\decdump\ctxholep{\decode{\stack}\ctxholep{\code\decstacktwop\const}}} & =\\
                        \decgenvp{\decdump\ctxholep{\decode{\decstacktwop\const\cons\stack}\ctxholep\code}} & =\\
                        \decode{(\dump,\code,\decstacktwop\const\cons\stack,\genv)}
    		\end{array}\]		

\item Case $\tomachse$:
		\[\begin{array}{lll}
        		\decode{(\dump,\var,\stack,\genv_1\esub\var\codetwo\genv_2)} & =\\
                        \decgenvpx{\genv_1\esub\var\codetwo\genv_2}{\decdump\ctxholep{\decstackp{\var}}} & =\\
                        \decgenvpx{\genv_2}{\decgenvpx{\genv_1}{\decdump\ctxholep{\decstackp{\var}}}\esub\var\codetwo} & \togense & \mbox{ (by \reflemma{GLAMInvariants}.\ref{p:GLAMInvariants-three},\ref{p:GLAMInvariants-four}})\\
			\decgenvpx{\genv_2}{\decgenvpx{\genv_1}{\decdump\ctxholep{\decstackp{{\rename\codetwo}}}}\esub\var\codetwo} & =\\
                        \decgenvpx{\genv_1\esub\var\codetwo\genv_2}{\decdump\ctxholep{\decstackp{\rename{\codetwo}}}} & =\\
                        \decode{(\dump,\rename{\codetwo},\stack,\genv_1\esub\var\codetwo\genv_2)}
    		\end{array}\]		
  Note that the exponential step is justified by points \ref{p:GLAMInvariants-three} and \ref{p:GLAMInvariants-four} of \reflemma{GLAMInvariants}, for which $\codetwo$ is a \fireball{} and $\decode\genv_2$ and $\decode{\genv_1}\ctxholep{\decdump\ctxholep{\decstack}}$ are evaluation contexts.\qedhere
\end{itemize}
\end{IEEEproof}

~

Next we prove progress. We first need to redefine the size of the machine state
to ignore the new environment component:

\begin{definition}\label{def:sizetwo}
   $\size{\glam\genv\dump\code\stack} \defeq \size\code + \Sigma_{\nfnst\codetwo\stack \in \dump} \size\codetwo$
\end{definition}

\begin{lemma}[Termination] $\tomacha$ is terminating
\end{lemma}
\begin{IEEEproof}
just reuse the proof of \refcorollary{trivial-bound1}.
\end{IEEEproof}

\begin{lemma}[Determinism]\label{l:glam-deterministic} The transition relation $\tomach$ of the \GLAM\ is deterministic.
\end{lemma}
\begin{IEEEproof}
a simple inspection of the transitions show no critical pairs.
\end{IEEEproof}

\begin{lemma}[Progress]
\label{l:progress}
if $\state$ is reachable, $\admnf{\state}=\state$ and $\decode\state\togenx\tm$ with $\mathtt{x}\in\set{\ssym\msym,\ssym\esym}$, then there exists $\statetwo$  such that $\state\tomachx\statetwo$, \ie, $\state$ is not final.
\end{lemma}
\begin{IEEEproof}
by \reflemma{glam-deterministic} and \reflemma{glam-distillation} it is sufficient to show that every
reachable stuck state decodes to a  normal form. The only stuck forms are:
\begin{itemize}
 \item $(\dump,\var,\stack,\genv)$ where $\var$ is not defined in $\genv$.
   The state is not reachable because it would violate the Closure
   invariant (\reflemma{GLAMInvariants}.\ref{p:GLAMInvariants-one}).
 \item $(\stempty,\l \var.\code,\stempty,\genv)$ that decodes to
       $\decgenvp{\l \var.\code}$, that by the contextual decoding invariant (\reflemma{GLAMInvariants}.\ref{p:GLAMInvariants-one}) is a normal form.
 \item $(\stempty,\const,\stack,\genv)$ that decodes to $\decgenvp{\decstackp\const}$, that by the contextual decoding invariant (\reflemma{GLAMInvariants}.\ref{p:GLAMInvariants-one}) is a normal form.~\qedhere
\end{itemize}
\end{IEEEproof}

\end{atend}

The invariants are used to prove the following theorem.

\begin{theorem}[\GLAM\ Distillation]\label{th:glam-refl-distillation}
	$(\mbox{\GLAM},\togens, \eqstruct, \decodefun)$ is a reflective explicit distillery. In particular, let $\state$ be a reachable state reachable:
		\begin{enumerate}
		\item \emph{Commutative}: if $\state\tomachcp{_{1,2,3}}\statetwo$ then $\decode\state=\decode\statetwo$;
		\item \emph{Multiplicative}: if $\state\tomachsm\statetwo$ then $\decode\state\togensm\eqstruct\decode\statetwo$;
		\item \emph{Exponential}: if $\state\tomachse\statetwo$ then $\decode\state\togense\decode\statetwo$.
	\end{enumerate}
\end{theorem}
\begin{proofatend}
it follows from \reflemma{glam-distillation} and \reflemma{progress}.
\end{proofatend}

Since the \ShFC\ suffers of size-explosion, an exponential step (and thus an exponential transition) may duplicate a subterm that is exponentially bigger than the input. Then $(\mbox{\GLAM},\togens, \eqstruct, \decodefun)$ does not satisfy bilinearity, for which every exponential transition has to have linear complexity in the size of the input.

\section{Interlude: Relative Unfoldings}
\label{sect:rel-unfolding}
\begin{atend}\section{Proofs Omitted From \refsect{rel-unfolding}\\ (Interlude II: Relative Unfoldings)}\end{atend}
Now we define some notions for weak contexts that will be implicitly instantiated to all kind of contexts in the paper. In particular, we define substitution over contexts, and then use it to define the unfolding of a context, and the more general notion of relative unfolding.

\emph{Implicit substitution on weak contexts $\wctx$} is defined by
\begin{center}
$\begin{array}{lllllllllllll}
	\ctxhole\isub\var\tmtwo 			& \defeq & \ctxhole\\
	(\tm \wctx)\isub\var\tmtwo 			& \defeq & \tm\isub\var\tmtwo \wctx\isub\var\tmtwo\\
	(\wctx \tm)\isub\var\tmtwo 			& \defeq & \wctx\isub\var\tmtwo \tm\isub\var\tmtwo\\
	\wctx \esub\vartwo\tm\isub\var\tmtwo	& \defeq & \wctx\isub\var\tmtwo \esub\vartwo{\tm\isub\var\tmtwo}\\
	\tm \esub\vartwo\wctx\isub\var\tmtwo	& \defeq & \tm\isub\var\tmtwo \esub\vartwo{\wctx\isub\var\tmtwo}
\end{array}$
\end{center}

\begin{lemma}
	\label{l:evctx-and-sub-commute}
	Let $\tm$ be a term and $\wctx$ a weak context. Then $\wctxp\tm\isub\var\tmtwo = \wctx\isub\var\tmtwo\ctxholep{\tm\isub\var\tmtwo}$.
\end{lemma}

\begin{proofatend}
	by induction on $\wctx$. 
\end{proofatend}

Now, we would like to extend the unfolding to contexts, but in order to do so we have to restrict the notion of context. Indeed, whenever the hole of a context is inside an ES, the unfolding may erase or duplicate the hole, producing a term or a multi-context, which we do not want. Thus, we turn to (weak) \emph{shallow contexts}, defined by:
$$
\tctx,\tctxtwo,\tctxthree \grameq \ctxhole\mid \tctx \tm \mid\tm\tctx \mid \tctx\esub{\var}{\tm}.
$$
(note the absence of the production $\tm\esub{\var}{\tctx}$).

Now, we define in \reftab{unfolding} \emph{context unfolding} $\unf\tctx$, \emph{unfolding $\relunf\tm\tctx$ of a term $\tm$ relative to a shallow context $\tctx$} and \emph{unfolding $\relunf\tctxtwo\tctx$ of a shallow context $\tctxtwo$ relative to a shallow context $\tctx$}.

Relative unfoldings have a number of properties, summed up in the appendix (page \pageref{l:relunf-properties}). Last, a definition that will be important in the next section.

\begin{definition}[Applicative Context]
\label{def:appl-ctx}
  A shallow context $\tctx$ is \emph{applicative} when its hole is applied to a sub term $\tmtwo$, \ie\ if $\tctx = \tctxtwop{\sctx \tmtwo}$. 
\end{definition}

\begin{atend}
\begin{lemma}[Properties of Relative Unfoldings]
	\label{l:relunf-properties} 
	Let $\tm$ and $\tmtwo$ be terms and $\tctx$ be a shallow context.
	\begin{enumerate}
		\item \label{p:relunf-properties-zero} \emph{Commutation}: $\relunf{(\la\var\tm)}\tctx = \relunf{\la\var\tm}\tctx$, 
		$\relunf{(\tm\tmtwo)}\tctx = \relunf\tm\tctx \relunf\tmtwo\tctx$, $\relunf{\tm\isub\var\tmtwo}\tctx = \relunf\tm\tctx \isub\var{\relunf\tmtwo\tctx}$, $\relunf{\tm\isub\var\tmtwo}\tctx = \relunf{\tm\esub\var\tmtwo}\tctx$, 
$\unf\tctx\isub{\var}{\unf\tm} = \unf{\tctx\isub{\var}{\unf\tm}}$, and $\relunf\tm{\tctx\esub\var\tmtwo} = \relunf{\tm\isub\var{\unf\tmtwo}}{\tctx\isub\var{\unf\tmtwo}}$.

		\item \label{p:relunf-properties-one} \emph{Freedom}:
		if $\tctx$ does not capture any free variable of $\tm$ then $\relunf\tm\tctx = \unf\tm$.
		
		\item \label{p:relunf-properties-two} \emph{Relativity}:
		if $\unf\tm =\unf\tmtwo$ then $\relunf\tm\tctx = \relunf\tmtwo\tctx$.
		
		\item \label{p:relunf-properties-three} \emph{Applicativity}:
		if $\tctx$ is applicative then $\unf\tctx$ is applicative.

		\item \label{p:relunf-properties-five} \emph{Splitting}:
		$ \relunf\tm{\tctxp\tctxtwo} = \relunf{\relunf\tm\tctxtwo}{\tctx}$.
		
		\item \label{p:relunf-properties-four} \emph{Factorisation}:
		$ \relunf{\tctxtwop\tm}\tctx = \relunf\tctxtwo\tctx \ctxholep{\relunf\tm{\tctxp\tctxtwo}}$, in particular $ \unf{\tctxp\tm} = \unf\tctx \ctxholep{\relunf\tm{\tctx}}$ and $ \relunf{\sctxp\tm}\tctx = \relunf\tm{\tctxp\sctx}$.

	\end{enumerate}
\end{lemma}

\begin{IEEEproof}
  Routine inductions on $\tctx$ or $\tctxtwo$.
\end{IEEEproof}
\end{atend}


\section{Introducing Useful Sharing}
	\label{sect:intro-useful}
\textbf{\textit{Beware}}: this and the next sections will heavily use contexts and notions about them as defined in \refsect{fireb-end-es}  and \refsect{rel-unfolding}, in particular the notions of \emph{shallow} context, \emph{applicative context}, and \emph{relative unfolding}.

\paragraph*{Introducing Useful Reduction} note that the substitution steps in the size exploding family do not create redexes. We want to restrict the calculus so that these \emph{useless} steps are avoided. The idea of useful sharing, is to trigger an exponential redex only if it will somehow contribute to create a multiplicative redex. Essentially, one wants only the exponential steps
$$\evctxp\var\esub\var{\sctxp{\fire}} \togene  \sctxp{\evctxp\fire\esub\var\fire}$$
s.t. $\evctx$ is applicative and $\fire$ is a value, so that the firing creates a multiplicative redex. Such a change of approach, however, has consequences on the whole design of the system. Indeed, since some substitutions are delayed, the present requirements for the rules might not be met. Consider:
$$(\la\var\tm) \vartwo\esub\vartwo{\const\consttwo}$$
we want to avoid substituting $\const\consttwo$ for the argument $\vartwo$, but we also want that evaluation does not stop, \ie\ that $(\la\var\tm) \vartwo\esub\vartwo{\const\consttwo} \tom \tm\esub\var{\vartwo\esub\vartwo{\const\consttwo}}$.
To accomodate such a dynamics, our definitions have to be \emph{up to unfolding}, \ie\ \emph{fireballs} have to be replaced by \emph{terms unfolding to fireballs}. There are 4 subtle things about useful reduction.

\emph{1) Multiplicatives and Variables}. The idea is that the multiplicative rule becomes
$$\begin{array}{rcl}
    \sctxp{\l\var.\tm}\sctxtwop\tmtwo &\rtodbv&  \sctxp{\tm\esub\var{\sctxtwop\tmtwo}}
\end{array}$$
where it is the unfolding $\unf{\sctxtwop\tmtwo}$ of the argument $\sctxtwop\tmtwo$ that is a fireball, and not necessarily $\sctxtwop\tmtwo$ itself. Note that sometimes variables are valid arguments of multiplicative redexes, and consequently substitutions may contain variables.

\emph{2) Exponentials and Future Creations}. The exponential rule involves contexts, and is trickier to make it useful. A first approximation of useful exponential step is
$$\begin{array}{rcl}    
    \evctxp\var\esub\var{\sctxp\tmtwo} &\rtolsv  &\sctxp{\evctxp\tmtwo\esub\var\tmtwo}
\end{array}$$
where $\unf{\sctxp\tmtwo}$ is a \emph{value} (\ie\ it is not a inert) and $\evctx$ is applicative, so that---after eventually many substitution steps, when $\var$ becomes $\unf\tmtwo$---a multiplicative redex will pop out.

Note that an useful exponential step does not always \emph{immediately} create a multiplicative redex. Consider the following step (where $I$ is the identity): 
\begin{equation}
\label{eq:rel-useful-step}
 (\var I)\esub\var\vartwo\esub\vartwo{I} \togene (\vartwo I)\esub\var\vartwo\esub\vartwo{I}
 \end{equation}
No multiplicative redex has been created yet, but step \refeq{rel-useful-step} is useful because the \emph{next} exponential step creates a multiplicative redex:
$$  (\vartwo I)\esub\var\vartwo\esub\vartwo{I} \togene 
(I I)\esub\var\vartwo\esub\vartwo{I}
$$

\emph{3) Evaluation and Evaluable Contexts}. The delaying of useless substitutions impacts also on the notion of evaluation context $\evctx$, used in the exponential rule. For instance, the following exponential step should be useful
$$ ((\var I)\vartwo)\esub\var{I}\esub\vartwo{\const\consttwo} \togene ((I I)\vartwo)\esub\var{I}\esub\vartwo{\const\consttwo}$$
but the context $((\ctxhole I)\vartwo)\esub\var{I}\esub\vartwo{\const\consttwo}$ isolating $\var$ is not an evaluation context, it only unfolds to one. We then need a notion of evaluation context up to unfolding. The intuition is that a shallow context $\tctx$ is \emph{evaluable} if $\unf\tctx$ is an evaluation context (see \refsect{rel-unfolding} for the definition of context unfolding), and it is \emph{useful} if it is evaluable and applicative. The exponential rule then should rather be:
$$\begin{array}{rcl}    
    \tctxp\var\esub\var{\sctxp{\tmtwo}} &\rtolsv  &\sctxp{\tctxp\tmtwo\esub\var\tmtwo}
\end{array}$$
where $\unf\tmtwo$ is a \emph{value} and $\tctx$ is \emph{useful}.

\emph{4) Context Closure vs Global Rules}. Such a definition, while close to the right one, still misses a fundamental point, \ie\ the  \emph{global} nature of useful steps.
Evaluation rules are indeed defined by a further \emph{closure by contexts}, \ie\ a step takes place in a certain shallow context $\tctxtwo$. Of course, $\tctxtwo$ has to be evaluable, but there is more. Such a context, in fact, may also give an essential contribution to the usefulness of a step. Let us give an example. Consider the following exponential step
$$ (\var\var)\esub\var\vartwo \togene (\vartwo\var)\esub\var\vartwo$$
By itself it is not useful, since $\vartwo$ is not a value nor unfolds to one. If we plug that redex in the context $\tctx \defeq \ctxhole\esub\vartwo{I}$, however, then $\vartwo$ unfolds to a value in $\tctx$, as $\relunf\vartwo\tctx = \relunf\vartwo{\ctxhole\esub\vartwo{\la\varthree\varthree}} = \la\varthree\varthree$, and the step becomes:
\begin{equation}
\label{eq:rel-useful-step-two}
 (\var\var)\esub\var\vartwo\esub\vartwo{\la\varthree\varthree} \togene (\vartwo\var)\esub\var\vartwo\esub\vartwo{\la\varthree\varthree}
 \end{equation}
No multiplicative redex has been created yet, but step \refeq{rel-useful-step-two} is useful because it is essential for the creation given by the \emph{next} exponential step:
$$ (\vartwo\var)\esub\var\vartwo\esub\vartwo{\la\varthree\varthree} \togene ((\la\varthree\varthree)\var)\esub\var\vartwo\esub\vartwo{\la\varthree\varthree}$$
Note, indeed, that $(\la\varthree\varthree)\var$ gives a useful multiplicative redex, because $\var$ unfolds to a fireball in its context $\ctxhole\esub\var\vartwo\esub\vartwo{\la\varthree\varthree}$. 

Summing up, the useful or useless character of a step depends crucially on the surrounding context. Therefore useful rules have to be \emph{global}: rather than given as axioms closed by evaluable contexts, they will involve the surrounding context itself and impose conditions about it.\medskip

The \UsFC, presented in the next section, formalises these ideas. We will prove it to be a locally bounded implementation of $\tof$, obtaining our fist high-level implementation theorem.
\begin{table*}
\caption{Distilleries in the Paper + Rewriting Rules for the \UsFC}
\label{tab:pre-useful-rewritingrules}
\centering
\begin{tabular}{c|c}
\begin{tabular}{ccc}
\emph{Calculus} & \emph{Machine}\\\hline
\FC\ $\tof$ & \\
\ShFC\ $\togens$ & \GLAM\\
\UsFC\ $\togenu$ & \glamour\\
\FFC\ $\togenlu$ & \fast\ \glamour
\end{tabular}
&
$\begin{array}{c@{\hspace{1.5cm}}l}
  \textsc{Rule (Already Closed by Contexts)} & \textsc{Side Conditions} \\
  	\tctxp{\sctxp{\l\var.\tm}\tmtwo} \togenum \tctxp{\sctxp{\tm\esub\var{\tmtwo}}} &
	\mbox{$\tctxp{\sctx\tmtwo}$ is useful}\\\\

		\tctxp{\tctxtwop\var\esub\var{\sctxp{\tmtwo}}} \togenue  \tctxp{\sctxp{\tctxtwop\tmtwo\esub\var\tmtwo}} &
		\tctxp{\tctxtwo\esub\var{\sctxp{\tmtwo}}}\mbox{ is
		useful}\\
		&\mbox{$\tmtwo\neq\tmtwop \esub\vartwo\tmthree$ and $\relunf\tmtwo{\tctxp\sctx}= \val$ }\\
\end{array}$
\end{tabular}
\end{table*}

\begin{table}
\caption{Evaluable Shallow Contexts}
\label{tab:evaluable}
\centering
$\begin{array}{lllllllllllll}
	\AxiomC{}
	\RightLabel{}
	\UnaryInfC{$\ctxhole$ is evaluable}
	\DisplayProof
	&
	\AxiomC{$\tctx$ is eval.}
		\AxiomC{$\unf\tm$ is a \fireball}
	\RightLabel{}
	\BinaryInfC{$\tctx \tm$ is evaluable}
	\DisplayProof \\\\
	
	\AxiomC{$\tctx$ is evaluable}
	\RightLabel{}
	\UnaryInfC{$\tm \tctx$ is evaluable}
	\DisplayProof 
	&
	\AxiomC{$\tctx\isub\var{\unf\tm}$ is eval.}
		\AxiomC{$\unf\tm$ is a \fireball}
	\RightLabel{}
	\BinaryInfC{$\tctx \esub\var\tm$ is evaluable}
	\DisplayProof 
\end{array}
$
\end{table}
\begin{table*}
\caption{Transitions of the \glamour}
\label{tab:pre-useful-transitions}
  \centering
  ${\setlength{\arraycolsep}{1em}
 \begin{array}{c|c|c|ccc|c|c|ccc}
     \usefmaca{\genv}{\dump}{\code\codetwo}{\stack}
     &\tomachcone&
     \usefmaca{\genv}{\dump\cons \nfnst\code\stack}{\codetwo}{\stempty}\\
  	
     \usefmaca{\genv}{\dump}{\l\var.\code}{\stackitem^\lab\cons\stack}
     &\tomachum &
     \usefmaca{\econs{\esub\var{\stackitem^\lab}}{\genv}}{\dump}{\code}{\stack}\\

     \usefmaca{\genv}{\dump \cons \nfnst\code\stack}{\l\var.\codetwo}{\stempty}
     & \tomachctwo &
     \usefmaca{\genv}{\dump}{\code}{\herval{(\l\var.\codetwo)}\cons\stack}\\

     \usefmaca{\genv}{\dump\cons \nfnst\code\stack}{\const}{\stacktwo}
     & \tomachcthree &
     \usefmaca{\genv}{\dump}{\code}{\pair{\const}{\stacktwo}^\dead\cons\stack}\\

     \usefmaca{\genv_1\esub\var{\phi^\dead}\genv_2}{\dump\cons\nfnst\code\stack}{\var}{\stacktwo}
     & \tomachcfour &
     \usefmaca{\genv_1\esub\var{\phi^\dead}\genv_2}{\dump}{\code}{\pair{\var}{\stacktwo}^\dead\cons\stack}\\

     \usefmaca{\genv_1\esub\var{\herval{\codetwo}}\genv_2}{\dump\cons\nfnst\code\stack}{\var}{\stempty}
     & \tomachcfive &
     \usefmaca{\genv_1\esub\var{\herval{\codetwo}}\genv_2}{\dump}{\code}{\herval{\var}\cons\stack}\\

     \usefmaca{\genv_1\esub\var{\herval\codetwo}\genv_2}{\dump}{\var}{\stackitem^\lab\cons\stack}
     & \tomachue &
     \usefmaca{\genv_1\esub\var{\herval\codetwo}\genv_2}{\dump}{\rename{\codetwo}}{\stackitem^\lab\cons\stack}\\

  	\end{array}}$
\begin{minipage}{14cm}
~\\where $\rename{\codetwo}$ is any code $\alpha$-equivalent to $\codetwo$ that preserves well-naming of the machine.
\end{minipage}
\end{table*}

\section{The Useful Fireball Calculus}
\label{sect:useful}
\begin{atend}\section{Proofs Omitted From \refsect{useful}\\ (The Useful Fireball Calculus)}
\label{sect:app-useful}\end{atend}
For the \UsFC, terms, values, and substitution contexts
are unchanged (with respect to the \ShFC), and we use \emph{shallow contexts} $\tctx$ as defined in \refsect{rel-unfolding}. An \emph{initial term} is still a closed term with no explicit substitutions.  

The new key notion is that of \emph{evaluable} context. 

\begin{definition}[Evaluable and Useful Contexts]
 \emph{\Evaluable}\ (shallow) contexts are defined by the inference system in \reftab{evaluable}. A context is \emph{useful} if it is evaluable and applicative (being applicative is easily seen to be preserved by unfolding).
\end{definition}

\refpoint{evaluable-prop-one} of the following \reflemma{evaluable-prop} guarantees that evaluable contexts capture the intended semantics suggested in the previous section. \refpoint{evaluable-prop-five} instead provides an equivalent inductive formulation that does not mention relative unfoldings. The definition in \reftab{evaluable} can be thought has been \emph{from the outside}, while the lemma give a characterisation \emph{from the inside}, relating sub-terms to their surrounding sub-context.

\begin{lemma}\hfill
	\label{l:evaluable-prop} 
	\begin{enumerate}

		\item \label{p:evaluable-prop-one}
		If $\tctx$ is evaluable then $\unf\tctx$ is an evaluation context.
		
		\item \label{p:evaluable-prop-five}
		$\tctx$ is evaluable iff $\relunf\tmtwo\tctxtwo$ is a \fireball\ whenever  $\tctx = \tctxtwop{\tctxthree\tmtwo}$ or $\tctx = \tctxtwop{\tctxthree\esub\var\tmtwo}$.

	\end{enumerate}
\end{lemma}

\begin{proofatend}\hfill
	\begin{enumerate}

		\item
			By induction on the pair $(\mbox{number of ES in $\tctx$}, $\tctx$)$. Cases of $\tctx$:
			\begin{enumerate}
				\item \emph{Empty, \ie\ $\tctx = \ctxhole$}. Then $\unf\tctx = \unf\ctxhole = \ctxhole$ is an evaluation context.
				\item \emph{Right Application, \ie\ $\tctx = \tm\tctxtwo$}. Then $\unf\tctx = \unf\tm\unf\tctxtwo =_{\ih} \unf\tm\evctx$ is an evaluation context.
				\item \emph{Left Application, \ie\ $\tctx = \tctxtwo\tm$ with $\unf\tm$ a \fireball\ $\fire$}. Then $\unf\tctx = \unf\tctxtwo \unf\tm=_{\ih} \evctx\fire$ is an evaluation context.
				\item \emph{Substitution, \ie\ $\tctx = \tctxtwo\esub\var\tm$ with $\unf\tm$ a \fireball\ $\fire$ and $\tctxtwo\isub\var{\unf\tm}$ is evaluable}. Note that the number of ES in  $\tctxtwo\isub\var{\unf\tm}$ is strictly smaller than the number of ES in $\tctx$, because $\unf\tm$ has no ES. Then by \ih\ $\unf{\tctxtwo\isub\var{\unf\tm}}$ is an evaluation context. Now, $\unf\tctx = \unf{\tctxtwo\esub\var\tm} = \unf\tctxtwo\isub\var{\unf\tm} =_{\reflemmaeqp{relunf-properties}{zero}} \unf{\tctxtwo\isub\var{\unf\tm}}$ which is an evaluation context.
			\end{enumerate}
		
		\item By induction on the pair $(\mbox{number of ES in $\tctx$}, $\tctx$)$. Cases of $\tctx$:
		\begin{enumerate}
			\item \emph{Empty, \ie\ $\tctx = \ctxhole$}. Directions
			\begin{enumerate}
				\item \emph{$\Rightarrow$, \ie\ $\tctx$ is evaluable}. Nothing to prove.
				\item \emph{$\Leftarrow$}. Then $\tctx$ is evaluable.
			\end{enumerate}
			
			\item \emph{Right Application, \ie\ $\tctx = \tmthree\tctxfour$}. Note that $\tctxtwo$ cannot be empty, otherwise $\tctx = \tmthree\tctxfour = \tctxthree\tmtwo$ and $\tctx$ would have two holes. Then $\tctxtwo = \tmthree \tctxfive$ for some $\tctxfive$, and the statements follows from the \ih\ applied to $\tctxfour$ and $\tctxfive$.
			\item \emph{Left Application, \ie\ $\tctx = \tctxfour\tmthree$}. Directions:
			\begin{enumerate}
				\item \emph{$\Rightarrow$. Since $\tctx$ is evaluable, $\unf\tmthree$ is a \fireball, and $\tctxfour$ is evaluable}. Note that either $\tctxtwo$ is empty, and then $\tmtwo = \tmthree$ and the statement holds because $\relunf\tmthree\ctxhole=\unf\tmthree$ is a \fireball, or $\tctxtwo = \tctxfive \tmthree$ with $\tctxfive$ s.t.---say---$\tctxfour = \tctxfivep{\tctxthree\esub\var\tmtwo}$. Now, note that $\relunf\tmtwo\tctxtwo = \relunf\tmtwo{\tctxfive\tmthree} = \relunf\tmtwo{\tctxfive}$ and the statement follows by the \ih\ applied to $\tctxfour$.
				\item \emph{$\Leftarrow$}. By taking $\tctxtwo \defeq \ctxhole$, the hypothesis becomes $\relunf\tmthree\ctxhole = \unf\tmthree$ is a \fireball. We are left to show that $\tctxfour$ is evaluable, that is given by the \ih.
			\end{enumerate}
			
			\item \emph{Substitution, \ie\ $\tctxtwo = \tctxfour\esub\vartwo\tmthree$}.
			\begin{enumerate}
				\item $\Rightarrow$. \emph{Since $\tctx$ is evaluable,  $\unf\tmthree$ is a \fireball, and $\tctxfour\isub\vartwo{\unf\tmthree}$ is evaluable}. Note that either $\tctxtwo$ is empty, and then $\tmtwo = \tmthree$ and the statement holds because $\relunf\tmthree\ctxhole=\unf\tmthree$ is a \fireball, or $\tctxtwo = \tctxfive \esub\vartwo\tmthree$ for some $\tctxfive$ that is a prefix of $\tctxfour$, \ie\ s.t. $\tctxfour = \tctxfivep{\tctxthree\tmtwo}$ or $\tctxfour = \tctxfivep{\tctxthree\esub\var\tmtwo}$. Let's say that $\tctxfour = \tctxfivep{\tctxthree\tmtwo}$. Now, applying the \ih\ to
				$$\begin{array}{ll}\tctxfour\isub\vartwo{\unf\tmthree} & =\\
                                  \tctxfivep{\tctxthree\tmtwo}\isub\vartwo{\unf\tmthree} &=\\
                                  \tctxfive\isub\vartwo{\unf\tmthree}\ctxholep{\tctxthree\isub\vartwo{\unf\tmthree}\tmtwo\isub\vartwo{\unf\tmthree}}\end{array}$$
				we obtain that $\relunf{\tmtwo\isub\vartwo{\unf\tmthree}}{\tctxfive\isub\vartwo{\unf\tmthree}}$ is a \fireball. We conclude noting that $\relunf{\tmtwo\isub\vartwo{\unf\tmthree}}{\tctxfive\isub\vartwo{\unf\tmthree}} =_{\reflemmap{relunf-properties}{zero}} \relunf\tmtwo{\tctxfive\esub\vartwo\tmthree} = \relunf\tmtwo\tctx$ (the other case, $\tctxfour = \tctxfivep{\tctxthree\esub\var\tmtwo}$, uses the same reasoning).
				\item \emph{$\Leftarrow$}. By taking $\tctxtwo \defeq \ctxhole$, the hypothesis becomes $\relunf\tmthree\ctxhole = \unf\tmthree$ is a \fireball. We are left to show that $\tctxfour\isub\vartwo{\unf\tmthree}$ is evaluable, that is given by the \ih.
			\end{enumerate}

		\end{enumerate}
		
	\end{enumerate}
\end{proofatend}

\begin{atend}
The following technical lemma is very useful to decompose and construct
evaluation contexts compositionally.

\begin{lemma}\hfill
	\label{l:bis-evaluable-prop} 
	\begin{enumerate}

		\item \label{p:bis-evaluable-prop-two}
		if $\tctxp\tctxtwo$ is evaluable then $\tctx$ is evaluable and $\relunf\tctxtwo\tctx$ is an evaluation context.
		
		\item \label{p:bis-evaluable-prop-two-bis}
		if $\tctx$ is evaluable, $\relunf\tctxtwo\tctx$ is an evaluation context and $\tctxtwo$ is without ES then $\tctxp\tctxtwo$ is evaluable.
	\end{enumerate}
\end{lemma}

\begin{IEEEproof}
 \begin{enumerate}
		\item By induction on the pair $(\mbox{number of ES in $\tctx$}, $\tctx$)$. Cases of $\tctx$:
		\begin{enumerate}
			\item \emph{Empty, \ie\ $\tctx = \ctxhole$}. The hypothesis becomes that $\tctxtwo$ is evaluable, and so $\relunf\tctxtwo\tctx = \relunf\tctxtwo\ctxhole = \unf\tctxtwo$ is an evaluation context by \refpoint{evaluable-prop-one}. Clearly $\ctxhole$ is evaluable.
			
			\item \emph{Right Application, \ie\ $\tctx = \tm\tctxthree$}. By \ih, $\tctxthree$ is evaluable, that implies $\tctx$ evaluable. Moreover, $\relunf\tctxtwo\tctx = \relunf\tctxtwo{\tm\tctxthree} = \relunf\tctxtwo\tctxthree$ which is an evaluation context by \ih.
			
			\item \emph{Left Application, \ie\ $\tctx = \tctxthree\tmtwo$}. By \ih, $\tctxthree$ is evaluable. From the hypothesis that $\tctxp\tctxtwo = \tctxthreep\tctxtwo \tmtwo$ is evaluable it follows that $\unf\tmtwo$ is a \fireball, and so $\tctx$ is evaluable. Moreover, $\relunf\tctxtwo\tctx = \relunf\tctxtwo{\tctxthree\tmtwo} = \relunf\tctxtwo\tctxthree$ which is an evaluation context by \ih.
			
			\item \emph{Substitution, \ie\ $\tctx = \tctxthree\esub\var\tmtwo$}. From the hypothesis that $\tctxp\tctxtwo = \tctxthreep\tctxtwo \esub\var\tmtwo$ is evaluable it follows that $\unf\tmtwo$ is a \fireball. Since $\tctxthree\isub\var{\unf\tmtwo}$ has strictly less ES than $\tctx$ (because $\unf\tmtwo$ has none), the \ih\ gives that $\tctxthree\isub\var{\unf\tmtwo}$ is evaluable, and so $\tctx$ is evaluable. Then $\relunf\tctxtwo\tctx = \relunf\tctxtwo{\tctxthree\esub\var\tmtwo} =_{\reflemmaeqp{relunf-properties}{zero}} \relunf{\tctxtwo\isub\var{\unf\tmtwo}}{\tctxthree\isub\var{\unf\tmtwo}}$ that by \ih\ is an evaluation context.
		\end{enumerate}
		
				\item We prove that $\relunf\tmtwo\tctxthree$ is a \fireball\ whenever $\tctxp\tctxtwo = \tctxthreep{\tctxfour\tmtwo}$ or $\tctxp\tctxtwo = \tctxthreep{\tctxfour\esub\var\tmtwo}$, and conclude by applying \refpoint{evaluable-prop-five}.  Now, since $\tctxtwo$ has no ES, if $\tctxp\tctxtwo = \tctxthreep{\tctxfour\esub\var\tmtwo}$ then $\esub\var\tmtwo$ occurs in $\tctx$, and $\tctxthree$ is a prefix of $\tctx$. We obtain that $\relunf\tmtwo\tctxthree$ is a \fireball\ by applying \refpoint{evaluable-prop-five} to $\tctx$, that is evaluable by hypothesis. If $\tctxp\tctxtwo = \tctxthreep{\tctxfour\tmtwo}$ with $\tctxthree$ a prefix of $\tctx$ we reason similarly. Otherwise, the application $\tctxfour\tmtwo$ occurs in $\tctxtwo$, \ie\ there is a context $\tctxfive$ s.t. $\tctxthree = \tctxp\tctxfive$ and $\tctxtwo = \tctxfivep{\tctxfour\tmtwo}$. Then we have $\relunf{\tctxtwo}\tctx = \relunf{\tctxfivep{\tctxfour\tmtwo}}\tctx =_{\reflemmaeqp{relunf-properties}{four}} \relunf\tctxfive\tctx\ctxholep{\relunf{(\tctxfour\tmtwo)}{\tctxp\tctxfive}} =_{\reflemmaeqp{relunf-properties}{zero}} \relunf\tctxfive\tctx\ctxholep{\relunf\tctxfour{\tctxp\tctxfive}\relunf\tmtwo{\tctxp\tctxfive}}$, which by hypothesis is an evaluation context. Therefore, $\relunf\tmtwo{\tctxp\tctxfive}$ is a \fireball. We conclude with $ \relunf\tmtwo{\tctxp\tctxfive} =_{\reflemmaeqp{relunf-properties}{five}} \relunf{\relunf\tmtwo\tctxfive}{\tctx} =_{\reflemmaeqp{relunf-properties}{one}} \relunf{\tmtwo}{\tctx}$, where the last equality follows because $\tctxfive$, being a prefix of $\tctxtwo$, has no ES and so cannot capture the variables in $\tmtwo$.\qedhere
 \end{enumerate}
\end{IEEEproof}
\end{atend}

\paragraph*{Rewriting Rules} the two rewriting rules $\togenum$ and $\togenue$ are defined in \reftab{pre-useful-rewritingrules}, and we use $\togenu$ for $\togenum\cup\togenue$. The rules are \emph{global}, \ie\ they do not factor as a rule followed by a contextual closure. As already explained, the context has to be taken into account, to understand if the step is useful to multiplicative redexes.

In rule $\togenum$, the requirement that whole context around the abstraction is useful guarantees that the argument $\tmtwo$ unfolds to a fireball in its context. Note also that in $\togenue$ this is not enough, we have to be sure that such an unfolding is a value, otherwise it will not be useful to multiplicative redexes. Moreover, the rule requires $\tmtwo\neq\tmtwop \esub\vartwo\tmthree$, to avoid copying substitutions.

\begin{atend}
~

The next result to be proved is \refthm{quadratichlis} ($(\tof,\togenu)$ is a locally bounded high-level implementation system). We follow closely the
same approach used for the \ShFC\ in \refapp{app-fireb-end-es} and \refapp{app-high-level-impl}: first we define proper terms and the invariants of reduction; then we characterize normal forms; finally we prove
projection and we obtain the theorem as a corollary.

\begin{definition}[\Proper\ Term]
	A term $\tm$ is \emph{\proper} if 
	\begin{enumerate}
		\item \emph{Evaluability}: $\tm = \tctxp{\tmtwo}$ with $\tctx$ evaluable and $\tmtwo$ a $\l$-term (without ES);
		\item \emph{Value}: no value in $\tm$ contains ES.
	\end{enumerate}
\end{definition}

For instance, a proper term cannot have $\togenum$ redexes inside ES.

Note that initial terms are \proper\ and so the next lemma applies in particular when the starting term is initial.

\begin{lemma}[\Proper\ Invariant]
	Let $\tm$ be a \proper\ and closed term. If $\tm\togenu^*\tmtwo$ then $\tmtwo$ is \proper\ and closed.
\end{lemma}

\begin{IEEEproof}
	by induction on the length $k$ of the derivation $\tm\togenu^*\tmtwo$. The base case is trivial. For the step case, assume $\tm\togenu^{k-1}\tmthree\togenu\tmtwo$. By \ih\ $\tmthree$ is \proper\ and closed.
We distinguish two cases:
\begin{enumerate}
\item \caselight{Case $\tmthree = \tctxp{\sctxp{\l\var.\tmfour}\tmfive} \togenum \tctxp{\sctxp{\tmfour\esub\var{\tmfive}}} = \tmtwo$ where $\tctxp{\sctx\tmfive}$ is evaluable and applicative:}

$\tmtwo$ is closed because $\tmthree$ is. All values in $\tmtwo$ are
values in the proper term $\tmthree$, and therefore they have no ES.
Moreover $\tmfour$ is a sub-term of a value of $\tmthree$, and therefore
has no ES. Since $\tctxp{\sctx\tmfive}$ is evaluable,
$\relunf\tmfive\tctx$ is a fireball by \reflemmap{evaluable-prop}{five} and $\tctx$ and $\tctxp\sctx$ are evaluable by \reflemmap{bis-evaluable-prop}{two}. Therefore
$\tctxp{\sctxp{\ctxhole\esub\var\tmfive}}$ is evaluable too
by the other direction of \reflemmap{evaluable-prop}{five} and
the evaluability of $\tctxp{\sctx}$. Therefore $\tmtwo$ is proper.

\item \caselight{Case\\\mbox{$\tmthree = \tctxp{\tctxtwop\var\esub\var{\sctxp{\valt}}} \togenue  \tctxp{\sctxp{\tctxtwop\valt\esub\var\valt}} = \tmtwo$} where $\tctxp{\tctxtwo\esub\var{\sctxp{\valt}}}$ is evaluable and applicative and $\relunf\valt{\tctxp\sctx}= \la\vartwo\tmsix$:}

$\tmtwo$ is closed because $\tmthree$ is. All values in $\tmtwo$ are
values in the proper term $\tmthree$, and therefore they have no ES.
In particular, $\valt$ has no ES. Thus $\tmtwo$ is proper.\qedhere
\end{enumerate}
\end{IEEEproof}

\begin{lemma}[Normal Form Characterization]
	\label{l:useful-nf-char}
	Let $\tm = \tctxp\tmtwo$ be a proper and closed term s.t. $\tmtwo\not\togenu$ and $\tctx$ is evaluable. Then 
	\begin{enumerate}
		\item \label{p:useful-nf-char-fireball} either  $\relunf\tmtwo\tctx$ is a \fireball, 
		\item \label{p:useful-nf-char-redex} or $\relunf\tmtwo\tctx\tof$, more precisely exists $\tctxtwo$ s.t.
		\begin{enumerate}
			\item $\tmtwo = \tctxtwop\var$ with 
			\item $\var\in\fv\tmtwo$,
			\item $\tctxp\tctxtwo$ evaluable,
			\item $\relunf\var\tctx = \la\vartwo\tmthree$, and
			\item $\tctxtwo$ is applicative.
			
		\end{enumerate}
	\end{enumerate}
\end{lemma}

\myproof{
\begin{IEEEproof}
	first of all, let us show that the conditions on $\tctxtwo$ imply $\relunf\tmtwo\tctx\tof$. We have:
	$$\begin{array}{ll}\relunf\tmtwo\tctx & =_a\\
          \relunf{\tctxtwop\var}\tctx & =_{\reflemmaeqp{relunf-properties}{four}}\\
          \relunf{\tctxtwo}\tctx\ctxholep{\relunf\var{\tctxp\tctxtwo}} & =_{c\&\reflemmaeqp{bis-evaluable-prop}{two}}\\
          \evctxp{\relunf\var{\tctxp\tctxtwo}} & =_{b\&\reflemmaeqp{relunf-properties}{one}}\\
          \evctxp{\relunf\var\tctx} & =_{d}\\
          \evctxp{\la\vartwo\tmthree}\end{array}$$
	and $\evctx \defeq \relunf{\tctxtwo}\tctx$ is applicative, by $c$ and \reflemmaeqp{relunf-properties}{three}.
	Then note that  \refpoint{useful-nf-char-fireball} and \refpoint{useful-nf-char-redex} are mutually exclusive. Indeed, by \reflemma{gconst-or-fire-fnorm}, an unfolded term which is a \fireball\ is $\tof$-normal. So, if \refpoint{useful-nf-char-fireball} holds then \refpoint{useful-nf-char-redex} does not, and vice-versa. Therefore, in the following proof for we only have to prove that \refpoint{useful-nf-char-fireball} or \refpoint{useful-nf-char-redex} holds.

	By induction on $\tmtwo$. Cases:
	\begin{enumerate}
		\item \emph{Variable $\var$}. Since $\tm$ is proper and closed, $\relunf\tmtwo\tctx$ is a \fireball.
		\item \emph{\Symb} and \emph{Abstraction}. Note that by properness, the abstraction is an ordinary $\lambda$-term, \ie\ it does not contain ES. Then in both cases we can apply \reflemma{es-answers-are-normal}, obtaining that $\relunf\tmtwo\tctx$ is a \fireball.
		\item \emph{Application} $\tmtwo = \tmthree\tmfour$. Since $\tctxp{\tmthree\ctxhole}$ is an evaluable context, we can apply the \ih\ to $\tmfour$, ending in one of the following two cases:
		\begin{enumerate}
			\item \emph{$\relunf\tmfour\tctx$ is a \fireball}. Then $\tctxp{\ctxhole\tmfour}$ is an evaluable context and we can apply the \ih\ to $\tmthree$ obtaining two cases:
			\begin{enumerate}
				\item \emph{$\relunf\tmthree\tctx$ is a \fireball}. Two kinds of \fireball:
				\begin{itemize}
					\item \emph{$\relunf\tmthree{\tctxp{\ctxhole\tmfour}}$ is a inert $\gconst$}. Then $\relunf\tmtwo\tctx = \gconst \relunf\tmfour\tctx$ is a inert, \ie\ a \fireball.
					\item \emph{$\relunf\tmthree{\tctxp{\ctxhole\tmfour}}$ is an abstraction $\la\vartwo\tmfive$}. Then $\relunf\tmtwo\tctx$ reduces, indeed $\relunf\tmtwo\tctx = \relunf\tmthree{\tctxp{\ctxhole\tmfour}} \relunf\tmfour\tctx = (\la\vartwo\tmfive) \relunf\tmfour\tctx \tof$. In terms of contexts, note that $\tmthree$ is not itself an abstraction, otherwise $\tmtwo$ would be a $\togenum$-redex, \ie\ $\tmthree$ has the form $\sctxp\var$. Moreover, $\sctx$ does not capture $\var$, otherwise $\tmtwo = \tmthree\tmfour = \sctxp\vartwo\tmfour$ would have a $\togenue$-redex (because $\tm$ is proper and so the substitution on $\var$ in $\sctx$ can fire). Then $\var\in\fv\tmthree$ (and so $\var\in\fv\tmtwo$) and $\tctxtwo \defeq \sctx\tmfour$ satisfies points $a,b,d,e$ of the statement. For $c$, we only have to show that the content of every substitution in $\sctx$ unfolds to a \fireball\ in its context (by \reflemmap{evaluable-prop}{five}). Note that, since $\tm$ is proper, there is an evaluable context containing all the ES in $\tm$, \ie\ the content of \emph{every} substitution in $\tm$ unfolds in its context to a \fireball.
				\end{itemize}
				
				\item \emph{$\relunf\tmthree\tctx$ reduces, \ie\ $\relunf\tmthree\tctx\tof$}. We have $\relunf\tmtwo\tctx = \relunf\tmthree{\tctxp{\ctxhole\tmfour}} \relunf\tmfour\tctx = \relunf\tmthree\tctx \relunf\tmfour\tctx\tof$ because $\relunf\tmfour\tctx$ is a \fireball\ and so $\ctxhole \relunf\tmfour\tctx$ is an evaluation context. In terms of contexts, set $\tctxtwo \defeq \tctxthree\tmfour$, where $\tctxthree$ is he context given by the \ih. It is easily seen that $\tctxtwo$ satisfies the statement.
			\end{enumerate}
			\item \emph{$\relunf\tmfour\tctx$ reduces, \ie\ $\relunf\tmfour\tctx\tof$}. We have $\relunf\tmtwo\tctx = \relunf\tmthree{\tctxp{\ctxhole\tmfour}} \relunf\tmfour\tctx$ because $\relunf\tmthree{\tctxp{\ctxhole\tmfour}}\ctxhole$ is an evaluation context. In terms of contexts, set $\tctxtwo \defeq \tmthree\tctxthree$, where $\tctxthree$ is he context given by the \ih. It is easily seen that $\tctxtwo$ satisfies the statement.
		\end{enumerate}
		\item \emph{Substitution}  $\tmtwo = \tmthree\esub\var\tmfour$.  $\unf{\tctxp{\ctxhole\esub\var\tmfour}}$ is an evaluation context and we can apply the \ih\ to $\tmthree$. Note that since $\relunf\tmthree{\tctxp{\ctxhole\esub\var\tmfour}} =_{\reflemmaeqp{relunf-properties}{four}} \relunf{\tmthree\esub\var\tmfour}{\tctx} = \relunf\tmtwo\tctx$, this case reduces to the \ih. In terms of contexts (for \refpoint{useful-nf-char-redex}), note that the context $\tctxthree$ given by the \ih\ cannot expose an occurrence of $\var$, otherwise there would be a $\togenue$-redex in $\tmtwo$ (because $\tm$ is proper and so $\tmfour$ has the form $\sctxp\val$). Thus, the context $\tctxtwo \defeq \tctxthree\esub\var\tmfour$ is easily seen to satisfy the statement (inheriting the properties of $\tctxthree$).
	\end{enumerate}
\end{IEEEproof}
}

\begin{corollary}[Normal Forms Unfold to Normal Forms]
	\label{coro:useful-nfs-unfold-to-nfs-useful}
	Let $\tm$ be a closed proper term. If $\tm$ is $\togenu$-normal then $\unf\tm$ is $\tof$-normal.
\end{corollary}

\begin{IEEEproof}
note that applying \reflemma{useful-nf-char} with $\tctx\defeq\ctxhole$ and $\tmtwo\defeq\tm$ one obtains that $\unf\tm$ is a \fireball, because the second case cannot happen, given that $\tmtwo$ now is closed and so it cannot be written as $\tmtwo = \tctxtwop\var$ with $\var\in\fv\tmtwo$. By \reflemma{gconst-or-fire-fnorm}, $\unf\tm$ is $\tof$-normal.
\end{IEEEproof}

To prove the projection lemma we need to prove first as a technical lemma
another sufficient condition for a context to be evaluable. The condition
is based on the definition of position of a redex.

The \emph{position} of a redex is (the context $\tctx$ exposing) the application that makes applicative the evaluable context in the side condition. For a $\togenum$-redex, it is given by $\tctx$, while for a $\togenue$-redex one needs to do a case analysis, because the application may lie in $\tctx$ or in $\tctxtwo$. Note that such a notion of position for $\togenue$-redexes is different with respect to the one used in \refssect{explicit-nfs-det}.

\begin{lemma}
		\label{l:redex-pos-eval}
		If $\tctxp\tm$ has a redex having its position in $\tm$ then $\tctx$ is evaluable.
\end{lemma}

\begin{IEEEproof}
then the position of the redex has the form $\tctxp\tctxtwo$ for some context $\tctxtwo$. By the hypothesis on redexes and \reflemmap{bis-evaluable-prop}{two}, $\tctxp\tctxtwo$ is evaluable. By \reflemmap{bis-evaluable-prop}{two}, $\tctx$ is evaluable.
\end{IEEEproof}

\begin{lemma}[Projection]
	\label{l:useful-projection}
	Let $\tm = \tctxp\tmtwo \togenu \tctxtwop{\tmthree} = \tmfour$ by reducing a redex whose position lies in $\tmtwo$. If the redex is
	\begin{enumerate}
		\item \emph{Multiplicative}: then $\relunf\tmtwo\tctx \tof \relunf\tmthree\tctxtwo$ and $\unf\tm \tof \unf\tmfour$; 
		\item \emph{Exponential}: then $\relunf\tmtwo\tctx \tof$ and $\unf\tm = \unf\tmfour\tof$.
	\end{enumerate}
	In both cases $\relunf\tmtwo\tctx$ is not a \fireball.
\end{lemma}

\begin{IEEEproof}
	the fact that in both cases $\relunf\tmtwo\tctx$ is not a \fireball, follows from \reflemma{gconst-or-fire-fnorm} and the fact that $\relunf\tmtwo\tctx$ reduces. Cases:
	\begin{enumerate}
		\item \emph{Multiplicative}. Note that in this case $\tctxtwo = \tctx$. Then $\unf\tm \tof \unf\tmfour$ follows from $\relunf\tmtwo\tctx \tof \relunf\tmthree\tctx$. By \reflemma{redex-pos-eval}, $\tctx$ is evaluable, and by \reflemmap{evaluable-prop}{one} $\unf\tctx$ is an evaluation context, so:
		$$\begin{array}{ll}
                  \unf\tm & =\\
                  \unf{\tctxp\tmtwo} & =_{\reflemmaeqp{relunf-properties}{four}}\\
                  \unf\tctx\ctxholep{\relunf\tmtwo\tctx} & \tof\\
                  \unf\tctx\ctxholep{\relunf\tmthree\tctxtwo} & =_{\reflemmaeqp{relunf-properties}{four}} \\
                  \unf{\tctxp\tmthree} & =\\
                  \unf\tmfour\end{array}$$
		
		We now show $\relunf\tmtwo\tctx \tof \relunf\tmthree\tctx$. Since the redex lies in $\tmtwo$, we have $\tmtwo = \tctxtwop{\sctxp{\la\var\tmthree}\tmfour}$ and $\tm = \tctxp{\tctxtwop{\sctxp{\la\var\tmthree}\tmfour}}$ with $\tctxp{\tctxtwop{\ctxhole \tmfour}}$, and thus $\tctxp\tctxtwo$, evaluable. Moreover, by \reflemmap{bis-evaluable-prop}{two} $\relunf{\tmfour}{\tctxp{\tctxtwop{\sctxp{\la\var\tmthree}\ctxhole}}} = \relunf\tmfour{\tctxp\tctxtwo}$ is a \fireball\ and $\relunf\tctxtwo\tctx$ is an evaluation context. Then
		\[\begin{array}{lll}
			\relunf\tmtwo\tctx & =\\
                        \relunf{\tctxtwop{\sctxp{\la\var\tmthree}\tmfour}}\tctx & = &\mbox{(\reflemmap{relunf-properties}{four})}\\
			\relunf{\tctxtwo}\tctx\ctxholep{\relunf{(\sctxp{\la\var\tmthree}\tmfour)}{\tctxp\tctxtwo}} & =&\mbox{(\reflemmap{relunf-properties}{zero})}\\
			\relunf{\tctxtwo}\tctx\ctxholep{\relunf{\sctxp{\la\var\tmthree}}{\tctxp\tctxtwo}\relunf{\tmfour}{\tctxp\tctxtwo}} & = &\mbox{(\reflemmap{relunf-properties}{four})}\\
			\relunf{\tctxtwo}\tctx\ctxholep{\relunf{(\la\var\tmthree)}{\tctxp{\tctxtwop\sctx}}\relunf{\tmfour}{\tctxp\tctxtwo}} & =&\mbox{(\reflemmap{relunf-properties}{zero})}\\
			\relunf{\tctxtwo}\tctx\ctxholep{\la\var\relunf{\tmthree}{\tctxp{\tctxtwop\sctx}}\relunf{\tmfour}{\tctxp\tctxtwo}} & \tof &\mbox{($\relunf\tctxtwo\tctx$ is  an ev.}\\&&\mbox{cont. \& $\relunf\tmfour{\tctxp\tctxtwo}$}\\&&\mbox{is a \fireball)}\\
			\relunf{\tctxtwo}\tctx\ctxholep{\relunf{\tmthree}{\tctxp{\tctxtwop\sctx}}\isub\var{\relunf\tmfour{\tctxp\tctxtwo}}} & = &\mbox{(\reflemmap{relunf-properties}{zero})}\\
			\relunf{\tctxtwo}\tctx\ctxholep{\relunf{\tmthree\isub\var\tmfour}{\tctxp{\tctxtwop\sctx}}} & = &\mbox{(\reflemmap{relunf-properties}{zero})}\\

			\relunf{\tctxtwo}\tctx\ctxholep{\relunf{\tmthree\esub\var\tmfour}{\tctxp{\tctxtwop\sctx}}} & = &\mbox{(\reflemmap{relunf-properties}{four})}\\
			\relunf{\tctxtwo}\tctx\ctxholep{\relunf{\sctxp{\tmthree\esub\var\tmfour}}{\tctxp\tctxtwo}} & = &\mbox{(\reflemmap{relunf-properties}{four})}\\
			\relunf{\tctxtwo\ctxholep{\sctxp{\tmthree\esub\var\tmfour}}}\tctx & =\\
                        \relunf\tmthree\tctx\\

		\end{array}\]
		
		\item \emph{Exponential}. We take $\unf\tm = \unf\tmfour$ for granted, because a substitution step by definition does not change the unfolding.
		 Similarly to the previous case, $\unf\tm \tof $ follows from $\relunf\tmtwo\tctx \tof $. Indeed, by \reflemmap{bis-evaluable-prop}{two}, $\tctx$ is evaluable, and by \reflemmap{evaluable-prop}{one} $\unf\tctx$ is an evaluation context, so:
		$$\unf\tm = \unf{\tctxp\tmtwo} =_{\reflemmaeqp{relunf-properties}{four}} \unf\tctx\ctxholep{\relunf\tmtwo\tctx} \tof $$
		
		Now we prove $\unf\tmtwo \tof$. We have $\tmtwo = \tctxtwop{\sctxp\var\tmfour}$ and $\tm = \tctxp{\tctxtwop{\sctxp\var\tmfour}}$. In $\tm$ there is somewhere (in $\sctx$, $\tctxtwo$, or $\tctx$) a substitution $\esub\var{\sctxtwop\tmfive}$ with the hypothesis that $\tmfive$ relatively unfolds to some value $\la\vartwo\tmthree$ in its context. So, $\relunf\var{\tctxp{\tctxtwop{\sctx}}}= \la\vartwo\tmthree$. Moreover, by hypothesis $\tctxp\tctxtwo$ is evaluable, and so by \reflemmap{bis-evaluable-prop}{two} $\relunf\tctxtwo\tctx$ is an evaluation context. Finally, $\relunf\tmfour{\tctxp\tctxtwo}$ is a \fireball, because $\tctxp{\tctxtwop{\ctxhole\tmfour}}$ is evaluable. Then
		
		\[\begin{array}{lll}
			\relunf\tmtwo\tctx & =\\
                        \relunf{\tctxtwop{\sctxp\var\tmfour}}\tctx & = &\mbox{(\reflemmap{relunf-properties}{four})}\\
			\relunf{\tctxtwo}\tctx\ctxholep{\relunf{(\sctxp\var\tmfour)}{\tctxp\tctxtwo}} & = &\mbox{(\reflemmap{relunf-properties}{zero})}\\
			\relunf{\tctxtwo}\tctx\ctxholep{\relunf{\sctxp\var}{\tctxp\tctxtwo}\relunf{\tmfour}{\tctxp\tctxtwo}} & = &\mbox{(\reflemmap{relunf-properties}{four})}\\
			\relunf{\tctxtwo}\tctx\ctxholep{\relunf{\var}{\tctxp{\tctxtwop\sctx}}\relunf{\tmfour}{\tctxp\tctxtwo}} & = &\mbox{($\relunf\var{\tctxp{\tctxtwop{\sctx}}}= \la\vartwo\tmthree$)}\\
			\relunf{\tctxtwo}\tctx\ctxholep{(\la\vartwo\tmthree)\relunf{\tmfour}{\tctxp\tctxtwo}} & \tof &\mbox{($\relunf\tctxtwo\tctx$ is an ev. cont.}\\&&\mbox{\& $\relunf\tmfour{\tctxp\tctxtwo}$ a \fireball)}\\
			\relunf{\tctxtwo}\tctx\ctxholep{\tmthree\esub\vartwo{\relunf{\tmfour}{\tctxp\tctxtwo}}}\\
		\end{array}\]
	\end{enumerate}
\end{IEEEproof}

Determinism of $\togenu$ is the last ingredient to prove that
$(\tof,\togenu)$ is a locally bounded high-level implementation system.

\begin{lemma}[Determinism]
	\label{l:useful-determinism}
	Let $\tm$ be a term and $\tctxp\tctxONE$ and $\tctxp\tctxTWO$ positions of $\togenu$-redexes. Then $\tctxONE = \tctxTWO$.
\end{lemma}

\myproof{
\begin{IEEEproof}
	by induction on $\tctxONE$. Cases:
	\begin{enumerate}
		\item \emph{Empty} $\tctxONE = \ctxhole$. Cases:
		\begin{enumerate}
			\item \emph{Multiplicative Redex}, \ie\ $\tmtwo = \sctxp{\la\var\tmfour}\tmfive$ with $\relunf\tmfive\tctx$ a \fireball. Now, $\tctxTWO$ cannot lie in $\sctxp{\la\var\tmfour}$, otherwise by \reflemma{useful-projection} $\relunf{\sctxp{\la\var\tmfour}}{\tctxp\tctxTWO}$ would not be a \fireball, while by (properness and) \reflemma{es-answers-are-normal} it does. Nor $\tctxTWO$ can lie in $\tmfive$, otherwise again by \reflemma{useful-projection} $\relunf\tmfive\tctx$ would not be a \fireball. Then necessarily $\tctxTWO = \tctxONE = \ctxhole$.
			\item \emph{Exponential Redex}, \ie\ $\tmtwo = \tctxtwop{\sctxp\var \tmfour}$. Now, $\tctxTWO$ cannot lie in $\sctxp\var$, otherwise by \reflemma{useful-projection} $\relunf{\sctxp\var}{\tctxp\tctxTWO}$ would not be a \fireball, while by the hypothesis on the $\togenue$-redex it does (it is an abstraction). Nor $\tctxTWO$ can lie in $\tmfour$, otherwise again by \reflemma{useful-projection} $\relunf\tmfour\tctx$ would not be a \fireball, while by the hypothesis on the $\togenue$-redex it does. Then necessarily $\tctxTWO = \tctxONE = \ctxhole$.
		\end{enumerate}
			
		\item \emph{Right Application} $\tctxONE = \tmfour\tctxONEtwo$ and $\tm = \tmfour\tctxONEtwop\tmfive$. 	By \reflemma{useful-projection}, $\relunf{\tctxONEtwop\tmfive}{\tctxp{\tmfour\ctxhole}}$ has a $\tof$-redex and it is not a \fireball, so no redexes can lie to its left, in particular $\tctxTWO$ does not lie in $\tmfour$. By \reflemma{useful-projection}, $\relunf{\tctxONEtwop\tmfive}{\tctxp{\tmfour\ctxhole}}$ is not a \fireball, and so $\tctxTWO$ cannot be empty (\ie\ $\tmfour\tctxONEtwop\tmfive$ cannot be the position of a $\togenum$-redex). Then, $\tctxTWO = \tmtwo\tctxTWOtwo$, and the statement follows from the \ih\ applied to $\tctxONEtwo$ and $\tctxTWOtwo$.

		\item \emph{Left Application} $\tctxONE = \tctxONEtwo \tmfive$ and $\tm = \tctxONEtwop\tmfour \tmfive$. Note that $\tctxTWO$ cannot lie in $\tmfive$, otherwise by \reflemma{useful-projection} $\relunf\tmfive{\tctxp{\tctxONEtwop\tmfour\ctxhole}}$ has a $\tof$-redex and it is not a \fireball, and so no redexes---in particular the one of position $\tctxp\tctxONE$---can lie to its left, absurd. And $\tctxTWO$ cannot be empty (\ie\ the position of a $\togenum$-redex), because then $\tctxONEtwop\tmfour$ would have the form $\sctxp{\la\var\tmsix}$, which by \reflemma{useful-projection} cannot contain the position of a redex, because by \reflemma{es-answers-are-normal} $\relunf{\sctxp{\la\var\tmsix}}{\tctxp{\ctxhole\tmfive}}$ is a \fireball. Then, $\tctxTWO = \tctxTWOtwo\tmthree$, and the statement follows from the \ih\ applied to $\tctxONEtwo$ and $\tctxTWOtwo$. 
		
		\item \emph{Substitution} $\tctxONE = \tctxONEtwo\esub\var\tmthree$. Then necessarily $\tctxTWO = \tctxTWOtwo\esub\var\tmthree$ (remember the position of a $\togenue$-redex is an application) and the statement follows from the \ih.\qedhere
	\end{enumerate}
\end{IEEEproof}
}

\end{atend}

A detailed study of useful evaluation (in the appendix) shows that:
\begin{theorem}[Quadratic High-Level Implementation]\label{thm:quadratichlis}
	$(\tof,\togenu)$ is a locally bounded high-level implementation system, and so it has a quadratic overhead wrt $\tof$.
\end{theorem}
\begin{proofatend}
the pair $(\tof,\togenu)$ is an high-level implementation system because of
\reflemma{useful-determinism},
\reflemma{useful-projection} and
\refcoro{useful-nfs-unfold-to-nfs-useful}.

We deduce that the implementation system is locally bounded from
the corresponding bound (\reflemmap{uglaminvariants}{size}) on the abstract
machine that implements the calculus. An alternative, direct proof without any
reference to abstract machines is surely possible, but we would need to
establish first additional invariants on the ES that occur in the term.
Intuitively, anyway, the local bound follows mainly from acyclicity of the
explicit substitutions and the fact that only
multiplicative steps can create a new ES, while exponential steps never
duplicate terms containing ES.
\end{proofatend}

Moreover, the structural equivalence $\equiv$ is a strong bisimulation also with respect to $\togenu$.

\begin{proposition}[$\tostruct$ is a Strong Bisimulation wrt $\togenu$]
	\label{prop:useful-strong-bis}
	Let $\mathtt{x}\in\set{\mathtt{um},\mathtt{ue}}$. Then, $\tm\eqstruct\tmtwo$ and $\tm\togenx\tmp$ implies that there exists $\tmtwop$ such that $\tmtwo\togenx\tmtwop$ and $\tmp\eqstruct\tmtwop$.
\end{proposition}

\begin{proofatend}
 omitted. All postponement proofs are similar and lengthy.
In \refssect{app-struct-eq} of the Appendix we proved the lemma for the \ShFC.
Other examples can be found in the long version of~\cite{DBLP:conf/icfp/AccattoliBM14}.
\end{proofatend}

\section{The GLAMoUr Machine}\label{sect:glamour}
\begin{atend}\subsection{Proofs Omitted From \refsect{glamour}\\ (The GLAMoUr Machine)}\end{atend}
\begin{atend}
The aim of this section is to prove \refth{uglam-refl-distillation}
($(\glamour,\togenu, \eqstruct, \decodefun)$ is a reflective explicit distillery) and \refth{preusefulimpl}
(the useful implementation has bilinear low level and quadratic high level
complexity). We start by proving that the invariants of the machine holds.

\end{atend}
Here we refine the GLAM with a very simple tagging of stacks and environments, in order to implement useful sharing. The idea is that every term in the stack or in the environment carries a label $\lab\in\set{\alive,\dead}$ indicating if it unfolds (relatively to the environment) to a value or to a inert. 

The grammars are identical to the GLAM, up to labels:
$$\begin{array}{rclllrcllllllll}
        \lab & \grameq & \alive \mid \dead &&&&
	\genv,\genvtwo		& \grameq & \stempty \mid  \econsx{\esub\var{\stackitem^\lab}}{\genv}\\
	\stack,\stacktwo 	& \grameq & \stempty \mid \stackitem^\lab \cons \stack
\end{array}$$

The decoding of the various machine components is identical to that for the
GLAM, up to labels that are ignored. The state context, however, now is noted $\tctx_\state$, as it is not necessary an evaluation context. 

The transitions are in \reftab{pre-useful-transitions}. They are obtained from those
of the GLAM by:

\begin{enumerate}
\item \emph{Backtracking instead of performing a useless substitution}: there are two new backtracking cases $\tomachcfour$ and $\tomachcfive$ (that in the \GLAM\ were handled by the exponential transition), corresponding to avoided useless duplications: $\tomachcfour$ backtracks when the entry $\phi$ to substitute is marked $\dead$ (as it unfolds to a inert) and $\tomachcfive$ backtracks when the term is marked $\alive$ but the stack is empty (\ie\ the context is not applicative). 

\item \emph{Substituting only when it is useful}: the exponential transition is applied only when the term to substitute has label $\alive$ and the stack is non-empty.
\end{enumerate}


\begin{atend}
\begin{lemma}[Contextual Decoding]\label{l:contextualdecoding}
$\decgenv$ is a substitution context; $\decdump$ and $\decstack$ are
shallow contexts without ES.
\end{lemma}
\begin{IEEEproof}
by induction on $\genv$, $\dump$ and $\stack$.
\end{IEEEproof}
\end{atend}



\begin{lemma}[GLAMoUr Invariants]\label{l:uglaminvariants} 
	Let $\state = \usefmac{\genv}{\dump}{\codetwo}{\stack} $ be a state reachable from an initial code $\code$. Then:
	\begin{enumerate}
		\item \emph{Closure}:\label{p:uglaminvariants-closure} $\state$ is closed and well named;
		\item \emph{Value}:\label{p:uglaminvariants-value} values in components of $\state$ are sub-terms of $\code$;
		\item \emph{\Fireball}:\label{p:uglaminvariants-fireball} $\relunf\code{\decgenv}$ is a fireball (of kind $\lab$) for every code $\code^\lab$ in $\stack$, $\genv$, and in every stack of $\dump$;
		\item \emph{Evaluability}:\label{p:uglaminvariants-eval} $\decode\genv$, $\relunf{\decdump}{\decode\genv}$, $\relunf{\decstack}{\decode\genv}$, and $\tctx_\state$ are evaluable contexts;
		\item \emph{Environment Size}:\label{p:uglaminvariants-size} the length of the global environment $\genv$ is
bound by $\sizem\exec$.
	\end{enumerate}
\end{lemma}

\begin{proofatend}
by induction over the length of the execution.
The base case holds because $\code$ is initial.
The inductive step is by cases over the kind of transition.
All the verifications are trivial apart for \refpoint{uglaminvariants-eval}. For \refpoint{uglaminvariants-eval}, evaluability for $\decode\genv$, $\relunf{\decdump}{\decode\genv}$, $\relunf{\decstack}{\decode\genv}$ follows from \refpoint{uglaminvariants-fireball} and \reflemmap{evaluable-prop}{five}, while evaluability for $\decgenvpx{\genv}{\decdumpp{\decstack}}$ follows from them and \reflemmap{evaluable-prop}{five}.
\end{proofatend}

\begin{atend}
\begin{lemma}[Explicit Distillation]\label{l:uglam-distillation}
	Let $\state$ be a reachable state. Then:
		\begin{enumerate}
		\item \emph{Commutative}: if $\state\tomachcp{_{1,2,3,4,5}}\statetwo$ then $\decode\state = \decode\statetwo$;
		\item \emph{Multiplicative}: if $\state\tomachum\statetwo$ then $\decode\state\togenum\eqstruct\decode\statetwo$;
		\item \emph{Exponential}: if $\state\tomachue\statetwo$ then $\decode\state\togenue\decode\statetwo$.
	\end{enumerate}
\end{lemma}

\begin{IEEEproof}
we list the transition in the order they appear in the definition of the machine.
\begin{itemize}
\item
Case
     $\usefmac{\genv}{\dump}{\code\codetwo}{\stack}
     \tomachcone
     \usefmac{\genv}{\dump\cons\nfnst\code\stack}{\codetwo}{\stempty}$:
 $$\begin{array}{ll}
     \decode{\usefmac{\genv}{\dump}{\code\codetwo}{\stack}}
     &=\\
     \decgenvpx{\genv}{\decodep{\dump}{\decodestack{\code\codetwo}{\stack}}}
     &=\\
     \decgenvpx{\genv}{\decodep{\dump}{\decodestack{\code\ctxholep{\codetwo}}{\stack}}} 
     &=\\
     \decgenvpx{\genv}{\decodep{\dump\cons\nfnst\code\stack}{\codetwo}}
     &=\\
     \decgenvpx{\genv}{\decodep{\dump\cons\nfnst\code\stack}{\decodestack{\codetwo}{\stempty}}}
     &=\\
     \decode{\usefmac{\genv}{\dump\cons\nfnst\code\stack}{\codetwo}{\stempty}}
 \end{array}$$

\item
Case 
     $\usefmac{\genv}{\dump}{\l\var.\code}{\stackitem^\lab\cons\stack}
     \tomachum
     \usefmac{\esub\var{\stackitem^\lab}\genv}{\dump}{\code}{\stack}$:
 $$\begin{array}{lll}
     \decode{\usefmac{\genv}{\dump}{\l\var.\code}{\stackitem^\lab\cons\stack}}
     &=\\
     \decgenvpx{\genv}{\decodep{\dump}{\decodestack{\l\var.\code}{\stackitem^\lab\cons\stack}}}
     &=\\
     \decgenvpx{\genv}{\decodep{\dump}{\decodestack{(\l\var.\code)\decode{\stackitem}}{\stack}}}
     &\togenum&\mbox{(by \reflemmap{uglaminvariants}{eval}}\\&&\mbox{and \reflemmap{uglaminvariants}{fireball})}\\
     \decgenvpx{\genv}{\decodep{\dump}{\decodestack{\code\esub{\var}{\decode{\stackitem}}}{\stack}}}
     &\eqstruct&\mbox{(by \reflemma{ev-comm-struct})}\\
     \decgenvpx{\genv}{\decodep{\dump}{\decodestack{\code}{\stack}}\esub\var{\stackitem}}
     &=\\
     \decgenvpx{\esub\var{\stackitem^\lab}\genv}{\decodep{\dump}{\decodestack{\code}{\stack}}}
     &=\\
     \decode{\usefmac{\econs{\esub\var{\stackitem^\lab}}{\genv}}{\dump}{\code}{\stack}}
 \end{array}$$

\item
Case
     $\usefmac{\genv}{\dump\cons\nfnst\code\stack}{\l\var.\codetwo}{\stempty}
     \tomachctwo
     \usefmac{\genv}{\dump}{\code}{(\l\var.\codetwo)^\alive\cons\stack}:$
 $$\begin{array}{ll}
     \decode{\usefmac{\genv}{\dump\cons\nfnst\code\stack}{\l\var.\codetwo}{\stempty}}
     &=\\
     \decgenvpx{\genv}{\decodep{\dump\cons\nfnst\code\stack}{\decodestack{\l\var.\codetwo}{\stempty}}}
     &=\\
     \decgenvpx{\genv}{\decodep{\dump\cons\nfnst\code\stack}{\l\var.\codetwo}}
     &=\\
     \decgenvpx{\genv}{\decodep{\dump}{\decodestack{\code(\l\var.\codetwo)}{\stack}}}
     &=\\
     \decgenvpx{\genv}{\decodep{\dump}{\decodestack{\code}{(\l\var.\codetwo)^\alive\cons\stack}}}
     &=\\
     \decode{\usefmac{\genv}{\dump}{\code}{(\l\var.\codetwo)^\alive\cons\stack}}
 \end{array}$$

\item
Case
     $\usefmac{\genv}{\dump\cons\nfnst\code\stack}{\const}{\stacktwo}
     \tomachcthree
     \usefmac{\genv}{\dump}{\code}{\pair{\const}{\stacktwo}^\dead\cons\stack}$:
 $$\begin{array}{ll}
     \decode{\usefmac{\genv}{\dump\cons\nfnst\code\stack}{\const}{\stacktwo}}
     &=\\
     \decgenvpx{\genv}{\decodep{\dump\cons\nfnst\code\stack}{\decodestack{\decode{\const}}{\stacktwo}}}
     &=\\
     \decgenvpx{\genv}{\decodep{\dump}{\decodestack{\code\decodestack{\const}{\stacktwo}}{\stack}}}
     &=\\
     \decgenvpx{\genv}{\decodep{\dump}{\decodestack{\code}{\pair{\const}{\stacktwo}^\dead\cons\stack}}}
     &=\\
     \decode{\usefmac{\genv}{\dump}{\code}{\pair{\const}{\stacktwo}^\dead\cons\stack}}
 \end{array}$$

\item
Case
     $\usefmac{\genv_1\esub\var{\stackitem^\dead}\genv_2}{\dump\cons\nfnst\code\stack}{\var}{\stacktwo}
     \tomachcfour
     \usefmac{\genv_1\esub\var{\phi^\dead}\genv_2}{\dump}{\code}{\pair{\var}{\stacktwo}^\dead\cons\stack}$:
 $$\begin{array}{ll}
     \decode{\usefmac{\genv_1\esub\var{\stackitem^\dead}\genv_2}{\dump\cons\nfnst\code\stack}{\var}{\stacktwo}}
     &=\\
     \ldots&=\\
     \decode{\usefmac{\genv_1\esub\var{\phi^\dead}\genv_2}{\dump}{\code}{\pair{\var}{\stacktwo}^\dead\cons\stack}}
 \end{array}$$
The proof is the one for the previous case $\tomachcthree$,
by replacing $\const$ with $\var$ and instantiating $\genv$ with
${\genv_1\esub\var{\phi^\dead}\genv_2}$.

\item
Case
     $\usefmac{\genv_1\esub\var{\codetwo^\alive}\genv_2}{\dump\cons\nfnst\code\stack}{\var}{\stempty}
     \tomachcfive
     \usefmac{\genv_1\esub\var{\herval{\codetwo}}\genv_2}{\dump}{\code}{\var^\alive\cons\stack}$:
 $$\begin{array}{ll}
     \decode{\usefmac{\genv_1\esub\var{\codetwo^\alive}\genv_2}{\dump\cons\nfnst\code\stack}{\var}{\stempty}}
     &=\\
     \ldots&=\\
     \decode{\usefmac{\genv_1\esub\var{\herval{\codetwo}}\genv_2}{\dump}{\code}{\var^\alive\cons\stack}}
 \end{array}$$
The proof is the one for the previous case $\tomachcfour$,
by replacing $(\l\var.\codetwo)$ with $\var$ and instantiating
$\genv$ with ${\genv_1\esub\var{\herval{\codetwo}}\genv_2}$.

\item
Case
     $\usefmac{\genv_1\esub\var{\herval\codevalt}\genv_2}{\dump}{\var}{\stackitem^\lab\cons\stack}
     \tomachee
     \usefmac{\genv_1\esub\var{\herval\codevalt}\genv_2}{\dump}{\rename\codevalt}{\stackitem^\lab\cons\stack}$:
 $$\begin{array}{lll}
     \decode{\usefmac{\genv_1\esub\var{\herval\codevalt}\genv_2}{\dump}{\var}{\stackitem^\lab\cons\stack}}
     &=\\
     \decgenvpx{\genv_1\esub\var{\herval\codevalt}\genv_2}{\decodep{\dump}{\decodestack{\var}{\stackitem^\val\cons\stack}}}
     &\togenue&\mbox{(by \reflemmap{uglaminvariants}{eval}}\\&&\mbox{and \reflemmap{uglaminvariants}{fireball})}\\
     \decgenvpx{\genv_1\esub\var{\herval\codevalt}\genv_2}{\decodep{\dump}{\decodestack{\rename\codevalt}{\stackitem^\lab\cons\stack}}}
     &=\\
     \decode{\usefmac{\genv_1\esub\var{\herval\codevalt}\genv_2}{\dump}{\rename\codevalt}{\stackitem^\lab\cons\stack}}
 \end{array}$$
\end{itemize}
\end{IEEEproof}

The next lemma extends the notion of state size $\size\state$ given in
\refdef{sizetwo} by ignoring the new machine component $\genv$.
The precise definition is \refdef{presize}.

\begin{lemma}[Determinism]\label{l:uglam-deterministic} The transition relation $\tomach$ of the \glamour\ is deterministic.
\end{lemma}
\begin{IEEEproof}
a simple inspection of the transitions shows no critical pairs.
\end{IEEEproof}

\begin{lemma}[Progress]\label{l:preprogress}
if $\state$ is reachable, $\admnf{\state}=\state$ and $\decode\state\togenx\tm$ with $\mathtt{x}\in\set{\usym\msym,\usym\esym}$, then there exists $\statetwo$  such that $\state\tomachx\statetwo$, \ie, $\state$ is not final.
\end{lemma}
\begin{IEEEproof}
by \reflemma{uglam-deterministic} and \reflemma{uglam-distillation} it is sufficient to show that every reachable stuck state decodes to a normal form. The only stuck forms are:
\begin{itemize}
 \item \emph{Error states}. The state con only be $\usefmac{\genv}{\dump}{\var}{\stack}$ where $\var$ is not defined
  in $\genv$ or it is defined to be a $\herval\code$ where $\code$ is not
  a variable or a value.\\
   The state is not reachable because it would violate either the
   invariant in \reflemmap{uglaminvariants}{closure} or the invariant in
   \reflemmap{uglaminvariants}{fireball}.

 \item \emph{Final states}. Cases:
 \begin{enumerate}
  \item \emph{The result is/unfolds to a value}. The state is $\usefmac{\genv}{\dumpempty}{\code}{\stempty}$
   with $\code$ an abstraction or a variable bound in
   $\genv$ to a $\stackitem^\alive$.
   By \reflemma{contextualdecoding},
   $\decode{\usefmac{\genv}{\dumpempty}{\code}{\stempty}} =
    \decgenvpx{\genv}{\code} = \sctxp\code$ for some $\sctx$.
   Note that $\unf{\sctxp\code} = \relunf\code\sctx$ is a \fireball, indeed if $\code$ is an abstraction it is given by \reflemma{es-answers-are-normal} and if it as a variable it is given by \reflemmap{uglaminvariants}{fireball}.
   Thus by \reflemma{useful-projection}, $\sctxp\code$ is normal.

  \item \emph{The result is/unfolds to a inert}. The state is $\usefmac{\genv}{\dumpempty}{\code}{\stack}$
   with $\code$ a \symb $\const$ or a variable bound in $\genv$ to a $\stackitem^\dead$.\\
   By \reflemma{contextualdecoding},
   $\decode{\usefmac{\genv}{\dumpempty}{\code}{\stack}} = \decgenvpx{\genv}{\decodestack{\code}{\stack}} = 
    \sctxp{\decodestack{\code}{\stack}}$ for some $\sctx$.
   Note that $\unf{\sctxp\code} = \relunf\code\sctx$ is a \fireball, indeed if $\code$ is a \symb it is given by \reflemma{es-answers-are-normal} and if it as a variable it is given by \reflemmap{uglaminvariants}{fireball}.
   Moreover, by \reflemmap{uglaminvariants}{fireball},
   $\relunf{\decstack}\sctx$ has the form
   $\ctxhole\fire_1\ldots\fire_n$.
   Thus, by \reflemmap{relunf-properties}{zero} and the definition of fireballs,
   $\relunf{\decodestack{\code}{\stack}}{\sctx}$ is a fireball too.
   Therefore by \reflemma{useful-projection}, $\sctxp{\decodestack{\code}{\stack}}$ is normal.\qedhere
  \end{enumerate}
\end{itemize}
\end{IEEEproof}

\end{atend}

\begin{theorem}[\glamour\ Distillation]\label{th:uglam-refl-distillation}
	$(\mbox{\glamour},\togenu, \eqstruct, \decodefun)$ is a reflective explicit distillery. In particular, let $\state$ be a reachable state:
	\begin{enumerate}
	  \item \emph{Commutative}: if $\state\tomachcp{_{1,2,3,4,5}}\statetwo$ then $\decode\state = \decode\statetwo$;
	  \item \emph{Multiplicative}: if $\state\tomachum\statetwo$ then $\decode\state\togenum\eqstruct\decode\statetwo$;
	  \item \emph{Exponential}: if $\state\tomachue\statetwo$ then $\decode\state\togenue\decode\statetwo$.
	\end{enumerate}
\end{theorem}

\begin{proofatend}
the theorem follows from \reflemma{uglam-distillation}, \reflemma{uglam-deterministic} and \reflemma{preprogress}.
\end{proofatend}
 
 In fact, the distillery is even bilinear, as we now show. The proof employs
 the following definition of size of a state.

\begin{definition}
  \label{def:presize}
  The \emph{size} of codes and states is defined by:
$$\begin{array}{rclllrclllll}
      \size{\var} = \size{\const}  	& \defeq & 1&&
      \size{\code\codetwo}		 	& \defeq & \size{\code} + \size{\codetwo}+1\\
      \size{\l \var.\code} 		& \defeq & \size{\code} + 1&&

      \size{\glam\genv\dump\code\stack} & \defeq & \size\code + \Sigma_{\nfnst\codetwo\stack \in \dump} \size\codetwo
  \end{array}$$
\end{definition}

\begin{lemma}[Size Bounded]
  \label{l:useful-dump-size-bounded}
 Let $\state=\usefmac\genv\dump\codetwo\stack$ be a state reached by an execution $\exec$ of initial code $\code$. Then
 $\size{\state} \leq (1+\sizeue{\exec})\size{\code} - \sizecom{\exec}$.
\end{lemma}
\begin{IEEEproof}
by induction over the length of the derivation. The property trivially holds for the empty derivation. Case analysis over the
last machine transition. \emph{Commutative rule} $\tomachcone$: the rule splits the code $\code\codetwo$ between the dump and the code, and the measure---as well as the rhs of the formula---decreases by 1 because the rule consumes the application node. \emph{Commutative rules} $\tomachcp{_{2,3,4,5}}$: these rules consume the current code, so they decrease the measure of at least 1. \emph{Multiplicative}: it consumes the lambda abstraction.  \emph{Exponential}: it modifies the current code by replacing a variable (of size 1) with a value $\codeval$ coming from the environment. Because of \reflemmap{uglaminvariants}{value}, $\codeval$ is a sub-term of $\code$ and the dump size increment is bounded by $\size\code$.
\end{IEEEproof}

\begin{corollary}[Bilinearity of $\tomachc$]
\label{coro:preuseful-bound1}
 Let $\state$ be a state reached by an execution $\exec$ of initial code $\code$. Then
 $\sizecom{\exec} \leq (1+\sizee{\exec})\size{\code}$.
\end{corollary}

Finally, we obtain our first implementation theorem.
\begin{table*}
\caption{Identity, Chain, and Chain-Starting Context + Rewriting Rules of the \FFC}
\label{tab:useful-rewritingrules}
\centering
\begin{tabular}{r|lccc}
\begin{tabular}{c}
$\begin{array}{lllllllllllll}
	\ictx, \ictxtwo			& \grameq & \ctxhole \mid \ictxp\var\esub\var\ictxtwo \mid \ictx\esub\var\tm\\
	\pevctx,\pevctxtwo 	& \grameq & \tctxp\var\esub{\var}{\ictx} \mid \pevctxp\var\esub{\var}{\ictx} \mid \tctxp\pevctx\\\\\cline{1-3}\\
	\end{array}$
	\\
	$\begin{array}{lllllllllllll}

		\startctx{\tctxp\vartwo\esub{\vartwo}{\ictx}}{\var} & \defeq& \tctx\esub{\vartwo}{\ictxp\var}\\
		\startctx{\pevctxp\vartwo\esub{\vartwo}{\ictx}}{\var} & \defeq& \startctx{\pevctx}{\vartwo}\esub{\vartwo}{\ictxp\var}\\
		\startctx{\tctxp\pevctx}{\var} & \defeq& \tctxp{\startctx{\pevctx}{\var}}

	\end{array}$
	\end{tabular}
	&
$\begin{array}{c@{\hspace{0.6cm}}l}
  \textsc{Rule (Already Closed by Contexts)} & \textsc{Side Condition} \\
	\tctxp{\sctxp{\l\var.\tm}\tmtwo} \togenlum \tctxp{\sctxp{\tm\esub\var\tmtwo}} &
	\mbox{$\tctxp{\ctxhole\tmtwo}$ is useful}\\\\
    
			\tctxp{\tctxtwop\var\esub\var{\sctxp{\valp}}} \togenluee  \tctxp{\sctxp{\tctxtwop\valp\esub\var\valp}} &
			    \mbox{$\tctxp{\tctxtwo\esub\var{\sctxp{\valp}}}$ is 
			     useful}\\\\

		\tctxp{\pevctxp\var\esub\var{\sctxp\valp}} \togenluei  \tctxp{\sctxp{\pevctxp\valp\esub\var\valp}} &
 	\mbox{$\tctxp{\startctx\pevctx\var\esub\var{\sctxp\valp}}$ is
useful}
\end{array}$
\\

\end{tabular}
\end{table*}

\begin{theorem}[Useful Implementation]\label{th:preusefulimpl}\hfill
 \begin{enumerate}
  \item \emph{Low-Level Bilinear Implementation}: a $\togenu$-derivation $\deriv$ is implementable on RAM in $O((1+\size\deriv)\cdot \size\tm)$ (\ie\ bilinear) steps.
  \item \emph{Low + High Quadratic Implementation}: a $\tof$-derivation $\deriv$ is implementable on RAM in $O((1+\size\deriv^2)\cdot \size\tm)$ steps, \ie\ linear in the size of the initial term $\tm$ and quadratic in the length of the derivation $\size\deriv$.
 \end{enumerate}
 \end{theorem}

\begin{proofatend}
the proof follows from
\refthm{hlithm} applied to \refthm{quadratichlis},
and \refth{lowlevelimplth} applied to \refth{uglam-refl-distillation} and \refcoro{preuseful-bound1}.
Bi-linearity of the machine requires to show that the commutative steps are implementable in constant time, while the principal ones in time $O(\size\tm)$. The machine is meant to be implemented using a representation of codes using pointers, in particular for variables, so that the environment can be accessed in constant time. Assuming this, all rules except the exponential one evidently take constant time on a RAM machine, because they amount to moving pointers. The exponential rule requires $O(\size\tm)$ because it copies and $\alpha$-renames a value $\val$. Both these operations take time $O(\size\val)$. The value invariant (\reflemmap{uglaminvariants}{value}) guarantees $\size\val\leq\size\tm$. Additional considerations on the cost of similar rules can be found in \cite{DBLP:conf/icfp/AccattoliBM14} (page 9 and 11, paragraphs \emph{Abstract Considerations on Concrete Implementations}).
\end{proofatend}


\section{The \FFC}\label{sect:usefullazy}
\begin{atend}\section{Proofs Omitted From \refsect{usefullazy}\\ (Optimising Useful Reduction:\\ \FFC\ and the \fast\ \glamour)}\end{atend}
In this section we start by analysing why the \UsFC{} has a quadratic overhead. We then refine it, obtaining the \FFC, that we will prove to have only a linear overhead wrt the \FC. The optimisation has to do with the order in which chains of useful substitutions are performed.
\begin{atend}
We prove 
\reflemma{pseudo-trunks} first; then we address \refthm{lhli} ($(\tof,\togenlu)$ is a globally bounded high-level implementation system) and \refprop{lazy-useful-strong-bis} ($\tostruct$ is a Strong Bisimulation).

~
\end{atend}

\paragraph*{Analysis of Useful Substitution Chains}
in the \UsFC, whenever there is a situation like
$$(\var_1\gconst)\esub{\var_1}{\var_2} \ldots \esub{\var_{n-1}}{\var_{n}}\esub{\var_n}{\val}$$
the $\togenu$ strategy performs $n+1$ exponential steps $\togenue$ 
replacing $\var_1$ with $\var_2$, then $\var_2$ with $\var_3$, and so on, until $\val$ is finally substituted on the head
$$\begin{array}{rlll}
   (\var_n\gconst)\esub{\var_1}{\var_2} \ldots \esub{\var_{n-1}}{\var_{n}}\esub{\var_n}{\val} & \togenue\\ (\val\gconst)\esub{\var_1}{\var_2} \ldots \esub{\var_{n-1}}{\var_{n}}\esub{\var_n}{\val}
  \end{array}$$
and a multiplicative redex can be fired. Any later occurrence of $\var_1$ will
trigger the same chain of exponential steps again. Because the length $n$ of the
chain is bounded by the number of previous multiplicative steps (local bound property), the overall
complexity of the machine is quadratic in the number of multiplicative steps. 
In our previous work \cite{DBLP:conf/wollic/AccattoliC14}, we showed that to reduce the
complexity to linear it is enough to perform substitution steps in reverse order, modifying the chains while traversing them. The idea is that in the previous example one should rather have a smart reduction $\togenfe$ ($\osym$ stays for optimised, as $\usym$ is already used for useful reduction) following the chain of substitutions and performing:
$$\begin{array}{llll}
   (\var_1\gconst)\esub{\var_1}{\var_2} \ldots \esub{\var_{n-1}}{\var_{n}}\esub{\var_n}{\val} & \togenfe\\ 
   (\var_1\gconst)\esub{\var_1}{\var_2} \ldots \esub{\var_{n-1}}{\val}\esub{\var_n}{\val} & \togenfe\\
   \ldots\\
   (\var_1\gconst)\esub{\var_1}{\val} \ldots \esub{\var_{n-1}}{\val}\esub{\var_n}{\val} & \togenfe\\
   (\val\gconst)\esub{\var_1}{\val} \ldots \esub{\var_{n-1}}{\val}\esub{\var_n}{\val}\\
  \end{array}$$
Later occurrences of $\var_1$ will no longer trigger the chain, because it has been \emph{unchained} by traversing it the first time.

Unfortunately, introducing such an optimisation for useful reduction is hard. In the shown example, that has a very simple form, it is quite easy to define what \emph{following the chain} means. For the distillation machinery to work, however, we need our rewriting rules to be stable by structural equivalence, whose action is a rearrangement of substitutions through the term structure. Then the substitutions $\esub{\var_i}{\var_{i+1}}$ of the example can be spread all over the term, interleaved by applications and other substitutions, and even nested one into the other (like in $\esub{\var_i}{\var_{i+1}\esub{\var_{i+1}}{\var_{i+2}}}$). This makes the specification of optimised useful reduction a quite technical affair. 

\paragraph*{Chain Contexts} reconsider a term like in the example, $(\var\gconst)\esub{\var_1}{\var_2}\esub{\var_2}{\var_3}\esub{\var_3}{\var_4}\esub{\var_4}{\val}$. We want the next step to substitute on $\var_4$ so we should give a status to the context $\pevctx \defeq (\var\gconst)\esub{\var_1}{\var_2}\esub{\var_2}{\var_3}\esub{\var_3}{\ctxhole}$. The problem is that $\pevctx$ can be deformed by structural equivalence $\eqstruct$ as 
$$\pevctxtwo \defeq (\var\esub{\var_1}{\var_2\esub{\var_2}{\var_3}}\gconst)\esub{\var_3}{\ctxhole}$$
and so this context has to be caught too. We specify these context in \reftab{useful-rewritingrules} as \emph{chain contexts} $\pevctx$, defined using the auxiliary notion of \emph{identity context} $\ictx$, that captures a simpler form of chain (note that both notions are not shallow).

Given a chain context $\pevctx$, we will need to retrieve the point where the chain started, \ie\ the shallow context isolating the variable at the left end of the chain ($\var_1$ in the example). We are now going to define an operation associating to every chain context its \emph{chain-starting (shallow) context}. To see the two as contexts of a same term, we need also to provide the sub-term that we will put in $\pevctx$ (that will always be a variable). The chain-starting context $\startctx\pevctx\var$ associated to the chain context $\pevctx$ (with respect to $\var$) is defined in \reftab{useful-rewritingrules}.

For our example $\pevctx \defeq (\var\gconst)\esub{\var_1}{\var_2}\esub{\var_2}{\var_3}\esub{\var_3}{\ctxhole}$ we have  $\startctx{\pevctx}{\var_4} = (\ctxhole\gconst)\esub{\var_1}{\var_2}\esub{\var_2}{\var_3}\esub{\var_3}{\var_4}$, as expected.

\begin{atend}
%
%

For chain-starting contexts $\startctx\pevctx\var$, we need prove that their hole is indeed the left end of the chain, with the help of a preliminary lemma.

\begin{lemma}
	\label{l:inert-ctx-unf}
	Let $\ictxp\var$ s.t. $\ictx$ does not capture $\var$. Then $\unf{\ictxp\var} = \var$.
\end{lemma}

\begin{IEEEproof}
	by induction on $\ictx$. Cases:
	\begin{enumerate}
		\item \emph{Base $\ictx = \ctxhole$}. Then $\unf{\ictxp\var} = \unf\var = \var$.
		\item \emph{Inductive $\ictx = \ictxp\vartwo\esub\vartwo\ictxtwo$}. Then 
		$$\begin{array}{l}\unf{\ictxp\var} = \unf{\ictxp\vartwo\esub\vartwo{\ictxtwop\var}} = \unf{\ictxp\vartwo}\isub\vartwo{\unf{\ictxtwop\var}} =_{\ih}\\ \vartwo\isub\vartwo{\unf{\ictxtwop\var}} =_{\ih} \vartwo\isub\vartwo{\var} = \var\end{array}$$
		\item \emph{Closure $\ictx = \ictx\esub\vartwo\tm$}. Then 
		$$\unf{\ictxp\var} = \unf{\ictxp\var\esub\vartwo\tm} = \unf{\ictxp\var}\isub\vartwo{\unf\tm} =_{\ih} \var\isub\vartwo{\unf\tm} = \var$$
	\end{enumerate}
\end{IEEEproof}

\begin{lemma}
	\label{l:pseudo-trunks}
	Let $\pevctxp\var$ s.t. $\pevctx$ does not capture $\var$. Then there exists $\vartwo$ s.t. $\pevctxp\var = \tpctxp\pevctx\var\vartwo$ and $\relunf\vartwo{\tpctx\pevctx\var} = \var$.
\end{lemma}

\myproof{\begin{IEEEproof}
	by induction on $\pevctx$. Cases:
	\begin{enumerate}
		\item	\emph{Base, \ie\ $\pevctx = \tctxp\vartwo\esub{\vartwo}{\ictx}$}. Then 
		$$\pevctxp\var = \tctxp\vartwo\esub{\vartwo}{\ictxp\var} = \tpctxp{\tctxtwop\vartwo\esub{\vartwo}{\ictx}}{\var}\vartwo$$
		Now, $\relunf\vartwo{\tpctx\pevctx\var} = \relunf\vartwo{\tctx\esub{\vartwo}{\ictxp\var}} = \unf{\ictxp\var} =_{\reflemmaeq{inert-ctx-unf}} \var$
		
		\item	\emph{Inductive, \ie\ $\pevctx = \pevctxtwop\varthree\esub{\varthree}{\ictx}$}. Then 
		$$\begin{array}{ll}
                  \pevctxp\var &=\\
                  \pevctxtwop\varthree\esub{\varthree}{\ictxp\var} &=_{\ih}\\
                  \tpctxp{\pevctxtwo}\varthree\vartwo\esub{\varthree}{\ictxp\var} &=\\
                  \tpctxp{\pevctxtwop\varthree\esub{\varthree}{\ictx}}\var\vartwo\end{array}$$
		
		Now, 
		$$\begin{array}{ll}
                  \relunf\vartwo{\tpctx\pevctx\var} &=\\
                  \relunf\vartwo{\tpctx{\pevctxtwo}\varthree\esub{\varthree}{\ictxp\var}} &=\\
                  \relunf\vartwo{\tpctx{\pevctxtwo}\varthree}\isub{\varthree}{\unf{\ictxp\var}} &=_{\ih}\\
                  \varthree\isub{\varthree}{\unf{\ictxp\var}} = \unf{\ictxp\var} =_{\reflemmaeq{inert-ctx-unf}} \var\end{array}$$
		
		\item	\emph{Closure, \ie\ $\pevctx = \tctxtwop\pevctxtwo$}. Then 
		$$\pevctxp\var = \tctxtwop{\pevctxtwop\var} =_{\ih} \tctxtwop{\tpctxp{\pevctx}{\var}\vartwo} = \tpctxp{\tctxtwop\pevctx}{\var}\vartwo$$
		Now, $$\relunf\vartwo{\tpctx\pevctx\var} = \relunf\vartwo{\tctxtwop{\tpctx{\pevctx}{\var}}} =_{\reflemmaeqp{relunf-properties}{five}} \relunf{\relunf\vartwo{\tpctx{\pevctx}{\var}}}\tctxtwo =_{\ih} \relunf\var\tctxtwo = \var$$ because $\pevctx$, and thus $\tctxtwo$, does not capture $\var$.\qedhere
	\end{enumerate}
\end{IEEEproof}}
\end{atend}

\paragraph*{Rewriting Rules} the rules of the \FFC\ are in \reftab{useful-rewritingrules}. Note that the exponential rule splits in two, the ordinary \emph{shallow} case $\togenfes$ (now constrained to values) and the \emph{chain} case $\togenfec$ (where the new definition play a role). They could be merged, but for the complexity analysis and the relationship with the next machine is better to distinguish them. We use $\togenlue$ for $\togenluee\cup\togenluei$, and $\togenlu$ for $\togenlum\cup\togenlue$. Note the use of $\tpctx\pevctx\var$ in the third side condition. 

\subsection{Linearity: Multiplicative vs Exponential Analysis} 
To prove that $\togenlu$ implements $\tof$ with a global bound, and thus with a linear overhead, we need to show that the global number of exponential steps ($\togenfe$) in a $\togenlu$-derivation is bound by the number of multiplicative steps ($\togenfm$). We need the following invariant.

\begin{lemma}[Subterm Invariant]
	\label{l:useful-subterm}
	Let $\tm$ be a $\l$-term and $\deriv:\tm\togen^*\tmtwo$. Then every value in $\tmtwo$ is a value in $\tm$.
\end{lemma}

\begin{proofatend}
by induction over the length of the derivation. A simple inspection of the
rewriting rules shows that all values in the result of a reduction step are
copies of values in the term being reduced.
\end{proofatend}

A substitution $\tm\esub\var\tmtwo$ is \emph{\basic}\ if $\tmtwo$ has the form $\sctxp\vartwo$. The \emph{basic size} $\sizeb\tm$ of $\tm$ is the number of its  basic substitutions.

\begin{lemma}[Steps and Basic Size]
	\label{l:steps-basic-size} 
	\hfill
	\begin{enumerate}
		\item \label{p:steps-basic-size-one} If $\tm\togenluee\tmtwo$ then $\sizeb\tmtwo = \sizeb\tm$;
		\item \label{p:steps-basic-size-two} If $\tm\togenluei\tmtwo$ then $\sizeb\tm > 0$ and $\sizeb\tmtwo = \sizeb\tm - 1$;
		\item \label{p:steps-basic-size-three} If $\tm\togenlum\tmtwo$ then $\sizeb\tmtwo = \sizeb\tm$ or $\sizeb\tmtwo = \sizeb\tm +1$.
	\end{enumerate}
\end{lemma}

\begin{proofatend}
	using the sub-term property (\reflemma{useful-subterm}).
\end{proofatend}

\begin{lemma}
Let $\tm$ be initial and $\deriv:\tm\togenlu^*\tmtwo$. Then $\sizeb\tmtwo \leq \sizeum\deriv - \sizeuei\deriv$.
\end{lemma}

\begin{IEEEproof}
	by induction on $\size\deriv$. If $\size\deriv = 0$ the statement holds. If $\size\deriv>0$ consider the last step $\tmthree\togenlu\tmtwo$ of $\deriv$ and the prefix $\derivtwo:\tm\togenlu^*\tmthree$ of $\deriv$. By \ih, $\sizeb\tmthree \leq \sizeum\derivtwo - \sizeuei\derivtwo$. Cases of $\tmthree\togenlu\tmtwo$.\\ \emph{Shallow Exponential Step $\togenluee$}:
		$$\begin{array}{rllllll}
		\sizeb\tmtwo &\leq_{\reflemmaeqp{steps-basic-size}{one}}&
		 \sizeb\tmthree-1 &\\
		  &\leq_{\ih}&\sizeum\derivtwo - \sizeuei\derivtwo -1 &\\
		   &=&\sizeum\derivtwo - (\sizeuei\derivtwo +1) &=
		    &\sizeum\deriv - \sizeuei\deriv
		    \end{array}$$
 \emph{Chain Exponential Step $\togenluei$}:
		$$\sizeb\tmtwo =_{\reflemmaeqp{steps-basic-size}{two}} \sizeb\tmthree \leq_{\ih} \sizeum\derivtwo - \sizeuei\derivtwo =  \sizeum\deriv - \sizeuei\deriv$$
 \emph{Multiplicative Step $\togenlum$}:
		$$\begin{array}{lllllll}
		\sizeb\tmtwo & \leq_{\reflemmaeqp{steps-basic-size}{three}}&
		\sizeb\tmthree+1 \\
		& \leq_{\ih} & \sizeum \derivtwo - \sizeuei\derivtwo +1 \\
		& = & \derivtwo +1 - \sizeuei\derivtwo &= & \sizeum\deriv - \sizeuei\deriv\quad\qedhere\end{array}$$
\end{IEEEproof}

\begin{corollary}[Linear Bound on Chain Exponential Steps]
	\label{coro:linear-bound-internal-exp}
	Let $\tm$ be initial and $\deriv:\tm\togenlu^*\tmtwo$. Then $\sizeuei\deriv \leq \sizeum\deriv$.
\end{corollary}

Next, we bound shallow steps.

\begin{lemma}[Linear Bound on Shallow Exponential Steps]
	\label{l:linear-bound-external-exp}
	Let $\tm$ be initial and $\deriv:\tm\togenlu^*\tmtwo$. Then $\sizeuee\deriv \leq \sizeum\deriv$.
\end{lemma}

\begin{IEEEproof}
	first note that if $\tm\togenluee\tmtwo$ then $\tmtwo\togenlum\tmthree$, because by definition $\togenluee$ can fire only if it creates a $\togenlum$-redex. Such a fact and determinism of $\togenlu$ together imply $\sizeuee\deriv \leq \sizeum\deriv+1$, because every $\togenluee$ step is matched by the eventual $\togenlum$ steps that follows it immediately. However, note that in $\tm$ there are no explicit substitutions so that the first step is necessarily an unmatched $\togenlum$ step. Thus $\sizeuee\deriv \leq \sizeum\deriv$.
\end{IEEEproof}

\begin{theorem}[Linear Bound on Exponential Steps]
\label{tm:linear-exponentials}
	Let $\tm$ be initial and $\deriv:\tm\togenlu^*\tmtwo$. Then $\sizeoe\deriv \leq 2\cdot \sizeum\deriv$.
\end{theorem}

\begin{IEEEproof}
	by definition, $\sizeoe{\deriv} = \sizeuei{\deriv} + \sizeuee{\deriv}$. By \refcoro{linear-bound-internal-exp}, $\sizeuei{\deriv} \leq \sizeum{\deriv}$ and by \reflemma{linear-bound-external-exp} $\sizeuee{\deriv} \leq \sizeum{\deriv}$, and so $\sizeoe\deriv \leq 2\cdot \sizeum\deriv$.
\end{IEEEproof}

We presented the interesting bit of the proof of our improved high-level implementation theorem, which follows. The remaining details are in the appendix.

\begin{atend}
From now on we follow closely the same approach used for the \ShFC\ (\refapp{app-fireb-end-es} and \refapp{app-high-level-impl}) and the \UsFC\ (\refapp{app-useful}), without the need to define proper terms first: we start characterizing normal forms; then we prove projection and we obtain \refthm{lhli} ($(\tof,\togenlu)$ is a globally bounded high-level implementation system) as a corollary.

\begin{lemma}[Normal Form Characterization]
	\label{l:lazy-useful-nf-char}
	Let $\tm = \tctxp\tmtwo$ be a proper term s.t. $\tmtwo$ is $\togenlu$-normal and $\tctx$ is evaluable.
	\begin{enumerate}
		\item either  $\relunf\tmtwo\tctx$ is a \fireball, 
		\item \label{p:lazy-useful-nf-char-two} or $\relunf\tmtwo\tctx\tof$, more precisely exists $\tctxtwo$ s.t.
		\begin{enumerate}
			\item $\tmtwo = \tctxtwop\var$ with 
			\item $\relunf\var\tctxtwo = \vartwo$, 
			\item $\vartwo\in\fv\tmtwo$, 
			\item $\tctxp\tctxtwo$ evaluable,
			\item $\relunf\vartwo\tctx = \la\vartwo\tmthree$, and
			\item $\tctxtwo$ is applicative.
		\end{enumerate}
	\end{enumerate}
	Moreover, the context $\tctxtwo$ in \refpoint{lazy-useful-nf-char-two} is unique.
\end{lemma}

\myproof{
\begin{IEEEproof}
	 first of all, let us show that conditions \emph{2.a-f} imply $\relunf\tmtwo\tctx\togen$. Indeed,
	 $$\begin{array}{ll}
            \relunf\tmtwo\tctx &=_{a}\\
            \relunf{\tctxtwop\var}\tctx &=_{c \& \reflemmaeqp{relunf-properties}{four}}\\
            \relunf\tctxtwo\tctx \ctxholep{\relunf\var{\tctxp\tctxtwo}} &=_{d\&\reflemmaeqp{bis-evaluable-prop}{two}}\\
            \evctxp{\relunf\var{\tctxp\tctxtwo}} &=_{b}\\
            \evctxp{\relunf\vartwo{\tctx}} &=_{e}\\
            \evctxp{\la\vartwo\tmthree}\end{array}$$
	 and $\evctx \defeq \relunf{\tctxtwo}\tctx$ is applicative, by $d$ and \reflemmaeqp{relunf-properties}{three}.
	 
	 Now, we show that 1 and 2 are mutually exclusive. By \reflemma{gconst-or-fire-fnorm}, an unfolded term which is a \fireball\ is $\tof$-normal. Then if 1 hold then 2 does not, and if 2 holds 1 does not. Therefore, in the following proof we only prove that 1 or 2 holds.
	 
	 By induction on $\tmtwo$. Cases:
	 	\begin{enumerate}
		\item \emph{Variable $\var$}. Since $\tm$ is proper, $\relunf\tmtwo\tctx$ is a \fireball. 
		\item \emph{\Symb} and \emph{Abstraction}. Note that by properness, the abstraction is an ordinary $\lambda$-term, \ie\ it does not contain ES. Then in both cases we can apply \reflemma{es-answers-are-normal}, obtaining that $\relunf\tmtwo\tctx$ is a \fireball.
		\item \emph{Application} $\tmtwo = \tmthree\tmfour$. Since $\tmfour$ is normal and $\tctxp{\tmthree\ctxhole}$ is an evaluable context, we can apply the \ih\ to $\tmfour$, ending in one of the following two cases:
		\begin{enumerate}
			\item \emph{1 holds for $\tmfour$}. Then $\tctxp{\ctxhole\tmfour}$ is an evaluable context and we can apply the \ih\ to $\tmthree$ and obtain two cases:
			\begin{enumerate}
				\item \emph{1 holds for $\tmthree$}. Two cases:
				\begin{enumerate}
					\item $\relunf\tmthree{\tctxp{\ctxhole\tmfour}}= \gconst$. Then $\relunf\tmtwo\tctx$ is a inert, \ie\ a \fireball.
					\item $\relunf\tmthree{\tctxp{\ctxhole\tmfour}}= \la\vartwo\tmfive$. Note that $\tmthree$ cannot be itself an abstraction, otherwise $\tmtwo$ would not be normal. Then $\tmthree = \sctxp\vartwo$. Now, $\relunf\vartwo\sctx$ cannot be an abstraction, otherwise---again---$\tmtwo$ would not be normal. Then $\relunf\vartwo\sctx =\var$ for some $\var\in\fv\tmthree$ (possibly $\var=\vartwo$). Note that $\tctxtwo \defeq \sctx\tmfour$ is applicative and satisfies the other points of 2. For $c$, in particular, we only have to show that the content of every substitution in $\sctx$ unfolds to a \fireball\ in its context (by \reflemmap{evaluable-prop}{five}). Note that, since $\tm$ is proper, there is an evaluable context containing all the ES in $\tm$, \ie\ the content of \emph{every} substitution in $\tm$ unfolds in its context to a \fireball.
				\end{enumerate}
				
				\item \emph{2 holds for $\tmthree$}. Then 2 holds for $\tmtwo$ by taking $\tctxtwo \defeq \tctxthree \tmfour$ where $\tctxthree$ is the context given by the \ih, as all the conditions for $\tctxtwo$ follows from those for $\tctxthree$. Unicity follows from the \ih\ and the fact that no other such context can have its hole in $\tmfour$, because 2 does not hold for it.
			\end{enumerate}
			\item \emph{2 holds for $\tmfour$}. Then 2 holds for $\tmtwo$ by taking $\tctxtwo \defeq \tmthree\tctxthree$ where $\tctxthree$ is the context given by the \ih, as all the conditions for $\tctxtwo$ follows from those for $\tctxthree$. Unicity follows from the \ih\ and the fact that no other such context can have its hole in $\tmthree$, because 1 does not hold for $\tmfour$.
		\end{enumerate}
		\item \emph{Substitution}  $\tmtwo = \tmthree\esub\varthree\tmfour$. Then $\unf{\tctxp{\ctxhole\esub\varthree\tmfour}}$ is an evaluation context and we can apply the \ih\ to $\tmthree$. Two cases:
		
		\begin{enumerate}
			\item \emph{1 holds for $\tmthree$}. Note that since $\relunf\tmthree{\tctxp{\ctxhole\esub\varthree\tmfour}} =_{\reflemmaeqp{relunf-properties}{four}} \relunf{\tmthree\esub\varthree\tmfour}{\tctx} = \relunf\tmtwo\tctx$, then 1 holds for $\tmtwo$.
			\item \emph{2 holds for $\tmthree$}. Let $\vartwo\in\fv\tmthree$ be the variable and $\tctxtwo$ be the context given by the \ih. Then we have two cases:
			\begin{enumerate}
				\item \emph{$\vartwo = \varthree$}. Necessarily, $\tmfour$ has the form $\sctxp{\var'}$ with $\relunf{\var'}\sctx = \vartwo'$, otherwise $\tmtwo$ would not be $\togenu$-normal. Taking $\tctxthree \defeq \tctxtwo\esub\varthree\tmfour$ it is easily seen that 2 holds for $\tmtwo$ with respect to $\var$ and $\vartwo'$. Unicity follows from the \ih.
				\item \emph{$\vartwo \neq \varthree$}. Taking $\tctxthree \defeq \tctxtwo\esub\varthree\tmfour$ it is easily seen that 2 holds for $\tmtwo$ with respect to $\var$ and $\vartwo$. Unicity follows from the \ih.\qedhere
			\end{enumerate}
		\end{enumerate}
		\end{enumerate}
\end{IEEEproof}
}

\begin{corollary}[Normal Forms Unfold to Normal Forms]
	\label{coro:lazy-useful-nfs-unfold-to-nfs-useful}
	Let $\tm$ be a closed proper term. If $\tm$ is $\togenlu$-normal then $\unf\tm$ is $\tof$-normal.
\end{corollary}

\begin{IEEEproof}
note that taking $\tctx\defeq\ctxhole$ and $\tmtwo\defeq\tm$ and applying \reflemma{lazy-useful-nf-char} one obtains that $\unf\tm$ is a \fireball, because the second case cannot happen, given that $\tmtwo$ now is closed. By \reflemma{gconst-or-fire-fnorm}, $\unf\tm$ is $\tof$-normal.
\end{IEEEproof}

~

To prove the projection lemma we need to prove first as a technical lemma
another sufficient condition for a context to be evaluable. The condition
is based on the definition of position of a redex.

The \emph{position} of a $\togenlum$-redex is $\tctx$. The position of $\togenluei$ and $\togenluee$ redexes is the application that makes applicative the evaluable context in the side condition. Note that the position of a redex is always a context exposing an application constructor.

\begin{lemma}[Projection]
	\label{l:lazy-useful-projection}
	Let $\tm = \tctxp\tmtwo \togenlu \tctxtwop{\tmthree} = \tmfour$ by reducing a redex whose position lies in $\tmtwo$. If the redex is
	\begin{enumerate}
		\item \emph{Multiplicative}: then $\relunf\tmtwo\tctx \tof \relunf\tmthree\tctxtwo$ and $\unf\tm \tof \unf\tmfour$; 
		\item \emph{Shallow or Chain Exponential}: then $\relunf\tmtwo\tctx \tof$ and $\unf\tm = \unf\tmfour\tof$.
	\end{enumerate}
	In both cases $\relunf\tmtwo\tctx$ is not a \fireball.
\end{lemma}

\myproof{
\begin{IEEEproof}
	the fact that in both cases $\relunf\tmtwo\tctx$ is not a \fireball, follows from \reflemma{gconst-or-fire-fnorm} and the fact that $\relunf\tmtwo\tctx$ reduces. Cases:
	\begin{enumerate}
		\item \emph{Multiplicative}. Exactly as in the proof of \reflemma{useful-projection}.
		\item \emph{Exponential}. We take $\unf\tm = \unf\tmfour$ for granted, because a substitution step by definition does not change the unfolding.
		 Similarly to the previous case, $\unf\tm \tof $ follows from $\relunf\tmtwo\tctx \tof $. Indeed, by \reflemmap{bis-evaluable-prop}{two}, $\tctx$ is evaluable, and by \reflemmap{evaluable-prop}{one} $\unf\tctx$ is an evaluation context, so:
		$$\unf\tm = \unf{\tctxp\tmtwo} =_{\reflemmaeqp{relunf-properties}{four}} \unf\tctx\ctxholep{\relunf\tmtwo\tctx} \tof $$
		
		Now we prove $\unf\tmtwo \tof$. We have $\tmtwo = \tctxtwop{\sctxp\var\tmfour}$ and $\tm = \tctxp{\tctxtwop{\sctxp\var\tmfour}}$. Let us show that for both exponential redexes $\var$ unfolds to an abstraction. In $\tm$ there is somewhere (in $\sctx$, $\tctxtwo$, or $\tctx$) a substitution $\esub\var\tmfive$. Now, if $\tmfive = \sctxtwop\valp$ then we have a $\togenluee$-redex (because $\valp$ is an abstraction). If instead $\tmfive = \sctxtwop\vartwo$ then we have a $\togenluei$-redex and $\tm$ writes also as  $\tctxthreep{\pevctxp\vartwo\esub\vartwo{\sctxthreep\valp}}$ with $\pevctxp\vartwo = \tpctxp\pevctx\vartwo\var$, $\relunf\var{\tpctx\pevctx\vartwo}= \vartwo$ (by \reflemma{pseudo-trunks}), and s.t. the two contexts $\tctxp{\tctxtwop{\sctx\tmfour}}$ and $\tctxthreep{\tpctx\pevctx\vartwo\esub\vartwo{\sctxthreep\valp}}$ coincide.
		Then
		\begin{center}$\begin{array}{lll}
		\relunf\var{\tctxp{\tctxtwop{\sctx\tmfour}}} & =\\
                \relunf\var{\tctxthreep{\tpctx\pevctx\vartwo\esub\vartwo{\sctxthreep\valp}}}  & = &\mbox{(\reflemmap{relunf-properties}{five})}\\
		\relunf{\relunf\var{\tpctx\pevctx\vartwo\esub\vartwo{\sctxthreep\valp}}}\tctxthree  & =\\
		\relunf{\relunf\var{\tpctx\pevctx\vartwo}\isub\vartwo{\unf{\sctxthreep\valp}}}\tctxthree &  = &\mbox{($\relunf\var{\tpctx\pevctx\vartwo}= \vartwo$)}\\
		\relunf{\vartwo\isub\vartwo{\unf{\sctxthreep\valp}}}\tctxthree & = &\mbox{($\unf{\sctxthreep\valp}$ is an abst.)}\\
		\relunf{\vartwo\isub\vartwo{\valptwo}}\tctxthree & = \\
		\relunf{\valptwo}\tctxthree & = &\mbox{($\relunf{\valptwo}\tctxthree$ is an abst.)}\\
		\valpthree$$
		\end{array}$\end{center}

	Summing up, $\relunf\var{\tctxp{\tctxtwop{\sctx}}} = \la\vartwo\tmthree$. Moreover, by hypothesis $\tctxp\tctxtwo$ is evaluable, and so by \reflemmap{bis-evaluable-prop}{two} $\relunf\tctxtwo\tctx$ is an evaluation context. Finally, $\relunf\tmfour{\tctxp\tctxtwo}$ is a \fireball, because $\tctxp{\tctxtwop{\ctxhole\tmfour}}$ is evaluable. Then
		
		\[\begin{array}{lll}
			\relunf\tmtwo\tctx & =\\
                        \relunf{\tctxtwop{\sctxp\var\tmfour}}\tctx & = &\mbox{(\reflemmap{relunf-properties}{four})}\\
			\relunf{\tctxtwo}\tctx\ctxholep{\relunf{(\sctxp\var\tmfour)}{\tctxp\tctxtwo}} & = &\mbox{(\reflemmap{relunf-properties}{zero})}\\
			\relunf{\tctxtwo}\tctx\ctxholep{\relunf{\sctxp\var}{\tctxp\tctxtwo}\relunf{\tmfour}{\tctxp\tctxtwo}} & = &\mbox{(\reflemmap{relunf-properties}{four})}\\
			\relunf{\tctxtwo}\tctx\ctxholep{\relunf{\var}{\tctxp{\tctxtwop\sctx}}\relunf{\tmfour}{\tctxp\tctxtwo}} & = &\mbox{($\relunf\var{\tctxp{\tctxtwop{\sctx}}}= \la\vartwo\tmthree$)}\\
			\relunf{\tctxtwo}\tctx\ctxholep{(\la\vartwo\tmthree)\relunf{\tmfour}{\tctxp\tctxtwo}} & \tof &\mbox{($\relunf\tctxtwo\tctx$ is an ev. cont.}\\&&\mbox{\& $\relunf\tmfour{\tctxp\tctxtwo}$ a \fireball)}\\
			\relunf{\tctxtwo}\tctx\ctxholep{\tmthree\esub\vartwo{\relunf{\tmfour}{\tctxp\tctxtwo}}}\\
		\end{array}\]
	\end{enumerate}
\end{IEEEproof}
}

\begin{lemma}[Positional Determinism]
	\label{l:lazy-useful-determinism}
	Let $\tm$ be a term and $\tctxp\tctxONE$ and $\tctxp\tctxTWO$ positions of $\togenlu$-redexes. Then $\tctxONE = \tctxTWO$.
\end{lemma}

\myproof{
\begin{IEEEproof}
	by induction on $\tctxONE$. Cases:
	\begin{enumerate}
		\item \emph{Empty} $\tctxONE = \ctxhole$. Cases:
		\begin{enumerate}
			\item \emph{Multiplicative Redex}, \ie\ $\tmtwo = \sctxp{\la\var\tmfour}\tmfive$ with $\relunf\tmfive\tctx$ a \fireball. Now, $\tctxTWO$ cannot lie in $\sctxp{\la\var\tmfour}$, otherwise by \reflemma{useful-projection} $\relunf{\sctxp{\la\var\tmfour}}{\tctxp\tctxTWO}$ would not be a \fireball, while by \reflemma{es-answers-are-normal} it does. Nor $\tctxTWO$ can lie in $\tmfive$, otherwise again by \reflemma{useful-projection} $\relunf\tmfive\tctx$ would not be a \fireball. Then necessarily $\tctxTWO = \tctxONE = \ctxhole$.
			\item \emph{Exponential Redex}, \ie\ $\tmtwo = \tctxtwop{\sctxp\var \tmfour}$. Now, $\tctxTWO$ cannot lie in $\sctxp\var$, otherwise by \reflemma{useful-projection} $\relunf{\sctxp\var}{\tctxp\tctxTWO}$ would not be a \fireball, while by the hypothesis on the $\togene$-redex it does (it is an abstraction). Nor $\tctxTWO$ can lie in $\tmfour$, otherwise again by \reflemma{useful-projection} $\relunf\tmfour\tctx$ would not be a \fireball, while by the hypothesis on the $\togene$-redex it does. Then necessarily $\tctxTWO = \tctxONE = \ctxhole$.
		\end{enumerate}
			
		\item \emph{Right Application} $\tctxONE = \tmfour\tctxONEtwo$ and $\tm = \tmfour\tctxONEtwop\tmfive$. 	By \reflemma{useful-projection}, $\relunf{\tctxONEtwop\tmfive}{\tctxp{\tmfour\ctxhole}}$ has a $\tof$-redex and it is not a \fireball, so no redexes can lie to its left, in particular $\tctxTWO$ does not lie in $\tmfour$. By \reflemma{useful-projection}, $\relunf{\tctxONEtwop\tmfive}{\tctxp{\tmfour\ctxhole}}$ is not a \fireball, and so $\tctxTWO$ cannot be empty (\ie\ $\tmfour\tctxONEtwop\tmfive$ cannot be the position of a $\togenm$-redex). Then, $\tctxTWO = \tmtwo\tctxTWOtwo$, and the statement follows from the \ih\ applied to $\tctxONEtwo$ and $\tctxTWOtwo$.

		\item \emph{Left Application} $\tctxONE = \tctxONEtwo \tmfive$ and $\tm = \tctxONEtwop\tmfour \tmfive$. Note that $\tctxTWO$ cannot lie in $\tmfive$, otherwise by \reflemma{useful-projection} $\relunf\tmfive{\tctxp{\tctxONEtwop\tmfour\ctxhole}}$ has a $\tof$-redex and it is not a \fireball, and so no redexes---in particular the one of position $\tctxp\tctxONE$---can lie to its left, absurd. And $\tctxTWO$ cannot be empty (\ie\ the position of a $\togenm$-redex), because then $\tctxONEtwop\tmfour$ would have the form $\sctxp{\la\var\tmsix}$, which by \reflemma{useful-projection} cannot contain the position of a redex, because by \reflemma{es-answers-are-normal} $\relunf{\sctxp{\la\var\tmsix}}{\tctxp{\ctxhole\tmfive}}$ is a \fireball. Then, $\tctxTWO = \tctxTWOtwo\tmthree$, and the statement follows from the \ih\ applied to $\tctxONEtwo$ and $\tctxTWOtwo$. 
		
		\item \emph{Substitution} $\tctxONE = \tctxONEtwo\esub\var\tmthree$. Then necessarily $\tctxTWO = \tctxTWOtwo\esub\var\tmthree$ (remember the position of a $\togene$-redex is an application) and the statement follows from the \ih.\qedhere
	\end{enumerate}
\end{IEEEproof}
}
Note that we did not yet prove determinism, as two redexes may a priori have the same position.

\begin{lemma}[Redexes Have Different Positions]
	\label{l:lazy-disjoint-pos}
	Any two $\togenlu$-redexes in a term $\tm$ have different positions.
\end{lemma}

\begin{IEEEproof}
	It is obvious that different multiplicative redexes have different positions, and that multiplicative and exponential redexes cannot have the same position. Now consider an exponential position $\tctxp{\sctxp\var\tm}$ and let $\esub\var\tmtwo$ be the substitution on $\var$ lying somewhere in $\tctx$ or $\sctx$. If $\tm$ has the form $\sctxtwop\valp$ then there is a $\togenluee$ redex, and obviously there cannot be other $\togenluee$ or $\togenluei$ redex with the same position. If instead $\tm$ has the form $\sctxtwop\vartwo$ then we start following the chain of substitutions leading to the abstraction. Note that there is no choice about the chain, so there can only by one $\togenluei$-redex with that position.
\end{IEEEproof}

\begin{corollary}[Determinism]
	$\togenlu$ is deterministic.
\end{corollary}

\begin{IEEEproof}
	it follows from \reflemma{lazy-useful-determinism} and \reflemma{lazy-disjoint-pos}.
\end{IEEEproof}

\end{atend}

\begin{theorem}[Linear High-Level Implementation]\label{thm:lhli}
	$(\tof,\togenlu)$ is a globally bounded high-level implementation system, and so it has a linear overhead wrt $\tof$.
\end{theorem}

\begin{proofatend}
the pair $(\tof,\togenlu)$ is an high-level implementation system because of
\reflemma{lazy-useful-determinism},
\reflemma{lazy-useful-projection} and
\refcoro{lazy-useful-nfs-unfold-to-nfs-useful}.
It is also globally bounded because we already proved the global linear bound
on exponential steps (\reftm{linear-exponentials}).
\end{proofatend}

%

Last, the structural equivalence $\equiv$ is a strong bisimulation also for the \FFC.

\begin{proposition}[$\tostruct$ is a Strong Bisimulation]	
	\label{prop:lazy-useful-strong-bis}
	Let $\mathtt{x}\in\set{\osym\msym,\osym\msym\ssym, \osym\msym\csym}$. Then, $\tm\eqstruct\tmtwo$ and $\tm\togenx\tmp$ implies that there exists $\tmtwop$ such that $\tmtwo\togenx\tmtwop$ and $\tmp\eqstruct\tmtwop$.
\end{proposition}

\begin{proofatend}
omitted. All postponement proofs are similar and lengthy.
In \refssect{app-struct-eq} of the Appendix we proved the lemma for the \ShFC.
Other examples can be found in the long version of~\cite{DBLP:conf/icfp/AccattoliBM14}.
\end{proofatend}
\begin{table*}
\caption{Transitions of the \fast\ \glamour}
\label{tab:useful-transitions}
\centering
  ${\setlength{\arraycolsep}{1em}
  \begin{array}{c|c|c|c|ccc|c|c|c|c}
     \maca{\genv}{\dump}{\heapempty}{\code\codetwo}{\stack}
     &\tomachcone&
     \maca{\genv}{\dump\cons \nfnst\code\stack}{\heapempty}{\codetwo}{\stempty}\\
  	
     \maca{\genv}{\dump}{\heapempty}{\l\var.\code}{\stackitem^\lab\cons\stack}
     &\tomachom &
     \maca{\econs{\esub\var{\stackitem^\lab}}{\genv}}{\dump}{\heapempty}{\code}{\stack}\\

     \maca{\genv}{\dump \cons \nfnst\code\stack}{\heapempty}{\l\var.\codetwo}{\stempty}
     & \tomachctwo &
     \maca{\genv}{\dump}{\heapempty}{\code}{\herval{(\l\var.\codetwo)}\cons\stack}\\

     \maca{\genv}{\dump\cons \nfnst\code\stack}{\heapempty}{\const}{\stacktwo}
     & \tomachcthree &
     \maca{\genv}{\dump}{\heapempty}{\code}{\pair{\const}{\stacktwo}^\dead\cons\stack}\\

     \maca{\genv_1\esub\var{\phi^\dead}\genv_2}{\dump\cons\nfnst\code\stack}{\heapempty}{\var}{\stacktwo}
     & \tomachcfour &
     \maca{\genv_1\esub\var{\phi^\dead}\genv_2}{\dump}{\heapempty}{\code}{\pair{\var}{\stacktwo}^\dead\cons\stack}\\

     \maca{\genv_1\esub\var{\herval{\codetwo}}\genv_2}{\dump\cons\nfnst\code\stack}{\heapempty}{\var}{\stempty}
     & \tomachcfive &
     \maca{\genv_1\esub\var{\herval{\codetwo}}\genv_2}{\dump}{\heapempty}{\code}{\herval{\var}\cons\stack}\\

     \maca{\genv_1\esub\var{\herval\codeval}\genv_2}{\dump}{\heapempty}{\var}{\stackitem^\lab\cons\stack}
     & \tomachee &
     \maca{\genv_1\esub\var{\herval\codeval}\genv_2}{\dump}{\heapempty}{\rename{\codeval}}{\stackitem^\lab\cons\stack}\\

     \maca{\genv_1\esub\var{\herval{\vartwo}}\genv_2}{\dump}{\heap}{\var}{\stackitem^\lab\cons\stack}
     & \tomachcsix &
     \maca{\genv_1\esub\var{\herval\vartwo}\genv_2}{\dump}{\heap\cons\var}{\vartwo}{\stackitem^\lab\cons\stack}\\

     \maca{\genv^\bullet}{\dump}{\heap\cons\vartwo}{\var}{\stackitem^\lab\cons\stack}
     & \tomachei &
     \maca{\genv^\circ}{\dump}{\heap}{\vartwo}{\stackitem^\lab\cons\stack}

  	\end{array}}$

\begin{minipage}{14cm}
 ~\\with $\genv^\bullet \defeq \genv_1\esub\var{\herval\codeval}\genv_2\esub\vartwo{\herval\var}\genv_3$, $\genv^\circ \defeq \genv_1\esub\var{\herval\codeval}\genv_2\esub\vartwo{\herval{\mbox{$\rename{\codeval}$}}}\genv_3$, and 
where $\rename{\codeval}$ is any code $\alpha$-equivalent to $\codeval$ that preserves well-naming of the machine.
\end{minipage}
\end{table*}


\section{\fast\ \glamour}\label{sect:lazyglamour}
\begin{atend}\section{Proofs Omitted From \refsect{lazyglamour}\\ (\fast\ \glamour)}\end{atend}
\begin{atend}
The aim of this section is to prove \refth{lazy-uglam-refl-distillation}
($(\mbox{\fast\ \glamour},\togenlu, \eqstruct, \decodefun)$ is a reflective explicit distillery) and the final result of the paper, \refth{usefulimpl}
(the useful implementation has bilinear low level and bilinear high level
complexity).

We follow closely the methodology of \refapp{glamour}. The first step is proving that the invariants of the machine holds.

\end{atend}
The \fast\ \glamour{} machine, in \reftab{useful-transitions}, behaves like the \glamour{} machine 
until the code is a variable $\var_1$ that is hereditarily bound in the global
environment to a value via the chain
$\esub{\var_1}{\var_2}^\alive \ldots \esub{\var_n}{\val}^\alive$. At this point the machine needs to traverse the chain
until it finds the final binding $\esub{\var_n}{\val}^\alive$, and then traverse
again the chain in the opposite direction replacing every
$\esub{\var_i}{\var_{i+1}}^\alive$ entry with $\esub{\var_i}{\val}^\alive$.

The forward traversal of the chain is implemented by a new commutative rule $\tomachcsix$
that pushes the variables encountered in the chain on a new machine component,
called the \emph{chain heap}. The backward traversal is driven by the next variable
popped from the heap, and it is implemented by a new exponential rule (the
\emph{chain} exponential rule, corresponding to that of the calculus). Most of the analyses performed on the
\glamour{} machine carry over to the \fast\ \glamour{} without modifications.

Every old grammar is as before, and heaps are simply lists of variables, \ie\ they are defined by $\heap  \grameq  \heapempty \mid \heap\cons\var$.

\paragraph*{Decoding and Invariants} because of chain heaps and chain contexts, the decoding is involved. 

First of all, note that there is a correlation between the chain and the environment, as the variables of a chain heap $\heap = \var_1\cons\ldots\cons\var_n$ need to have corresponding entries $\esub{\var_i}{\var_{i+1}^\alive}$. More precisely, we will show that the following notion of compatibility is an invariant of the machine.

\begin{definition}[Compatibility Heap-Environment]
 Let $\genv$ be an environment and $\heap = \var_1 \cons \ldots \cons \var_n$ be a heap. We say that $\heap$ is \emph{compatible} with $\genv$ if 
 either $\heap$ is empty or $\esub{\var_i}{\var_{i+1}^\alive} \in \genv$ for $i < n$, $\esub{\var_n}{\var^\alive} \in \genv$, and $\esub{\var}{\stackitem^\alive} \in \genv$ for some $\stackitem^\alive$.
 \end{definition}


Given a state $\state = \mac\genv\dump\heap\code\stack$, the dump, the stack and the environment provide a shallow context $\tctx_\state  \defeq  \decgenvpx{\genv}{\decdumpp{\decstack}}$ that will be shown to be evaluable, as for the \glamour. 

If the chain heap $\heap$ is not empty, the current code $\code$ is somewhere in the middle of a chain inside the environment, and it is not apt to fill the state context $\tctx_\state$. The right code is the variable $\var_1$ starting the chain heap $\heap = \var_1\cons\ldots\cons\var_n$, \ie:

\begin{center}
$\begin{array}{rclllrcllllllll}
        \decodeheap{\heapempty}{\code} & \defeq & \code&&&
        \decodeheap{\var_1\cons\ldots\cons\var_n}{\code} & \defeq & \var_1
\end{array}$
\end{center}

Finally, a state decodes to a term as follows: $\decode{\state}  \defeq  \tctx_\state\ctxholep{\decodeheap{\heap}{\code}}$.

\begin{atend}
\begin{lemma}\label{l:heapprop1}
$ \decodeheap{\heap\cons\var}{\vartwo} = \decodeheap{\heap}{\var}$
\end{lemma}
\begin{IEEEproof}
by induction over $\heap$.
\end{IEEEproof}

\begin{lemma}[Contextual Decoding]\label{l:lazy-contextualdecoding}
$\decgenv$ is a substitution context; $\decdump$ and $\decstack$ are
shallow contexts without ES.
\end{lemma}
\begin{IEEEproof}
by induction over $\genv$, $\dump$ and $\stack$.
\end{IEEEproof}
\end{atend}

\begin{atend}
\begin{remark}
\label{rem:heap-comp}
if $\heap\cons\var$ is compatible with $\genv$, then also $\heap$ is compatible with $\genv$. 
\end{remark}
\end{atend}

\begin{lemma}[\fast\ \glamour\ Invariants]\label{l:lazy-uglaminvariants} 
	Let $\state = \mac{\genv}{\dump}{\heap}{\codetwo}{\stack}$ be a state reachable from an initial code $\code$. 
	\begin{enumerate}
		\item \emph{Closure}:\label{p:lazy-uglaminvariants-closure} $\state$ is closed and $\state$ is well named;
		\item \emph{Value}:\label{p:lazy-uglaminvariants-value} values in components of $\state$ are sub-terms of $\code$;
		\item \emph{\Fireball}:\label{p:lazy-uglaminvariants-fireball} $\relunf\code{\decgenv}$ is a fireball (of kind $\lab$) for every code $\code^\lab$ in $\stack$ and $\genv$;
		\item \emph{Evaluability}:\label{p:lazy-uglaminvariants-eval} $\decode\genv$, $\relunf{\decdump}{\decode\genv}$, $\relunf{\decstack}{\decode\genv}$, and $\tctx_\state$ are evaluable cont.;

		\item \emph{Environment Size}:\label{p:lazy-uglaminvariants-size} the length of the global environment $\genv$ is
bound by $\sizem\exec$.
                \item \emph{Compatible Heap}:\label{p:lazy-uglaminvariants-heap} if $\heap \neq \heapempty$ then the stack is not empty, $\codetwo = \var$, and $\heap$ is compatible with $\genv$.
	\end{enumerate}
\end{lemma}

\begin{proofatend}
by induction over the length of the execution.
The base case holds because $\code$ is initial.
The inductive step is by cases over the kind of transition.
All the verifications are trivial. \refpoint{lazy-uglaminvariants-eval} is proved as in the useful case (see \reflemma{uglaminvariants}, page \pageref{l:uglaminvariants}).
\end{proofatend}

We need additional decodings to retrieve the chain-starting context $\pevctx$ in the side-condition of $\togenfec$ rule, that---unsurprisingly---is given by the chain heap. Let $\state = \mac{\genv}{\dump}{\heap\cons\vartwo}{\code}{\stack}$ be a state s.t. $\heap\cons\vartwo$ is compatible with $\genv$. Note that compatibility gives $\genv = \genv_1\esub\vartwo{\herval\code}\genv_2$. Define the chain context $\psctx{\state}$ and the substitution context $\ssctx{\state}$ as:
\begin{center}
$\begin{array}{lclllllllllll}
        \psctx{\state} & \defeq & 
        \decgenvpx{\genv_1}{\decdumpp{\decstackp{\decheapp\vartwo}}}\esub\vartwo\ctxhole&&&

	\ssctx{\state} & \defeq & 
        \decode{\genv_2}        
\end{array}$
\end{center}

The first point of the following lemma guarantees that $\psctx{\state}$ and $\ssctx{\state}$ are well defined. The second point proves that filling $\ssctxp{\state}{\psctx{\state}}$ with the current term gives exactly the decoding of the state $\decode\state = \tctx_\state\ctxholep{\decheapp\vartwo}$, and that moreover the chain starts exactly on the evaluable context given by the state, \ie\ that $\tctx_\state = \ssctxp{\state}{\tpctx{\psctx{\state}}\var} $.
 
\begin{lemma}[Heaps and Contexts]\label{l:heap-contexts}
	Let $\state = \mac{\genv}{\dump}{\heap\cons\vartwo}{\var}{\stack}$ be a state s.t. $\heap\cons\vartwo$ is compatible with $\genv$. Then:
	\begin{enumerate}
		\item $\ssctx{\state}$ is a substitution context and $\psctx{\state}$ is a chain context		
		\item s. t. $\decode\state = \tctx_\state\ctxholep{\decheapp\vartwo} = \ssctxp{\state}{\psctxp{\state}\var}$ with $\tctx_\state = \ssctxp{\state}{\tpctx{\psctx{\state}}\var} $
	\end{enumerate}
\end{lemma}

\begin{proofatend}
	the first point is trivial, we prove the other two. By induction on the length $k$ of $\heap$. Cases:
	 \begin{itemize}
	  \item \empty{$\heap$ is empty}, \ie\ $\heap=\heapempty$. By \reflemmap{lazy-uglaminvariants}{heap} we have $\genv \defeq \genv_1\esub\vartwo{\herval\var}\genv_2$. Let also $\tctx \defeq \decgenvp{\decdumpp{\decstack}}$. We have $ \decode\state =  \decgenvp{\decdumpp{\decstackp{\decheappp\vartwo{\heapempty\cons\vartwo}}}} = \tctxp{\decheappp\vartwo{\heapempty\cons\vartwo}}$ and
	 \begin{enumerate}
	 \item $\ssctx{\state} = \decode{\genv_2}$ and $\psctx{\state} = \decgenvpx{\genv_1}{\decdumpp{\decstackp{\vartwo}}}\esub\vartwo\ctxhole$, that (by \reflemma{lazy-contextualdecoding}) has the form $\tctxp\vartwo\esub\vartwo\ictx$, and so it is a chain context,
	 
	 \item Now, 
	 \begin{center}$\begin{array}{ll}
	 \decode\state & =\\
	 \tctx_\state\ctxholep{\decheappp\vartwo{\heapempty\cons\vartwo}} & = \\
         \decgenvp{\decdumpp{\decstackp{\decheappp\vartwo{\heapempty\cons\vartwo}}}} & =\\ 
	 \decgenvp{\decdumpp{\decstackp{\vartwo}}} & =\\
	 \decgenvpx{\genv_1\esub\vartwo{\herval\var}\genv_2}{\decdumpp{\decstackp{\vartwo}}} & =\\
	 \decgenvpx{\genv_2}{\decgenvpx{\genv_1}{\decdumpp{\decstackp{\vartwo}}}\esub\vartwo\var} & =\\
	 \ssctxp{\state}{\decgenvpx{\genv_1}{\decdumpp{\decstackp{\vartwo}}}\esub\vartwo\var} & =\\
         \ssctxp{\state}{\psctxp{\state}\var} 
	 \end{array}$\end{center}
	 
	 and
	 
	 \begin{center}$\begin{array}{ll}
	 \ssctxp{\state}{\tpctx{\psctx{\state}}\var}  & =\\
         \ssctxp{\state}{\tpctx{\decgenvpx{\genv_1}{\decdumpp{\decstackp{\vartwo}}}\esub\vartwo\ctxhole}\var} & = \\
	 \ssctxp{\state}{\decgenvpx{\genv_1}{\decdumpp{\decstack}}\esub\vartwo\var} & =\\ 
	 \decgenvpx{\genv_2}{\decgenvpx{\genv_1}{\decdumpp{\decstack}}\esub\vartwo\var} & =\\ 
	 \decgenvpx{\genv_1\esub\vartwo{\herval\var}\genv_2}{\decdumpp{\decstack}}& =\\
	 \decgenvpx{\genv}{\decdumpp{\decstack}} &=\\
         \tctx_\state\\
	 \end{array}$\end{center}
	 \end{enumerate}
	 
	 \item \empty{Non-empty}, \ie\ $\heap = \heaptwo\cons\varthree$. By \reflemmap{lazy-uglaminvariants}{heap} we have $\genv = \genv_1\esub\varthree{\herval\vartwo}\genv_2\esub\vartwo{\herval\var}\genv_3$ and $\tctx \defeq \decgenvp{\decdumpp{\decstack}}$, so that $\decode\state = \decgenvp{\decdumpp{\decstackp{\decheappp\vartwo{\heaptwo\cons\varthree\cons\vartwo}}}} = \tctxp{\decheappp\vartwo{\heaptwo\cons\varthree\cons\vartwo}}$. Note that by \refrem{heap-comp} we can apply the \ih\ to the state $\statetwo = \mac{\genv}{\dump}{\heaptwo\cons\varthree}{\vartwo}{\stack}$, and we will do it in the following points.
	 
	 Now,
	 \begin{enumerate}
	 \item $\ssctx{\state} = \decode{\genv_3}$ and for $\psctx{\state}$, note that we have $$\psctx\statetwo = \decgenvpx{\genv_1}{\decdumpp{\decstackp{\decheappp\vartwo{\heaptwo\cons\varthree}}}}\esub\varthree\ctxhole$$ and that by \ih\ $\psctx\statetwo$ is a chain context. Then
	 
	 \begin{center}$\begin{array}{lll}
	 \psctx{\state} & =\\
         \decgenvpx{\genv_1\esub\varthree{\herval\vartwo}\genv_2}{\decdumpp{\decstackp{\decheappp\vartwo{\heaptwo\cons\varthree}}}}\esub\vartwo\ctxhole & =\\
	 \decgenvpx{\genv_2}{\decgenvpx{\genv_1}{\decdumpp{\decstackp{\decheappp\vartwo{\heaptwo\cons\varthree}}}}\esub\varthree\vartwo}\esub\vartwo\ctxhole & =\\
         \decgenvpx{\genv_2}{\psctxp\statetwo\vartwo}\esub\vartwo\ctxhole
	 \end{array}$\end{center}
	 
	 and so $\psctx{\state}$ is a chain context.

	 \item Note that $\ssctx\statetwo = \decode{\genv_2\esub\vartwo{\herval\var}\genv_3}$, and so
	 \begin{center}$\begin{array}{lll}
	  \ssctxp{\state}{\psctxp{\state}\var} & =\\
	    \decgenvpx{\genv_3}{\decgenvpx{\genv_2} {\psctxp\statetwo\vartwo}\esub\vartwo\var} & =\\
	   \decgenvpx{\genv_2\esub\vartwo{\herval\var}\genv_3}{\psctxp\statetwo\vartwo} & =\\
	   \ssctxp\statetwo{\psctxp\statetwo\vartwo} & =_{\ih}\\
          \decode\state
	 \end{array}$\end{center}
	 
	 Then note that 
	 \begin{center}$\begin{array}{lll}
	 \tpctx{\psctx{\state}}\var  & =\\
         \tpctx{\decgenvpx{\genv_2}{\psctxp\statetwo\vartwo}\esub\vartwo\ctxhole}{\var} & =\\
	 \tpctx{\decgenvpx{\genv_2}{\psctx\statetwo}}\vartwo\esub\vartwo\var & =\\
         \decgenvpx{\genv_2}{\tpctx{\psctx\statetwo}\vartwo}\esub\vartwo\var
	 \end{array}$\end{center}
	 
	 Now we conclude with
	 \begin{center}$\begin{array}{lll}
	 \ssctxp{\state}{\tpctx{\psctx{\state}}\var}  & =\\
	 \decgenvpx{\genv_3}{\tpctx{\psctx{\state}}\var}  & =\\
	 \decgenvpx{\genv_3}{\decgenvpx{\genv_2}{\tpctx{\psctx\statetwo}\vartwo}\esub\vartwo\var} & = \\
	 \decgenvpx{\genv_2\esub\vartwo{\herval\var}\genv_3}{\tpctx{\psctx\statetwo}\vartwo} & = \\
	 \ssctxp\statetwo{\tpctx{\psctx\statetwo}\vartwo} & =_{\ih}\\
         \tctx_{\statetwo} & = \\
         \tctx_{\state}\\
	 \end{array}$\end{center}
	 \end{enumerate}
	 \end{itemize}
\end{proofatend}

We can now sum up.

\begin{atend}
\begin{lemma}[\fast\ \glamour\ Distillation]\label{l:lazy-uglam-distillation}
	Let $\state$ be a reachable state. Then:
		\begin{enumerate}
		\item \emph{Commutative}: if $\state\tomachcp{_{1,2,3,4,5}}\statetwo$ then $\decode\state = \decode\statetwo$;
		\item \emph{Multiplicative}: if $\state\tomachum\statetwo$ then $\decode\state\togenum\eqstruct\decode\statetwo$;
		\item \emph{Shallow Exponential}: if $\state\tomachee\statetwo$ then $\decode\state\togenluee\decode\statetwo$;
		\item \emph{Chain Exponential}: if $\state\tomachei\statetwo$ then $\decode\state\togenluei\decode\statetwo$.
	\end{enumerate}
\end{lemma}

\begin{IEEEproof}
we list the transition in the order they appear in the definition of the machine.
\begin{itemize}
\item
Case
     $\mac{\genv}{\dump}{\heapempty}{\code\codetwo}{\stack}
     \tomachcone
     \mac{\genv}{\dump\cons\nfnst\code\stack}{\heapempty}{\codetwo}{\stempty}$:
 $$\begin{array}{ll}
     \decode{\mac{\genv}{\dump}{\heapempty}{\code\codetwo}{\stack}}
     &=\\
     \decgenvpx{\genv}{\decodep{\dump}{\decodestack{\decodeheap{\heapempty}{(\code\codetwo)}}{\stack}}}
     &=\\
     \decgenvpx{\genv}{\decodep{\dump}{\decodestack{\code\codetwo}{\stack}}}
     &=\\
     \decgenvpx{\genv}{\decodep{\dump}{\decodestack{\code\ctxholep{\codetwo}}{\stack}}}
     &=\\
     \decgenvpx{\genv}{\decodep{\dump\cons\nfnst\code\stack}{\codetwo}}
     &=\\
     \decgenvpx{\genv}{\decodep{\dump\cons\nfnst\code\stack}{\decodestack{\codetwo}{\stempty}}}
     &=\\
     \decgenvpx{\genv}{\decodep{\dump\cons\nfnst\code\stack}{\decodestack{\decodeheap{\heapempty}{\codetwo}}{\stempty}}}
     &=\\
     \decode{\mac{\genv}{\dump\cons\nfnst\code\stack}{\heapempty}{\codetwo}{\stempty}}
 \end{array}$$

\item
Case 
     $\mac{\genv}{\dump}{\heapempty}{\la\var\code}{\stackitem^\lab\cons\stack}
     \tomachom
     \mac{\genv\esub\var{\stackitem^\lab}}{\dump}{\heapempty}{\code}{\stack}$:
 $$\begin{array}{ll}
     \decode{\mac{\genv}{\dump}{\heapempty}{\la\var\code}{\stackitem^\lab\cons\stack}}
     &=\\
     \decgenvpx{\genv}{\decodep{\dump}{\decodestack{\decodeheap{\heapempty}{\la\var\code}}{\stackitem^\lab\cons\stack}}}
     &=\\
     \decgenvpx{\genv}{\decodep{\dump}{\decodestack{\la\var\code}{\stackitem^\lab\cons\stack}}}
     &=\\
     \decgenvpx{\genv}{\decodep{\dump}{\decodestack{(\la\var\code)\decode{\stackitem}}{\stack}}}
     &\togenum\\
     \decgenvpx{\genv}{\decodep{\dump}{\decodestack{\code\esub{\var}{\decode{\stackitem}}}{\stack}}}
     &\eqstruct\\
     \decgenvpx{\genv}{\decodep{\dump}{\decodestack{\code}{\stack}}\esub\var{\stackitem}}
     &=\\
     \decgenvpx{\genv\esub\var{\stackitem^\lab}}{\decodep{\dump}{\decodestack{\code}{\stack}}}
      &=\\
     \decgenvpx{\genv\esub\var{\stackitem^\lab}}{\decodep{\dump}{\decodestack{\decodeheap{\heapempty}{\code}}{\stack}}}
     &=\\
     \decode{\mac{\genv\esub\var{\stackitem^\lab}}{\dump}{\heapempty}{\code}{\stack}}
 \end{array}$$
The multiplicative step is justified by \reflemmap{lazy-uglaminvariants}{eval} and \reflemmap{lazy-uglaminvariants}{fireball}. The bisimulation step is justified by \reflemma{ev-comm-struct}.

\item
Case
     $$\mac{\genv}{\dump\cons\nfnst\code\stack}{\heapempty}{\la\var\codetwo}{\stempty}
     \tomachctwo
     \mac{\genv}{\dump}{\heapempty}{\code}{(\la\var\codetwo)^\alive\cons\stack}$$
We have
 $$\begin{array}{ll}
     \decode{\mac{\genv}{\dump\cons\nfnst\code\stack}{\heapempty}{\la\var\codetwo}{\stempty}}
     &=\\
     \decgenvpx{\genv}{\decodep{\dump\cons\nfnst\code\stack}{\decodestack{\decodeheap{\heapempty}{\la\var\codetwo}}{\stempty}}}
     &=\\
     \decgenvpx{\genv}{\decodep{\dump\cons\nfnst\code\stack}{\decodestack{\la\var\codetwo}{\stempty}}}
     &=\\
     \decgenvpx{\genv}{\decodep{\dump\cons\nfnst\code\stack}{\la\var\codetwo}}
     &=\\
     \decgenvpx{\genv}{\decodep{\dump}{\decodestack{\code(\la\var\codetwo)}{\stack}}}
     &=\\
     \decgenvpx{\genv}{\decodep{\dump}{\decodestack{\code}{(\la\var\codetwo)^\alive\cons\stack}}}
     &=\\
     \decgenvpx{\genv}{\decodep{\dump}{\decodestack{\decodeheap{\heapempty}{\code}}{(\la\var\codetwo)^\alive\cons\stack}}}
     &=\\
     \decode{\mac{\genv}{\dump}{\heapempty}{\code}{(\la\var\codetwo)^\alive\cons\stack}}
 \end{array}$$

\item
Case
     $$\mac{\genv}{\dump\cons\nfnst\code\stack}{\heapempty}{\const}{\stacktwo}
     \tomachcthree
     \mac{\genv}{\dump}{\heapempty}{\code}{\pair{\const}{\stacktwo}^\dead\cons\stack}$$
We have
 $$\begin{array}{ll}
     \decode{\mac{\genv}{\dump\cons\nfnst\code\stack}{\heapempty}{\const}{\stacktwo}}
     &=\\
     \decgenvpx{\genv}{\decodep{\dump\cons\nfnst\code\stack}{\decodestack{\decodeheap{\heapempty}{\decode{\const}}}{\stacktwo}}}
      &=\\
     \decgenvpx{\genv}{\decodep{\dump\cons\nfnst\code\stack}{\decodestack{\decode{\const}}{\stacktwo}}}
     &=\\
     \decgenvpx{\genv}{\decodep{\dump}{\decodestack{\code\decodestack{\const}{\stacktwo}}{\stack}}}
     &=\\
     \decgenvpx{\genv}{\decodep{\dump}{\decodestack{\code}{\pair{\const}{\stacktwo}^\dead\cons\stack}}}
     &=\\
     \decgenvpx{\genv}{\decodep{\dump}{\decodestack{\decodeheap{\heapempty}{\code}}{\pair{\const}{\stacktwo}^\dead\cons\stack}}}
     &=\\
     \decode{\mac{\genv}{\dump}{\heapempty}{\code}{\pair{\const}{\stacktwo}^\dead\cons\stack}}
 \end{array}$$

\item
Case
     $$\begin{array}{l}\mac{\genv_1\esub\var{\stackitem^\dead}\genv_2}{\dump\cons\nfnst\code\stack}{\heapempty}{\var}{\stacktwo}
     \tomachcfour\\
     \mac{\genv_1\esub\var{\phi^\dead}\genv_2}{\dump}{\heapempty}{\code}{\pair{\var}{\stacktwo}^\dead\cons\stack}\end{array}$$
We have
 $$\begin{array}{ll}
     \decode{\mac{\genv_1\esub\var{\stackitem^\dead}\genv_2}{\dump\cons\nfnst\code\stack}{\heapempty}{\var}{\stacktwo}}
     &=\\
     \ldots&\\
     \decode{\mac{\genv_1\esub\var{\phi^\dead}\genv_2}{\dump}{\heapempty}{\code}{\pair{\var}{\stacktwo}^\dead\cons\stack}}
 \end{array}$$
The proof is the one for the previous case $\tomachcthree$,
by replacing $\const$ with $\var$ and instantiating $\genv$ with
${\genv_1\esub\var{\phi^\dead}\genv_2}$.

\item
Case
     $$\begin{array}{l}\mac{\genv_1\esub\var{\codetwo^\alive}\genv_2}{\dump\cons\nfnst\code\stack}{\heapempty}{\var}{\stempty}
     \tomachcfive\\
     \mac{\genv_1\esub\var{\herval{\codetwo}}\genv_2}{\dump}{\heapempty}{\code}{\var^\alive\cons\stack}\end{array}$$
We have
 $$\begin{array}{ll}
     \decode{\mac{\genv_1\esub\var{\codetwo^\alive}\genv_2}{\dump\cons\nfnst\code\stack}{\heapempty}{\var}{\stempty}}
     &=\\
     \ldots&=\\
     \decode{\mac{\genv_1\esub\var{\herval{\codetwo}}\genv_2}{\dump}{\heapempty}{\code}{\var^\alive\cons\stack}}
 \end{array}$$
The proof is the one for the previous case $\tomachcfour$,
by replacing $(\la\var\codetwo)$ with $\var$ and instantiating
$\genv$ with ${\genv_1\esub\var{\herval{\codetwo}}\genv_2}$.

\item
Case
     $$\begin{array}{l}\mac{\genv_1\esub\var{\herval\codeval}\genv_2}{\dump}{\heapempty}{\var}{\stackitem^\lab\cons\stack}
     \tomachee\\
     \mac{\genv_1\esub\var{\herval\codeval}\genv_2}{\dump}{\heapempty}{\rename\codeval}{\stackitem^\lab\cons\stack}\end{array}$$
We have
 $$\begin{array}{lll}
     \decode{\mac{\genv_1\esub\var{\herval\codeval}\genv_2}{\dump}{\heapempty}{\var}{\stackitem^\lab\cons\stack}}
     &=\\
     \decgenvpx{\genv_1\esub\var{\herval\codeval}\genv_2}{\decodep{\dump}{\decodestack{\decodeheap{\heapempty}{\var}}{\stackitem^\val\cons\stack}}}
       &=\\
     \decgenvpx{\genv_1\esub\var{\herval\codeval}\genv_2}{\decodep{\dump}{\decodestack{\var}{\stackitem^\val\cons\stack}}}
     &\togenluee & \mbox{(by \reflemmaeqp{lazy-uglaminvariants}{eval})}\\
     \decgenvpx{\genv_1\esub\var{\herval\codeval}\genv_2}{\decodep{\dump}{\decodestack{\rename\codeval}{\stackitem^\lab\cons\stack}}}
     &=\\
     \decgenvpx{\genv_1\esub\var{\herval\codeval}\genv_2}{\decodep{\dump}{\decodestack{\decodeheap{\heapempty}{\rename\codeval}}{\stackitem^\lab\cons\stack}}}
     &=\\
     \decode{\mac{\genv_1\esub\var{\herval\codeval}\genv_2}{\dump}{\heapempty}{\rename\codeval}{\stackitem^\lab\cons\stack}}
 \end{array}$$

\item
Case
     $$\begin{array}{l}\mac{\genv_1\esub\var{\herval{\vartwo}}\genv_2}{\dump}{\heap}{\var}{\stackitem^\lab\cons\stack}
     \tomachcsix\\
     \mac{\genv_1\esub\var{\herval\vartwo}\genv_2}{\dump}{\heap\cons\var}{\vartwo}{\stackitem^\lab\cons\stack}\end{array}$$
 $$\begin{array}{ll}
   \decode{\mac{\genv_1\esub\var{\herval{\vartwo}}\genv_2}{\dump}{\heap}{\var}{\stackitem^\lab\cons\stack}}
   &=\\
   \decgenvpx{\genv_1\esub\var{\herval{\vartwo}}\genv_2}{\decodep{\dump}{\decodestack{\decodeheap{\heap}{\var}}{\stackitem^\lab\cons\stack}}}
   &=_\reflemmaeq{heapprop1}\\
   \decgenvpx{\genv_1\esub\var{\herval\vartwo}\genv_2}{\decodep{\dump}{\decodestack{\decodeheap{\heap\cons\var}{\vartwo}}{\stackitem^\lab\cons\stack}}}
   &=\\
   \decode{\mac{\genv_1\esub\var{\herval\vartwo}\genv_2}{\dump}{\heap\cons\var}{\vartwo}{\stackitem^\lab\cons\stack}}
 \end{array}$$

\item
Case
     $\statetwo \defeq \mac{\genv^\bullet}{\dump}{\heap\cons\vartwo}{\var}{\stackitem^\lab\cons\stack}
     \tomachei
     \mac{\genv^\circ}{\dump}{\heap}{\vartwo}{\stackitem^\lab\cons\stack} = \state$, where $\genv^\bullet = \genv_1\esub\vartwo{\herval\var}\genv_2\esub\var{\herval\codevalp}\genv_3$, and 
     $\genv^\circ = \genv_1\esub\vartwo{\herval{\mbox{$\rename\codevalp$}}}\genv_2\esub\var{\herval\codeval}\genv_3$. Note that we have:
     \begin{enumerate}
      \item $\ssctx{\statetwo,\vartwo} = \decode{\genv_2\esub\var{\herval\codevalp}\genv_3}$
      \item $\psctx{\statetwo,\vartwo} = \decgenvpx{\genv_1}{\decodep{\dump}{\decodestack{\decodeheap{\heap}{\vartwo}}{\stackitem^\val\cons\stack}}}\esub\vartwo\ctxhole $
     \end{enumerate}
Then,
 $$\begin{array}{ll}
     \decode{\mac{\genv^\bullet}{\dump}{\heap\cons\vartwo}{\var}{\stackitem^\lab\cons\stack}}
     &=\\
     \decgenvpx{\genv^\bullet}{\decodep{\dump}{\decodestack{\decodeheap{\heap\cons\vartwo}{\var}}{\stackitem^\val\cons\stack}}}
     &=\\
     \decgenvpx{\genv^\bullet}{\decodep{\dump}{\decodestack{\decodeheap{\heap}{\vartwo}}{\stackitem^\val\cons\stack}}}
     &=_\reflemmaeq{heap-contexts} \\
     \ssctxp{\statetwo,\vartwo}{\psctxp{\statetwo,\vartwo}\var}
     &=\\
     \decgenvpx{\genv_2\esub\var{\herval\codevalp}\genv_3}{\psctxp{\statetwo,\vartwo}\var}
     &=\\
     \decgenvpx{\genv_3}{\decgenvpx{\genv_2}{{\psctxp{\statetwo,\vartwo}\var}}\esub\var{\codevalp}}
     & \togenluei \\
     \decgenvpx{\genv_3}{\decgenvpx{\genv_2}{{\psctxp{\statetwo,\vartwo}{\rename\codevalp}}}\esub\var{\codevalp}}
     & = \\
     \decgenvpx{\genv_3}{\decgenvpx{\genv_2}{{\decgenvpx{\genv_1}{\decodep{\dump}{\decodestack{\decodeheap{\heap}{\vartwo}}{\stackitem^\val\cons\stack}}}\esub\vartwo{\rename\codevalp}}}\esub\var{\codevalp}}
     & = \\
     \decgenvpx{\genv^\circ}{\decodep{\dump}{\decodestack{\decodeheap{\heap}{\vartwo}}{\stackitem^\val\cons\stack}}}
     & =\\
     \decode{\mac{\genv^\circ}{\dump}{\heap}{\vartwo}{\stackitem^\lab\cons\stack}}     
 \end{array}$$
 The chain exponential step is justified because 
 \begin{enumerate}
  \item $\heap\cons\vartwo$ is compatible with $\genv^\bullet$, and so we can apply \reflemmaeq{heap-contexts}, obtaining $\ssctxp{\statetwo,\vartwo}{\tpctx{\psctx{\statetwo,\vartwo}}\var} = \decgenvpx{\genv^\bullet}{\decdumpp{\decode{\stackitem^\lab\cons\stack}}}$ 
  \item \reflemmap{lazy-uglaminvariants}{eval} guarantees that such a context---which is the context in the side-condition of the rule--- is evaluable. It is also obviously applicative (because the stack has the form $\stackitem^\lab\cons\stack$).\qedhere
 \end{enumerate}

\end{itemize}
\end{IEEEproof}

\begin{lemma}[Determinism]\label{l:lazy-uglam-deterministic} The transition relation $\tomach$ of the \fast\ \glamour\ is deterministic.
\end{lemma}
\begin{IEEEproof}
a simple inspection of the reduction rules shows no critical pairs.
\end{IEEEproof}

\begin{lemma}[Progress]\label{l:lazyprogress} if $\state$ is reachable, $\admnf{\state}=\state$ and $\decode\state\togenx\tm$ with $\mathtt{x}\in\set{\osym\msym,\osym\esym\ssym,\osym\esym\csym}$, then there exists $\statetwo$  such that $\state\tomachx\statetwo$, \ie, $\state$ is not final.
\end{lemma}
\begin{IEEEproof}
by \reflemma{lazy-uglam-deterministic} and \reflemma{lazy-uglam-distillation} it is sufficient to show that every reachable stuck state decodes to a normal form. The only stuck forms are:
\begin{itemize}
 \item \emph{Error states}. 
 \begin{enumerate}
  \item \emph{Problem with the heap}. $\mac{\genv}{\dump}{\heap\cons\vartwo}{\code}{\stack}$
   when $\code$ is not a variable bound in $\genv$ to a~$\stackitem^\alive$
   or $\stack$ is empty or $\vartwo$ is not bound to $\code$ in $\genv$.
   The state is not reachable because it would violate the
   invariant \reflemmap{lazy-uglaminvariants}{heap}.
   \item \emph{Problem with the environment}. The state is $\mac{\genv}{\dump}{\heap}{\var}{\stack}$ where $\var$ is not defined
  in $\genv$ or it is defined to be a $\herval\code$ where $\code$ is not
  a variable or a value.\\
   The state is not reachable because it would violate either the
   invariant in \reflemmap{lazy-uglaminvariants}{closure} or the invariant in
   \reflemmap{lazy-uglaminvariants}{fireball}.
 \end{enumerate}

 \item \emph{Final states}. Cases:
 \begin{enumerate}
  \item \emph{The result is/unfolds to a value}. The state is $\mac{\genv}{\dumpempty}{\heapempty}{\code}{\stempty}$
   with $\code$ an abstraction or a variable bound in
   $\genv$ to a $\stackitem^\alive$.
   By \reflemma{contextualdecoding},
   $\decode{\mac{\genv}{\dumpempty}{\heapempty}{\code}{\stempty}} =
    \decgenvpx{\genv}{\code} = \sctxp\code$ for some $\sctx$.
   Note that $\unf{\sctxp\code} = \relunf\code\sctx$ is a \fireball, indeed if $\code$ is an abstraction it is given by \reflemma{es-answers-are-normal} and if it as a variable it is given by \reflemmap{lazy-uglaminvariants}{fireball}.
   Thus by \reflemma{lazy-useful-projection}, $\sctxp\code$ is normal.

  \item \emph{The result is/unfolds to a inert}. The state is $\mac{\genv}{\dumpempty}{\heapempty}{\code}{\stack}$
   with $\code$ a \symb $\const$ or a variable bound in $\genv$ to a $\stackitem^\dead$.\\
   By \reflemma{contextualdecoding},
   $\decode{\mac{\genv}{\dumpempty}{\heapempty}{\code}{\stack}} = \decgenvpx{\genv}{\decodestack{\code}{\stack}} = 
    \sctxp{\decodestack{\code}{\stack}}$ for some $\sctx$.
   Note that $\unf{\sctxp\code} = \relunf\code\sctx$ is a \fireball, indeed if $\code$ is a \symb it is given by \reflemma{es-answers-are-normal} and if it as a variable it is given by \reflemmap{lazy-uglaminvariants}{fireball}.
   Thus by \reflemma{lazy-useful-projection}, $\sctxp\code$ is normal.\qedhere
  \end{enumerate}
\end{itemize}
\end{IEEEproof}

\end{atend}

\begin{theorem}[\fast\ \glamour\ Distillation]\label{th:lazy-uglam-refl-distillation}
	$(\mbox{\fast\ \glamour},\togenlu, \eqstruct, \decodefun)$ is a reflective explicit distillery. In particular, let $\state$ be a reachable state:
	\begin{enumerate}
		\item \emph{Commutative}: if $\state\tomachcp{_{1,2,3,4,5,6}}\statetwo$ then $\decode\state = \decode\statetwo$;
		\item \emph{Multiplicative}: if $\state\tomachom\statetwo$ then $\decode\state\togenlum\eqstruct\decode\statetwo$;
		\item \emph{Shallow Exponential}: if $\state\tomachee\statetwo$ then $\decode\state\togenluee\decode\statetwo$;
		\item \emph{Chain Exponential}: if $\state\tomachei\statetwo$ then $\decode\state\togenluei\decode\statetwo$.
	\end{enumerate}
\end{theorem}
\begin{proofatend}
the theorem follows from \reflemma{lazy-uglam-distillation}, \reflemma{lazy-uglam-deterministic} and \reflemma{lazyprogress}.
\end{proofatend}

\subsection{Bilinearity: Principal vs Commutative Analysis}
\label{ssect:bilinearity-analysis}
\begin{atend}\subsection{Proofs Omitted From \refssect{bilinearity-analysis}\\ (Bilinearity: Principal vs Commutative Analysis)}\end{atend}

Bilinearity wrt $\tomachcp{_{1,2,3,4,5}}$ is identical to that of the \glamour, thus we omit it and focus on $\tomachcsix$.

\begin{atend}
~

In the remaining of the appendix we prove bilinearity of $\tomachcp{}$.
We begin redoing the proof for $\tomachcp{_{1,2,3,4,5}}$, that is almost
identical to that of the \glamour.

\begin{lemma}[Size Bounded]
 Let $\state=\usefmac\genv\dump\codetwo\stack$ be a state reached by an execution $\exec$ of initial code $\code$. Then
 $\size{\state} \leq (1+\sizeuee{\exec})\size{\code} - \sizecomphone{\exec}$.
\end{lemma}
\begin{IEEEproof}
the same reasoning as for the useful case (\reflemma{useful-dump-size-bounded}) provides the proof for $\tomachm, \tomachee, \tomachcp{_{1,2,3,4,5}}$, while for the new transitions $\tomachcsix$ and $\tomachei$ it is enough to note that they do not change the size of the state.
\end{IEEEproof}

\begin{corollary}[Termination and Bilinearity of $\tomachcp{_{1,2,3,4,5}}$]
\label{coro:lazy-useful-bound1}
 Let $\state$ be a state reached by an execution $\exec$ of initial code $\code$. Then
 $\sizecomphone{\exec} \leq (1+\sizee{\exec})\size{\code} = O(\sizep{\exec}\cdot\size{\code})$. In particular, $\tomachcp{_{1,2,3,4,5}}$ terminates.
\end{corollary}

\end{atend}
The size $\size\heap$ of a chain heap is its length as a list.

\begin{lemma}[Linearity of $\tomachcsix$]
\label{l:heap-bound} %
  Let $\state=\mac\genv\dump\heap\code\stack$ be a state reached by an execution $\exec$. Then 
  \begin{enumerate}
   \item \label{p:heap-bound-comm} $\sizecomsix\exec = \size\heap + \sizeuei\exec$.
   \item \label{p:heap-bound-heap} $\size\heap \leq \sizem\exec$.
   \item \label{p:heap-bound-final} $\sizecomsix\exec \leq \sizem\exec + \sizeuei\exec = O(\sizep{\exec})$.
  \end{enumerate}

\end{lemma}
\begin{IEEEproof}
\emph{1)} By induction over $\size\exec$ and analysis of the
last machine transition.
The $\tomachcsix$ steps increment the size of the heap. The $\tomachei$ steps
decrement it. All other steps do not change the heap. \emph{2)} By the compatible heap invariant (\reflemmap{lazy-uglaminvariants}{heap}), $\size\heap \leq \size\genv$. By the environment size invariant (\reflemmap{lazy-uglaminvariants}{size}), $\size\genv \leq \sizem\exec$. Then $\size\heap \leq \sizem\exec$. \emph{3)}  Plugging \refpoint{heap-bound-heap} into \refpoint{heap-bound-comm}.
\end{IEEEproof}

\begin{corollary}[Bilinearity of $\tomachc$]
\label{coro:useful-bound1}
 Let $\state$ be a state reached by an execution $\exec$ of initial code $\code$. Then
 $\sizecom{\exec} \leq (1+\sizee{\exec})\size{\code} + \sizem\exec + \sizeuei\exec = O((1+\sizep{\exec})\cdot\size{\code})$.
\end{corollary}
\begin{proofatend}
combining \refcoro{lazy-useful-bound1} with \reflemma{heap-bound}.
\end{proofatend}

Finally, we obtain the main result of the paper.

\begin{theorem}[Useful Implementation]\label{th:usefulimpl}\hfill
 \begin{enumerate}
  \item \emph{Low-Level Bilinear Implementation}: a $\togenlu$-derivation $\deriv$ is implementable on RAM in $O((1+\size\deriv)\cdot \size\tm)$  steps.
  \item \emph{Low + High Bilinear Implementation}: a $\tof$-derivation $\deriv$ is implementable on RAM in $O((1+\size\deriv)\cdot \size\tm)$ steps.
 \end{enumerate}
\end{theorem}

%
%

\begin{proofatend}
the proof follows from
\refthm{hlithm} applied to \refthm{lhli},
and \refth{lowlevelimplth} applied to \refth{lazy-uglam-refl-distillation} and \refcoro{useful-bound1}.
For the implementability of the steps we refer to the proof of \refth{preusefulimpl}.
\end{proofatend}


Let us conclude with a  remark. For our results to hold, the output of the computation has to be given in compact form, \ie\ with ES. The unfolding a term $\tm$ with ES may have size exponential in the size of $\tm$. It is important to show, then, that the common operations on $\lambda$-terms, and in particular equality checking (up to $\alpha$-conversion), can be implemented efficiently on the shared representation, avoiding unfolding. In other words, we want to prove that ES are \emph{succinct data structures}, in the sense of Jacobson~\cite{Jacobson:1988:SSD:915547}.

Despite quadratic and quasi-linear recent algorithms \cite{DBLP:conf/rta/AccattoliL12,DBLP:conf/icfp/GrabmayerR14} for testing equality of terms with ES, we discovered that a linear algorithm can be obtained slightly
modifying an algorithm already known quite some time before (1976!): the Paterson-Wegman linear unification algorithm~\cite{Paterson:1976:LU:800113.803646}
(better explained in~\cite{Champeaux:1986:PLU:9134.9137}). The algorithm works
on first order terms represented as DAGs, and unification boils down to equality
checking when no metavariable occurs in the involved terms.

To apply the Paterson-Wegmar algorithm, we need to overcome two difficulties.
The first one is that ES implement sharing explicitly: to represent the term
$\tm\tm$ sharing the two occurrences of $\tm$ we need to introduce a variable
and an ES, obtaining $\var\var\esub\var\tm$. On the contrary, the input to
Paterson-Wegmar should be a DAG where the application node points directly
twice to the root of $\tm$. The required change in representation can be easily
computed in linear time in the size of the input. The second difficulty is that
Paterson-Wegmar works on first-order terms, while we want to consider
$\alpha$-conversion. If we assume that occurrences of $\lambda$-bound variables point to their binder, two variables are $\alpha$-equivalent when they point
to nodes that have already been determined to be candidates for equality.  
The details of the adaptation of Paterson-Wegmar are left to a forthcoming publication.

\ignore{
\claudio{\section{Explicit Substitutions are Succinct}
For our results to hold, the output of the computation has to be given in compact form, \ie\ with ES. The unfolding a term $\tm$ with ES may have size exponential in the size of $\tm$. It is important to show, then, that the common operations on $\lambda$-terms, like equality checking (up to $\alpha$-conversion), can be implemented efficiently on the shared representation, avoiding unfolding. In other words, we want to prove that ES are \emph{succinct data structures}, in the sense of Jacobson~\cite{Jacobson:1988:SSD:915547}.

We only consider here equality checking and we recall that the problem of testing equality of strong normal forms can be reduced to computing weak normal forms and checking equality of prefixes of $\lambda$-terms~\cite{DBLP:conf/icfp/GregoireL02}. The reduction consists in reducing the terms to be compared by levels. At each iteration the two terms are put in weak normal form and the two normal forms are recursively checked for equality until two values or a value and an inert are found. In the latter case, $\eta$-expansion is applied to the inert to reduce the problem to the comparison of values. The latter is solved applying the two values to a fresh symbol and iterating the algorithm.

What is the complexity of testing equality of terms with ES? Despite quadratic and quasi-linear recent algorithms \cite{DBLP:conf/rta/AccattoliL12,DBLP:conf/icfp/GrabmayerR14}, we discovered that a linear algorithm can be obtained slightly
modifying an algorithm already known quite some time before (1976!): the Paterson-Wegman linear unification algorithm~\cite{Paterson:1976:LU:800113.803646}
(better explained in~\cite{Champeaux:1986:PLU:9134.9137}) works on
first order terms represented as DAGs, and unification boils down to equality
checking when no metavariable occurs in the involved terms.

To apply the Paterson-Wegmar algorithm, we need to overcome two difficulties.
The first one is that ES implement sharing explicitly: to represent the term
$\tm\tm$ sharing the two occurrences of $\tm$ we need to introduce a variable
and an ES, obtaining $\var\var\esub\var\tm$. On the contrary, the input to
Paterson-Wegmar should be a DAG where the application node points directly
twice to the root of $\tm$. The required change in representation can be easily
computed in linear time in the size of input. The second difficulty is that
Paterson-Wegmar works on first-order terms, while we want to consider
$\alpha$-conversion. If we assume that abstraction nodes point directly to
the shared node for all the occurrences of the bound variable, we can easily
modify the Paterson-Wegmar algorithm to check for $\alpha$-conversion.
We now sketch both solutions.

\subsection{Change of representation}
To apply the Paterson-Wegmar algorithm, we need to turn the two $\lambda$-terms with ES that we need to compare into first order terms with sharing. The transformation assumes that closed terms with ES are implemented with DAGS as
follows:
\begin{enumerate}
\item A symbol $\const$ is encoded with a node labelled with $\const$ that
 has no outgoing edges.
\item An application $\tm\tmtwo$ is encoded with a node labelled
 with $\appl$ that has two outgoing edges to the graphs that encode $\tm$ and
 $\tmtwo$.
\item An abstraction $\l\var.\tm$ is encoded with a node labelled with
 $\abst$ that has two outgoing edges: the first points to an unlabelled node
 $\node$ with no outgoing edges that encode $\var$; the second points to the graph that
 encodes $\tm$. All the occurrence of $\var$ in $\tm$ are encoded as pointers
 to $\node$.
\item A substitution $\esub\var\tm\tmtwo$ is encoded by the graph that encodes
 $\tmtwo$ and an additional node $\node$ labelled with $\es$ that encodes $\var$ and has an outgoing edge pointing to the encoding of $\tm$. All occurrences of $\var$ in $\tmtwo$ are encoded as pointers to $\node$.
\end{enumerate}

The transformation to change the representation performs a post-visit of
the DAG erasing all substitution nodes marked with $\es$. A substitution node has several incoming edges and exactly one outgoing edge that points to a node
$\node$. The erasure of the node makes all incoming edges point directly to
$\node$.

Note that renaming chains in the initial DAG are unchained by the
translation.

The resulting graph is still a DAG that encodes the initial term
as a first order term with sharing. For example, the term
$\l\var.\l\vartwo.(\var (\varthree\esub\varthree\varfour))\esub\varfour\vartwo$
is transformed in the DAG whose pre-order visit is $\abst(\var,\abst(\vartwo,\appl(\var,\vartwo)))$ (the labels $x$ and $y$ stand here for the corresponding unlabeled symbols).

The transformation takes time linear in the size of the input
because each edge is traversed exactly once and its endpoint can be changed at most once.

Note that, differently from $\l$, $\abst$ is not a binder, but a symbol.

\subsection{Modified Paterson-Wegmar}
The Paterson-Wegmar takes in input two DAGs that represent first order terms,
possibly containing metavariables, and computes their most general unifier.
Because our terms do not contain metavariables, the algorithm will either fail
or succeed returning the trivial unifier.
}}

\section*{Acknowledgements}
A special acknowledgement to Ugo Dal Lago, to whom we owe the intuition that using labels may lead to a local and efficient implementation of useful sharing. We are also grateful to Fran\c{c}ois Pottier, whose comments on a draft helped to improve the terminology and the presentation.
\bibliographystyle{IEEEtran}
\bibliography{IEEEabrv,\macrospath/biblio}

%
%
%

\newpage
\appendices
\ifthenelse{\boolean{withproofs}}{
\printproofs
}{}

\end{document}